\documentclass[11pt]{article}
\usepackage{amssymb}
\usepackage{graphicx}
\usepackage{theorem}
\usepackage{times}

\theoremheaderfont{\scshape}
\theorembodyfont{\upshape}
\newtheorem{axiom}{Axiom}

\newcommand{\be}{\begin{equation}}
\newcommand{\ee}{\end{equation}}
\newcommand{\bea}{\begin{eqnarray}}
\newcommand{\eea}{\end{eqnarray}}
\newcommand{\E}{{\rm E}}
\newcommand{\Cov}{{\rm Cov}}
\newcommand{\Var}{{\rm Var}}

\begin{document}

\title{ Multi-Moments Method for Portfolio
      Management:\\ Generalized Capital Asset Pricing Model\\ in
      Homogeneous and Heterogeneous markets
      \footnote{We acknowledge helpful discussions and
exchanges with J.V. Andersen, J.P. Laurent and V. Pisarenko. We are grateful to
participants of the workshop on ``Multi-moment Capital Asset Pricing
Models and Related Topics'', ESCP-EAP European School of Management, Paris,
April,19, 2002, and in particular to Philippe Spieser, for their comments.
This work was partially supported by
the James S. Mc Donnell Foundation 21st century scientist award/studying
complex system.}}
\author{\bf Y. Malevergne$^{1,2}$ and D. Sornette$^{1,3}$\\
$^1$ Laboratoire de Physique de la Mati\`{e}re Condens\'{e}e\\ CNRS UMR6622 and
Universit\'{e} de Nice-Sophia Antipolis\\ Parc
Valrose, 06108 Nice Cedex 2, France \\
$^2$ Institut de Science Financi\`ere et d'Assurances - Universit\'e Lyon I\\
43, Bd du 11 Novembre 1918, 69622 Villeurbanne Cedex\\
$^3$ Institute of Geophysics and
Planetary Physics and Department of Earth and Space Science\\
University of California, Los Angeles, California 90095\\
e-mails: Yannick.Malevergne@unice.fr and sornette@unice.fr
}

\maketitle

\begin{abstract}
\noindent
We introduce a new set of consistent measures of risks, in terms of the
semi-invariants
of pdf's, such that the centered moments and the cumulants of the portfolio
distribution of returns that put more emphasis on the tail the distributions.
We derive generalized efficient frontiers, based
on these novel measures of risks and present
the generalized CAPM,
both in the cases of homogeneous and heterogeneous markets. Then,
using a family of modified Weibull distributions,
encompassing both sub-exponentials and super-exponentials, to parameterize
the marginal distributions of asset returns and their natural multivariate
generalizations, we offer exact formulas for the
moments and cumulants of the
distribution of returns of a portfolio made of an arbitrary 
composition of these
assets. Using combinatorial and hypergeometric functions,
we are in particular able to extend previous results to the case
where the exponents of the Weibull distributions are different from
asset to asset and
in the presence of dependence between assets. In this parameterization, we
treat in details the problem of risk minimization using the cumulants as
measures of risks for a portfolio made of two assets and compare the
theoretical predictions with direct empirical data. Our extended formulas
enable us to determine analytically the conditions under which
it is possible to ``have your cake and eat it too'', i.e.,
to construct a portfolio with both larger return and smaller ``large risks''.

\end{abstract}

\section{Introduction}

The Capital Asset Pricing Model (CAPM) is still the most widely used approach
to
relative asset evaluation, although its empirical roots are been found weaker
and weaker
in recent years. This asset
valuation model describing the relationship between expected risk and expected
return for marketable assets is strongly entangled with the Mean-Variance
Portfolio Model. Indeed both of them fundamentally rely on the description
of the probability distribution function (pdf) of asset returns in terms of
Gaussian functions. The Mean-Variance description is thus at the basis of
Markovitz's portfolio theory \cite{Markovitz} and of the CAPM (see
for instance \cite{Merton}).

Otherwise, the determination of the risks and returns associated with
a given
portfolio constituted
of $N$ assets is completely embedded in the knowledge of their
multivariate distribution
of returns. Indeed,
the dependence between random variables is completely described by
their joint distribution.
This remark entails the two major problems of portfolio theory: 1) determine
the multivariate distribution function of asset returns; 2) derive
from it useful measures of
portfolio risks and use them to analyze and optimize portfolios.

The variance (or volatility) of  portfolio returns provides the simplest way
to quantify its fluctuations and is at the fundation of the \cite{Markovitz}'s
portfolio selection theory.  Nonetheless, the variance of a portfolio offers
only a limited quantification of incurred risks (in terms of fluctuations), as
the empirical distributions of returns have ``fat tails''
\cite[among many others]{Lux,Gopikrishnan} and the dependences between assets
are only imperfectly accounted for by the covariance matrix \cite{Litterman}.
It is thus essential to extend portfolio theory and the CAPM to tackle these
empirical facts.

The Value-at-Risk \cite{Jorion} and many other measures of risks
\cite{A_etal98,Sornettepre,A_etal99,Bouchaudetal,Sornette1} have then been
developed to account for the larger moves allowed by non-Gaussian distributions
and non-linear correlations but they mainly allow for the assessment of
down-side risks. Here, we consider both-side risk and define general
measures of fluctuations. It is the first goal of this article. Indeed,
considering the minimum set of properties a fluctuation measure must fulfil, we
characterize these measures. In particular, we show that any absolute central
moments and some cumulants satisfy these requirement as well as do
any combination
of these quantities. Moreover, the weights involved in these combinations can
be interpreted in terms of the portfolio
manager's aversion against large fluctuations.

Once the definition of the fluctuation measures have been set, it is possible
to classify the assets and portfolios using for instance a risk adjustment
method \cite{S94, Dowd00} and to develop a portfolio selection and
optimization approach. It is the second goal of this article.

Then a new model of market equilibrium can be derived, which generalizes the
usual Capital Asset Pricing Model (CAPM). This is the third goal of
our paper. This improvement is necessary since, although the use of
the CAPM is still widely spread, its empirical justification has been found
less and less convincing in the past years \cite{Lim89, HS00}.

The last goal of this article is to present an efficient parametric method
allowing for the estimation of the centered moments and cumulants, based upon a
maximum entropy principle. This parameterization of the problem is necessary in
order to obtain accurate estimates of the high order moment-based quantities
involved the portfolio optimization problem with our generalized measures of
fluctuations.

The paper is organized as follows.

Section 2 presents a new set of consistent measures of risks, in terms of the
semi-invariants  of pdf's, such as the centered moments and the cumulants of
the portfolio distribution of returns, for example.

Section 3 derives the generalized efficient frontiers, based
on these novel measures of risks. Both cases with and without risk-free asset
are analyzed.

Section 4 offers a generalization of the Sharpe ratio and thus provides new
tools to classify assets with respect to their risk adjusted performance. In
particular, we show that this classification may depend on the choosen risk
measure.

Section 5 presents the generalized CAPM based on these new measures of risks,
both in the cases of homogeneous and heterogeneous agents.

Section 6 introduces a novel general parameterization of the multivariate
distribution of
returns based on two steps: (i) the projection of the empirical
marginal distributions
onto Gaussian laws via nonlinear mappings; (ii) the use of an entropy
maximization
to construct the corresponding most parsimonious
representation of the multivariate distribution.

Section 7 offers a specific parameterization
of marginal distributions in terms of so-called modified Weibull
distributions, which are
essentially exponential of minus a power law.
Notwithstanding their possible fat-tail nature, all their moments and cumulants
are finite and can be calculated.
We present empirical calibration of the two key parameters of the
modified Weibull
distribution, namely the exponent $c$ and the characteristic scale $\chi$.

Section 8 provides the analytical expressions of the cumulants of the
distribution of
portfolio returns for the parameterization
of marginal distributions in terms of so-called modified Weibull
distributions, introduced in section 6.
Empirical tests comparing the direct numerical evaluation of
the cumulants
of financial time series to the values predicted from our analytical formulas
find a good consistency.

Section 9 uses these two sets of results to illustrate how portfolio
optimization works in this context. The main novel result
is an analytical understanding of the conditions under which it is possible to
simultaneously increase the portfolio return and decreases its large
risks quantified
by large-order cumulants. It thus appears that the multidimensional
nature of risks
allows one to break the stalemate of no better return without more risks, for
some special kind of rational agents.

Section 10 concludes.

Before proceeding with the presentation of our results, we set the notations to
derive the basic problem addressed in this paper, namely to study the
distribution
of the sum of weighted random variables with arbitrary marginal
distributions and
dependence.  Consider a portfolio with $n_i$ shares of asset
$i$ of price $p_i(0)$ at time $t=0$ whose initial wealth is
\be
W(0) = \sum_{i=1}^N n_i p_i(0)~.
\ee
A time $\tau$ later, the wealth has become $W(\tau) = \sum_{i=1}^N n_i
p_i(\tau)$ and the wealth variation is
\be
\delta_{\tau} W \equiv W(\tau) -W(0) = \sum_{i=1}^N n_i p_i(0)
{{p_i(\tau) - p_i(0)} \over p_i(0)}
= W(0) ~\sum_{i=1}^N w_i r_i(t,\tau) ,
\ee
where
\be
w_i = {n_i p_i(0)  \over \sum_{j=1}^{N} n_j p_j(0)}~
\ee
is the fraction in capital invested in the $i$th asset at time $0$ and
the return
$r_i(t,\tau)$ between time $t-\tau$ and $t$ of asset $i$ is defined as:
\be
   r_i(t,\tau) = {p_i(t)-p_i(t-\tau) \over p_i(t-\tau)} ~.
   \label{jhghss}
\ee
Using the definition (\ref{jhghss}),
this justifies us to write the return $S_{\tau}$ of the portfolio over a time
interval $\tau$ as the weighted sum of the returns $r_i(\tau)$ of the assets
$i=1,...,N$ over the time interval $\tau$
\be
S_{\tau} = {\delta_{\tau} W \over W(0)} = \sum_{i=1}^N w_i ~r_i(\tau)~.
\label{jjkmmq}
\ee
In the sequel, we shall thus consider asset returns as the
fundamental variables
(denoted $x_i$ or $X_i$ in the sequel)
and study their aggregation properties, namely how the distribution
of portfolio
return equal to their weighted sum derives for their multivariable
distribution.
We shall consider a single time scale $\tau$ which can be chosen arbitrarily,
say equal to one day. We shall thus drop the dependence on $\tau$,
understanding
implicitely that all our results hold for returns estimated over the time
step $\tau$.

\section{Measuring large risks of a portfolio}
\label{sec:cum}

The question on how to assess risk is recurrent in
finance (and in many other fields) and
has not yet received a general solution. Since the middle of the
twentieth century, several paths have been explored. The pioneering
work by \cite{VNM44} has given birth to the mathematical definition of
the expected utility function which provides interesting insights on
the behavior of a rational economic agent and formalized the concept
of risk aversion. Based upon the properties of the utility function,
\cite{RS70} and \cite{RS71} have attempted to define the notion of increasing
risks. But, as revealed by \cite{A53,A90}, empiric investigations
has proven that the postulates chosen by \cite{VNM44} are actually
often violated. Many generalizations have been proposed for curing the
so-called Allais' Paradox, but up to now, no generally accepted procedure
has been found in this way.

Recently, a theory due to \cite{A_etal98, A_etal99} and its generalization
by \\
\cite{FS02a, FS02b}, have appeared. Based on a series of postulates that
are quite natural, this theory
allows one to build  coherent (convex) measures of risks. In fact, this theory
seems well-adapted to the assessment of the needed economic capital, that is,
of the fraction of capital a company must keep as risk-free assets in order to
face its commitments and thus avoid ruin. However, for the purpose of
quantifying the fluctuations of the asset returns and of developing a theory of
portfolios, this approach does not seem to be operational.
Here, we shall rather revisit  \cite{Markovitz}'s approach to investigate how
its extension to higher-order moments or cumulants, and any combination
of these quantities, can be used operationally to account for large risks.

\subsection{Why do higher moments allow to assess larger risks?}

In principle, the complete description of the fluctuations of an asset
at a given time scale is given by the knowledge of the
probability distribution function (pdf) of its returns. The pdf
encompasses all the risk dimensions
associated with this asset. Unfortunately, it is impossible to classify or
order
the risks described by the entire pdf, except in special cases where the
concept of stochastic dominance applies. Therefore, the
whole pdf can not provide an adequate measure of risk, embodied
by a single variable. In order to perform a selection among
a basket of assets and construct optimal portfolios, one needs
measures given as real numbers, not functions, which can be ordered
according to the natural ordering of real numbers on the line.

In this vein,
\cite{Markovitz} has proposed to summarize the risk of an asset by
the variance of its pdf of returns
(or equivalently by the corresponding standard deviation). It is clear that
this
description of risks is fully satisfying only for
assets with Gaussian pdf's. In any other case, the variance generally
provides a very poor estimate of the real risk. Indeed, it is a
well-established empirical fact that the pdf's of asset returns has fat tails
\cite{Lux, P96, Gopikrishnan}, so that the Gaussian approximation
underestimates significantly the large prices movements frequently observed on
stock markets. Consequently, the variance can not be taken as a suitable
measure of risks, since it only accounts for the smallest contributions
to the fluctuations of the assets returns.

\begin{figure}[!h]
\begin{center}
\includegraphics[width=15cm]{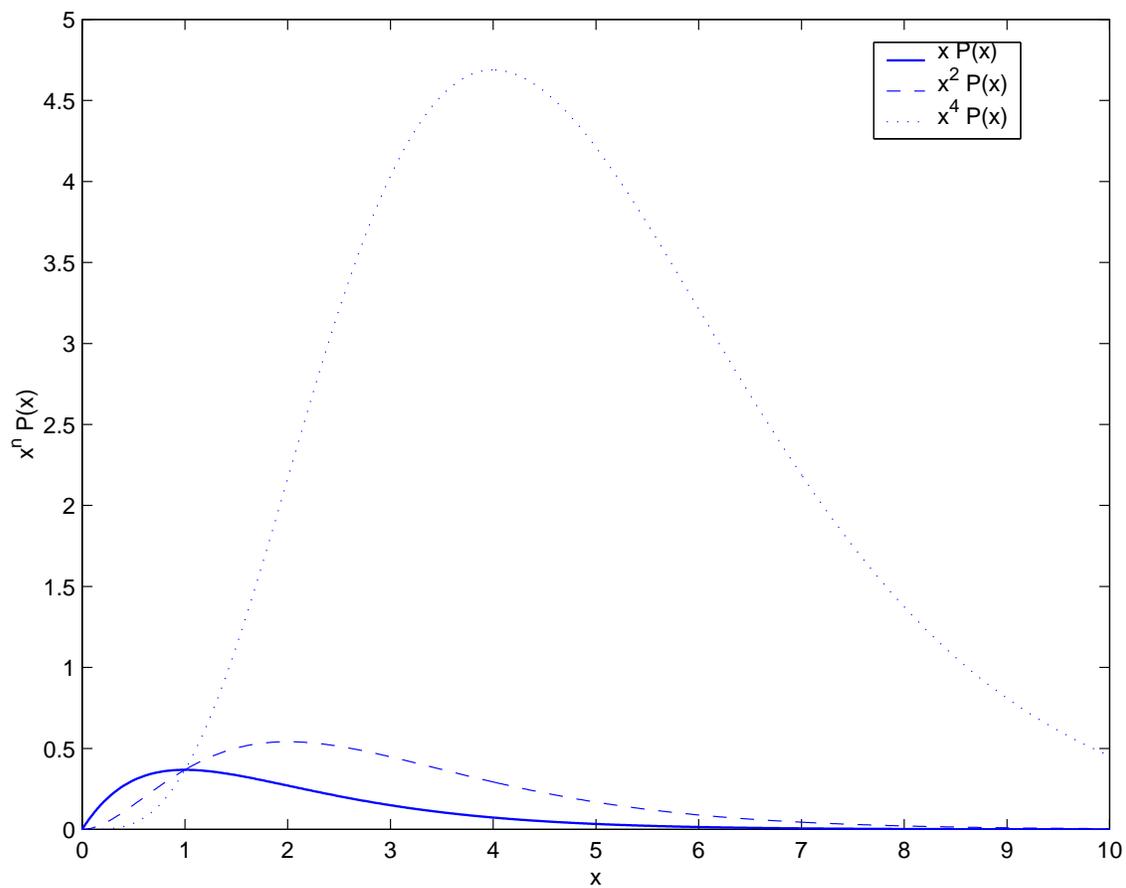}
\end{center}
\caption{\label{fig:mom} This figure represents the function $x^n \cdot e^{-x}$
for $n=1,2$ and $4$. It shows the typycal size of the fluctuations involved
in the moment of order $n$.}
\end{figure}

The variance of the return $X$ of an asset involves its second
moment $\E[X^2]$ and, more precisely, is equal to its second centered moment
(or moment about the mean) $\E\left[ \left( X - \E[X] \right)^2
\right]$. Thus, the weight of a given fluctuation $X$
entering in the definition of the variance of the returns
is proportional to its square. Due to the decay of the
pdf  of $X$ for large $X$ bounded from above by $\sim 1/|X|^{1+\alpha}$
with $\alpha>2$, the largest fluctuations do not
contribute significantly to this expectation. To increase their
contributions, and in this way
to account for the largest fluctuations, it is natural to invoke
higher order moments of order $n>2$. The large $n$ is, the larger is
the contribution of the rare and large returns in the tail of the pdf.
This phenomenon is demonstrated
in figure \ref{fig:mom}, where we can observe the evolution of the
quantity $x^n \cdot P(x)$ for $n=1,2$ and $4$, where $P(x)$, in this example,
is the standard exponential distribution $e^{-x}$. The expectation $\E[X^n]$
is then simply represented geometrically as equal
to the area below the curve $x^n \cdot P(x)$. These curves provide
an intuitive illustration of the fact that
the main contributions to the moment $\E[X^n]$ of order $n$ come
from values of $X$ in the vicinity of the maximum of $x^n \cdot P(x)$
which increases fast with
the order $n$ of the moment we consider, all the more so, the fatter is
the tail of the pdf of the returns $X$.
For the exponential distribution chosen to construct figure \ref{fig:mom},
the value of $x$ corresponding to the maximum of $x^n \cdot P(x)$
is exactly equal to $n$. Thus, increasing the order of the moment
allows one to sample larger fluctuations of the asset prices.

\subsection{Quantifying the fluctuations of an asset}

Let us now examine what should be the properties that coherent measures
of risks adapted to the portfolio problem
must satisfy in order to best quantify the asset price
fluctuations. Let us consider an asset denoted $X$, and let ${\cal G}$ be the
set of all the risky assets available on the market. Its profit and loss
distribution is the distribution of $\delta X=X(\tau)-X(0)$, while the return
distribution is given by the distribution of $\frac{X(\tau)-X(0)}{X(0)}$.
The risk measures will be defined for the profit and loss
distributions and then shown to be equivalent to another definition
applied to the return distribution.

Our first
requirement is that the risk measure $\rho(\cdot)$, which is a functional on
${\cal G}$, should always remain positive
\begin{axiom}
\hspace{5cm}
$
\forall X \in {\cal G},~~~~~~~\rho(\delta X) \ge 0~,
\label{mngjkwlw}
$
\end{axiom}
where the equality holds if and only if $X$ is certain. Let
us now add to this asset a given amount $a$ invested in the
risk free-asset whose return is $\mu_0$ (with therefore no randomness in its
price trajectory) and define the new asset $Y=X+a$. Since $a$ is
non-random, the fluctuations of $X$ and $Y$ are the same. Thus, it is desirable
that $\rho$ enjoys the property of {\it translational invariance}, whatever the
asset $X$ and the non-random coefficient $a$ may be:
\begin{axiom}
\hspace{4cm}
$\forall X \in {\cal G},~ \forall a \in {\mathbb R},~~~~~~~~~~~  \rho(\delta X
+\mu \cdot a)= \rho(\delta X).$
\end{axiom}

We also require that our risk measure increases with the quantity  of
assets held in the portfolio. {\it A priori}, one should expect that the
risk of a position is proportional to its size. Indeed, the fluctuations
associated with the variable $2 \cdot X$ are naturally twice larger as the
fluctuations of $X$. This is true as long as we can consider that a large
position can be liquidated as easily as a smaller one. This is obviously not
true, due to the limited liquidity of real markets. Thus, a large
position in a given asset is more risky than the sum of the
risks associated with the many smaller positions which add up
to the large position. To account for this point, we assume that $\rho$
depends on the size of the position in the same manner for all assets.
This assumption is slightly
restrictive but not unrealistic for companies with comparable properties
in terms of market capitalization or sector of activity. This
requirement reads
\begin{axiom}
\hspace{4cm}
$ \forall X \in {\cal G},~ \forall \lambda \in {\mathbb R}_+,~~~~~~~~~~~
\rho(\lambda \cdot \delta X) = f(\lambda) \cdot \rho(\delta X),
$
\end{axiom}
where the function $f :
{\mathbb R}_+ \longrightarrow {\mathbb R}_+$ is increasing and convex
to account for liquidity risk. In fact, it is straightforward to show
\footnote{using the trick $\rho(\lambda_1 \lambda_2 \cdot \delta X)
= f(\lambda_1) \cdot \rho(\lambda_2 \cdot \delta X) =
f(\lambda_1) \cdot f(\lambda_2) \cdot \rho(\delta X)
= f(\lambda_1 \cdot \lambda_2) \cdot \rho(\delta X)$ leading
to $f(\lambda_1 \cdot \lambda_2) = f(\lambda_1) \cdot f(\lambda_2)$. The
unique increasing convex solution of this functional equation is
$f_\alpha(\lambda)=\lambda^\alpha$ with $\alpha \ge 1$.}
that the only functions statistying this axiom are the fonctions
$f_\alpha(\lambda)=\lambda^\alpha$ with $\alpha \ge 1$, so that axiom 3 can be
reformulated in terms of positive homogeneity of degree $\alpha$:
\begin{axiom}
\hspace{4cm}
\be
  \forall X \in {\cal G},~ \forall \lambda \in {\mathbb R}_+,~~~~~~~~~~~
\rho(\lambda \cdot \delta X) = \lambda^\alpha \cdot \rho(\delta X).
\label{mgmld}
\ee
\end{axiom}
Note that the case of liquid markets is recovered by $\alpha=1$ for which the
risk is directly proportionnal to the size of the position.

These axioms, which define our risk measures for profit and
loss can easily be extended to the returns of the assets.
Indeed, the return is nothing but the profit and loss divided by the
initial value $X(0)$ of the asset. One can thus easily check that the risk
defined on the profit and loss distribution is $X(0)^\alpha$ times the risk
defined on the return distribution. In the sequel, we will only consider this
later definition, and, to simplify the notations since we will only 
consider the returns and not the profit and loss, the notation $X$ will be used
to denote the asset and its return as well.

We can remark that the risk measures $\rho$ enjoying the two properties
defined by the axioms 2 and 4 are known as the {\it semi-invariants} of
the distribution of the profit and loss / returns of $X$ (see \cite[p
86-87]{SO}). Among the large familly of semi-invariants, we can cite the
well-known centered moments and cumulants of $X$.

\subsection{Examples}

The set of risk measures obeying axioms 1-4 is huge since
it includes all the homogeneous functionals of $(X-\E[X])$, for instance.
The centered moments (or moments about the mean) and the cumulants are two
well-known classes of semi-invariants. Then, a given value of $\alpha$ can be
seen as nothing but a specific choice of the order $n$ of the centered moments
or of the cumulants. In this case, our risk measure defined via these
semi-invariants fulfills the two following conditions:
\bea
\rho(X +\mu)&=& \rho(X),\\
\rho(\lambda \cdot X) &=& \lambda^n \cdot \rho(X).
\eea

In order to satisfy the positivity condition (axiom \ref{mngjkwlw}), we need to
restrict the
set of values taken by $n$. By construction, the centered
moments of even order are always positive while the odd order
centered moments can be negative. Thus, only the even order centered
moments are acceptable risk measures. The situation is not so clear for
the cumulants, since the even order cumulants, as well as the odd
order ones, can be negative. In full generality, only the centered moments
provide reasonable risk measures satifying our
axioms. However, for a large class of distributions,
even order cumulants remain positive, especially for
fat tail distributions (eventhough there are simple but
somewhat artificial counter-examples). Therefore, cumulants of even order
can be useful risk measures when restricted to these distributions.

Indeed, the cumulants enjoy a property which can be considered as a
natural requirement for a risk measure. It can be desirable that
the risk associated with a portfolio made of independent assets is
exactly the sum of the risk associated with each individual
asset. Thus, given $N$ independent
assets $\{ X_1, \cdots, X_N \}$, and the portfolio $S_N = X_1 + \cdots
+ X_N$, we wish to have
\be
\rho_n(S_N) = \rho_n(X_1) + \cdot + \rho_n(X_N)~.  \label{nbnmk}
\ee
This property is verified for all cumulants while is not true
for centered moments. In addition, as seen from their definition
in terms of the characteristic function
(\ref{ghlqjfjgf}), cumulants of order larger than $2$ quantify
deviation from the
Gaussian law, and thus large risks beyond the variance (equal to the
second-order cumulant).

Thus, centered moments of even orders possess all the minimal
properties required for a suitable portfolio risk measure.
Cumulants fulfill these requirement only for well
behaved distributions, but have an additional advantage compared to the
centered moments, that is, they fulfill the condition (\ref{nbnmk}).
For these reasons, we shall consider below both the centered moments and the
cumulants.

In fact, we can be more general. Indeed, as we have written, the
centered moments or the cumulants of order $n$ are homogeneous functions of
order $n$, and due to the positivity requirement, we have to restrict
ourselves to even order centered moments and cumulants. Thus, only homogeneous
functions of order $2n$ can be considered.
Actually, this restrictive constraint can be relaxed by recalling that,
given any homogeneous function $f(\cdot)$ of order $p$, the
function $f(\cdot)^q$ is also homogeneous of order $p \cdot q$.
This allows us to decouple the order of the
moments to consider, which quantifies the impact of the large fluctuations,
from the influence of the size of the positions held, measured by the degres of
homogeneity of $\rho$.  Thus, considering any even order centered moments, we
can build a risk measure $\rho(X) = \E \left[ (X- \E[X])^{2n}
\right]^{\alpha/2n}$ which account for the fluctuations measured by 
the centered
moment of order $2n$ but with a degree of homogeneity equal to $\alpha$.

A further generalization is possible to odd-order moments. Indeed,
the {\it absolute} centered moments satisfy our three
axioms for any odd or even order. We can go one step further and
use non-integer order absolute centered moments, and
define the more general risk measure
\be
\rho(X) = \E \left[ |X- \E[X]|^{\gamma} \right]^{\alpha/\gamma},
\ee
where $\gamma$ denotes any positve real number.

These set of risk measures are very interesting since, due to the Minkowsky
inegality, they are convex for any $\alpha$ and $\gamma$ larger than 1~:
\be
\rho(u \cdot X + (1-u) \cdot Y) \le u \cdot \rho(X) + (1-u) \cdot \rho(Y),
\ee
which ensures that aggregating two risky assets lead to diversify their risk.
In fact, in the special case $\gamma=1$, these measures enjoy the stronger
sub-additivity property.

Finally, we should stress that any discrete or continuous (positive) 
sum of these
risk measures, with the same degree of homogeneity is again a risk 
measure. This
allows us to define ``spectral measures of fluctuations'' in the same 
spirit as in
\cite{Acerbi02}:
\be
\label{eq:spf}
\rho(X) = \int d\gamma ~ \phi (\gamma) ~  \E \left[ (X- \E[X])^{\gamma}
\right]^{\alpha/\gamma},
\ee
where $\phi$ is a positive real valued function defined on any subinterval of
$[1, \infty)$ such that the integral in (\ref{eq:spf}) remains finite. It
is interesting to restrict oneself to the functions $\phi$ whose integral sums
up to one: $ \int d\gamma ~ \phi(\gamma) = 1$, which is always possible, up to
a renormalization. Indeed, in such a case, $\phi(\gamma)$ represents the
relative weight attributed to the fluctuations measured by a given
moment order. Thus, the function $\phi$ can be
considered as a measure of the risk aversion of the risk manager with respect
to the large fluctuations.

Let us stress that the variance, originally used in
\cite{Markovitz}'s portfolio theory, is nothing but the second
centered moment, also equal to the second order cumulant (the three first
cumulants and centered moments are equal). Therefore,
a portfolio theory based on the centered
moments or on the cumulants automatically contain
\cite{Markovitz}'s theory as a special case,  and thus offers a natural
generalization emcompassing large risks of this masterpiece of the financial
science. It also embodies several other generalizations where homogeneous
measures of risks are considered, a for instance in \cite{HS99}.

\section{The generalized efficient frontier and some of its properties}

We now address the problem of the portfolio selection and optimization, based
on the risk measures introduced in the previous section. As we have already
seen, there is a large choice of relevant risk measures from which 
the portfolio
manager is free to choose as a function of his
own aversion to small versus large risks. A strong risk aversion to large risks
will lead him to choose a risk measure which puts the emphasis on the large
fluctuations. The simplest examples of such risk measures are provided by the
high-order centered moments or cumulants. Obviously, the utility
function of the fund manager plays a  central role in his choice of the risk
measure. The relation between the central moments and the utility function has
already been underlined by several authors such as \cite{R73} or \cite{JM02},
who have shown that an economic agent with a quartic utility function is
naturally sensitive to the first four moments of his expected wealth
distribution. But, as stressed before, we do not wish to consider the
expected utility formalism since our goal, in this paper, is not to study the
underlying behavior leading to the choice of any risk measure.

The choice of the risk measure also depends upon the time horizon of
investment. Indeed, as the time scale increases, the distribution of asset
returns progressively converges to the Gaussian pdf, so that only the variance
remains relevant for very long term investment horizons.  However, for shorter
time horizons, say, for portfolio rebalanced at a weekly, daily or intra-day
time scales, choosing a risk measure putting the emphasis on the large
fluctuations, such as the centered moments $\mu_6$ or $ \mu_8$ or the cumulants
$C_6$ or $C_8$ (or of larger orders), may be necessary to account for the
``wild'' price fluctuations usually observed for such short time scales.

Our present approach uses a single time scale over which the returns are
estimated, and is thus restricted to portfolio selection with a fixed
investment horizon.  Extensions to a portofolio analysis and optimization in
terms of high-order moments and cumulants performed simultaneously over
different time scales can be found in \cite{MuzyQF}.

\subsection{Efficient frontier without risk-free asset}

Let us consider $N$ risky assets, denoted by $X_1, \cdots, X_N$. Our goal is to
find the best possible allocation, given a set of constraints.The portfolio
optimization generalizing the approach of \cite{SorAnderSim2,AndersenRisk}
corresponds to accounting for large fluctuations of the assets through the risk
measures introduced above in the presence of a constraint on the return as well
as the ``no-short sells'' constraint:
\be
\label{eq:opt0}
\left\{
\begin{array}{l}
\inf_{w_i\in[0,1]} \rho_\alpha(\{w_i\})\\
\sum_{i \ge 1} w_i = 1 \\
\sum_{i \ge 1} w_i \mu(i) = \mu~, \\
w_i \ge 0,~~~~ \forall i>0,
\end{array}
\right.
\ee
where $w_i$ is the weight of $X_i$ and $\mu(i)$ its expected return. In all
the sequel, the subscript $\alpha$ in $\rho_\alpha$ will refer to the degree of
homogeneity of the risk measure.

This problem cannot be solved analytically (except in the Markovitz's case
where the risk measure is given by the variance). We need to perform numerical
calculations to obtain the shape of the efficient frontier. Nonetheless, when
the $\rho_\alpha$'s denotes the centered moments or any convex risk measure, we
can assert that this optimization problem is a convex optimization problem and
that it admits one and only one solution which can be easily determined
by standard numerical relaxation or gradient methods.

As an example, we have represented In figure \ref{fig:GEF}, the
mean-$\rho_\alpha$ efficient frontier for a portfolio made of seventeen assets
(see appendix \ref{app:DS} for details) in the plane ($\mu, \rho_\alpha
^{1/\alpha}$), where $\rho_\alpha$ represents the centered moments 
$\mu_{n=\alpha}$ of
order $n=2, 4, 6$ and $8$. The efficient frontier is concave, as expected from
the nature of the optimization problem (\ref{eq:opt0}). For a given value of
the expected return $\mu$,  we observe that  the amount of risk measured by
$\mu_n^{1/n}$ increases with $n$, so that there is an additional price to pay
for earning more: not only the $\mu_2$-risk increases, as usual according to
Markowitz's theory, but the large risks increases faster, the more so, the
larger $n$ is. This means that, in this example, the large risks increases more
rapidly than the small risks, as the required return increases. This is an
important empirical result that has obvious implications for portfolio
selection and risk assessment. For instance, let us consider an efficient
portfolio whose expected (daily) return equals 0.12\%, which gives an
annualized return equal to 30\%. We can see in table \ref{table:EF} that the
typical fluctuations around the expected return are about twice larger when
measured by $\mu_6$ compared with $\mu_2$ and that they are 1.5 larger when
measured with $\mu_8$ compared with $\mu_4$.

\subsection{Efficient frontier with a risk-free asset}

Let us now assume the existence of a risk-free asset $X_0$. The optimization
problem with the same set of constraints as previoulsy can be written as:
\be
\label{eq:opt}
\left\{
\begin{array}{l}
\inf_{w_i\in[0,1]} \rho_\alpha(\{w_i\})\\
\sum_{i \ge 0} w_i = 1 \\
\sum_{i \ge 0} w_i \mu(i) = \mu~, \\
w_i \ge 0,~~~~ \forall i>0,
\end{array}
\right.
\ee

This optimization problem can be solved exactly. Indeed, due to
existence of a risk-free asset, the normalization condition $\sum w_i
= 1$ is not-constraining since one can always adjust, by lending or
borrowing money, the fraction $w_0$ to a value satisfying the
normalization condition. Thus, as shown in appendix \ref{app:GEF},
the efficient frontier is a straight line in the plane $(\mu,
{\rho_\alpha}^{1/\alpha})$, with positive slope and whose intercept is given by
the value of the risk-free interest rate:
\be
\mu = \mu_0 + \xi \cdot {\rho_\alpha}^{1/\alpha}~,
\ee
where $\xi$ is a coefficient given explicitely below. This
result is very natural when $\rho_\alpha$ denotes the variance, since it is
then nothing but \cite{Markovitz}'s result. But in addition, it shows
that the mean-variance result can be generalized to every mean-$\rho_\alpha$
optimal portfolios.

We present in figure \ref{fig:GEF_RF} the results given by numerical
simulations. The set of assets is the same as before and the
risk-free interest rate has been set to $5\%$ a year. The optimization
procedure has been performed using a genetic algorithm on
the risk measure given by the centered moments $\mu_2, \mu_4, \mu_6$
and $\mu_8$.
As expected, we observe three increasing straight lines, whose slopes
monotonically decay with the order of the centered moment under
consideration. Below, we will discuss this property in greater detail.

\subsection{Two funds separation theorem}

The two funds separation theorem is a well-known result associated
with the mean-variance efficient portfolios. It results from the
concavity of the Markovitz's efficient frontier for portfolios made of
risky assets only. It states that, if the investors can choose between a
set of risky assets and a risk-free asset, they invest a fraction
$w_0$ of their wealth in the risk-free asset and the fraction $1-w_0$ in a
portfolio composed only with risky assets. This risky portofolio is
the same for all the investors and the fraction $w_0$ of wealth
invested in the risk-free asset depends on the risk aversion of the
investor or on the amount of economic capital an institution must 
keep aside due
to the legal requirements insuring its solvency at a given confidence level. We
shall see that this result can be generalized to any mean-$\rho_\alpha$
efficient portfolio.

Indeed, it can be shown (see appendix \ref{app:GEF}) that the weights of the
optimal portfolios that are solutions of (\ref{eq:opt}) are given by:
\bea
\label{eq:w0}
w_0^* &=& w_0,\\
w_i^* &=& (1-w_0)\cdot \tilde w_i, ~~~~~ i \ge 1,
\label{eq:wi}
\eea
where the $\tilde w_i$'s are constants such that $\sum \tilde w_i =
1$ and whose expressions are given appendix \ref{app:GEF}. Thus, denoting
by $\Pi$ the portfolio only made of risky assets whose weights are
the $\tilde w_i$'s, the optimal portfolios are the linear
combination of the risk-free asset, with weight $w_0$, and of the
portfolio $\Pi$, with weigth $1-w_0$. This result generalizes the
mean-variance two fund theorem to any mean-$\rho_\alpha$ efficient portfolio.

To check numerically this prediction, figure
\ref{fig:twofunds} represents the five largest weights of assets
in the portfolios
previously investigated as a function of the weight of the risk-free
asset, for the four risk measures given by the centered moments
$\mu_2, \mu_4, \mu_6$ and $\mu_8$.
One can observe decaying straight lines that intercept the
horizontal axis at $w_0=1$, as predicted by equations
(\ref{eq:w0}-\ref{eq:wi}).

In figure \ref{fig:GEF}, the straight lines representing
the efficient portfolios with a risk-free asset are also represented. They are
tangent
to the efficient frontiers without risk-free asset. This is natural since the
efficient portfolios with the risk-free asset are the weighted sum of the
risk-free asset and the optimal portfolio $\Pi$ only made of risky assets.
Since $\Pi$ also belongs to the efficient frontier without risk-free asset, the
optimum is reached when the straight line describing the efficient frontier
with a risk-free asset and the (concave) curve of the efficient frontier
without risk-free asset are tangent.

\subsection{Influence of the risk-free interest rate}

Figure \ref{fig:GEF_RF} has shown that the slope of the efficient
frontier (with a risk-free asset) decreases when the order $n$ of the centered
moment used to measure risks increases. This is an important qualitative
properties of the risk measures offered by the centered moments, as this means
that higher and higher large risks are sampled under increasing imposed return.

\begin{figure}[!h]
\begin{center}
\includegraphics[width=15cm]{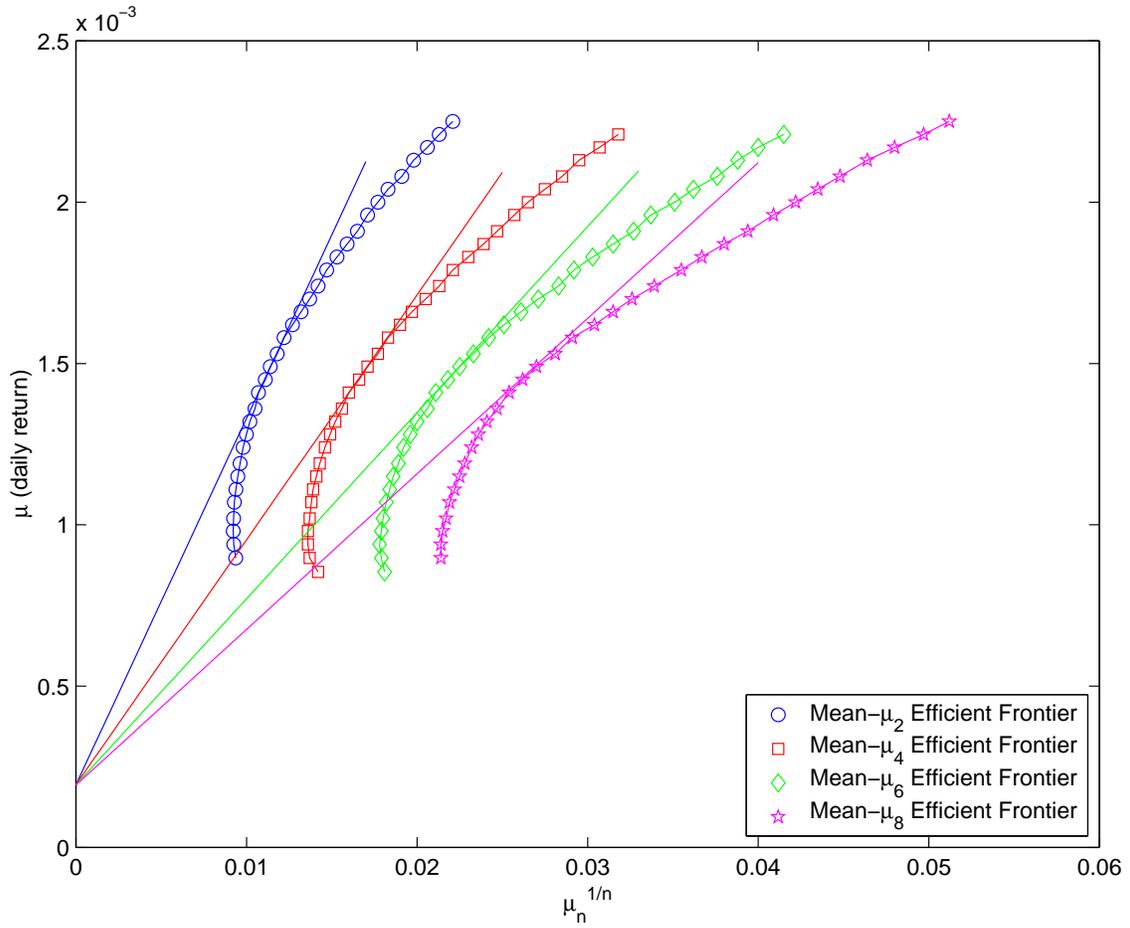}
\end{center}
\caption{\label{fig:GEF} This figure represents the generalized
    efficient frontier for a portfolio made of seventeen risky assets.  The
    optimization problem is solved numerically, using a genetic
    algorithm, with risk measures given respectively by the
    centered moments $\mu_2, \mu_4$, $\mu_6$ and $\mu_8$.
The straight lines are the efficient frontiers when we add to these assets a
risk-free asset whose interest rate is set to 5\% a year.} \end{figure}

\begin{figure}[!h]
\begin{center}
\includegraphics[width=15cm]{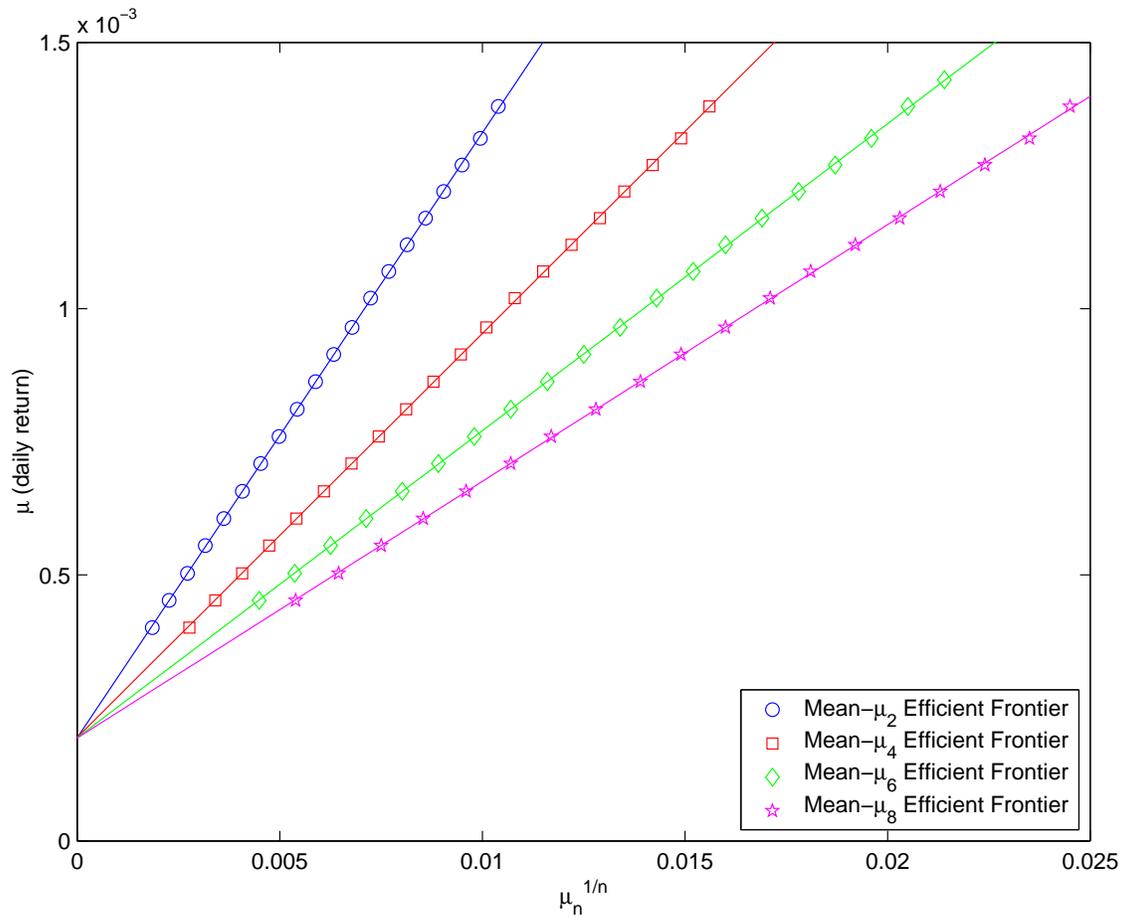}
\end{center}
\caption{\label{fig:GEF_RF} This figure represents the generalized
    efficient frontier for a portfolio made of seventeen risky assets and
    a risk-free asset whose interest rate is set to 5\% a year. The
    optimization problem is solved numerically, using a genetic
    algorithm, with risk measures given by the
    centered moments $\mu_2, \mu_4$, $\mu_6$ and $\mu_8$.}
\end{figure}

\begin{figure}[!h]
\begin{center}
\includegraphics[width=15cm]{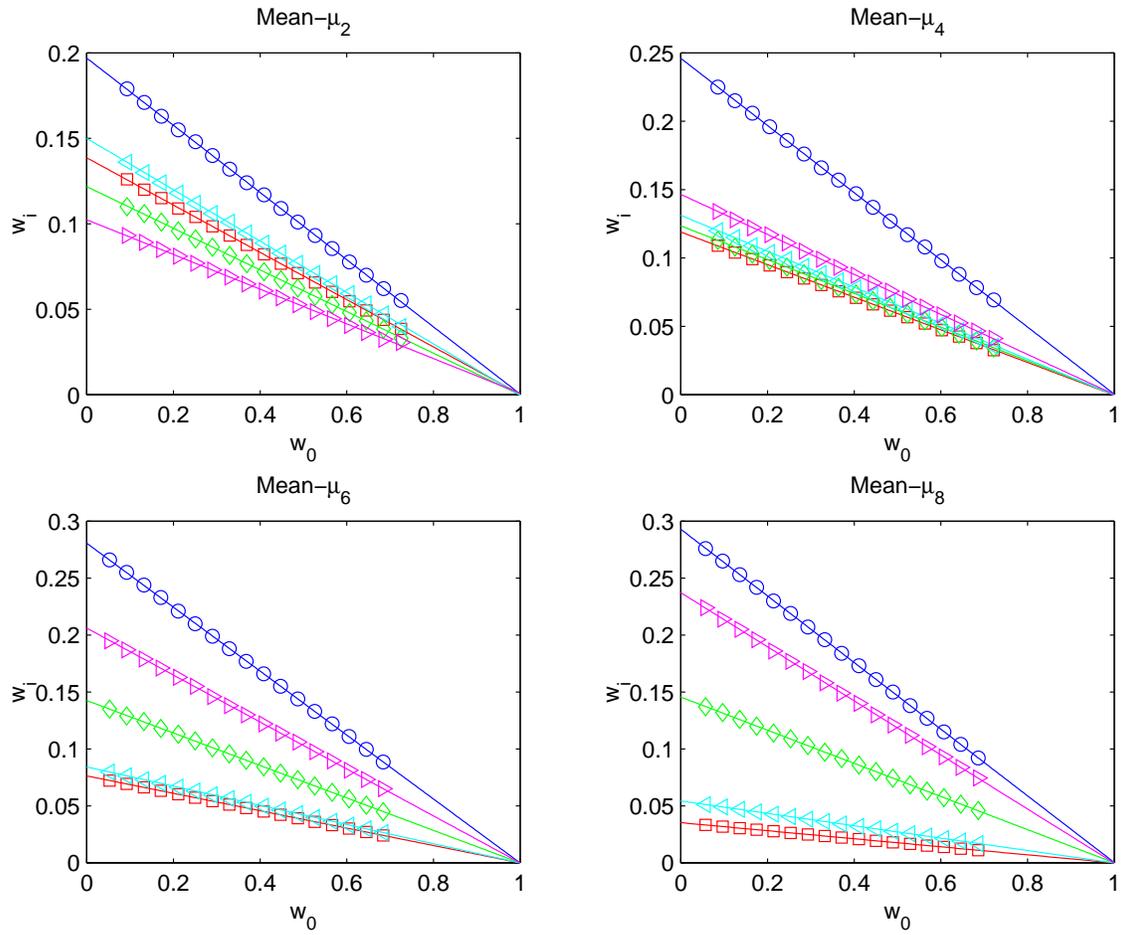}
\end{center}
\caption{\label{fig:twofunds} Dependence of the
five largest weights of risky assets in the efficient portfolios found in
figure \ref{fig:GEF_RF}
as a function of the weight  $w_0$ invested in the risk-free asset, for
the four risk measures given by the centered moments $\mu_2, \mu_4$, $\mu_6$
and $\mu_8$.
The same symbols always represent the same asset.}
\end{figure}

Is it possible that the largest risks captured by the
high-order centered moments could increase at a slower rate
than the small risks embodied in the small-order centered cumulants?
For instance, is it possible for the slope of the mean-$\mu_6$ efficient
frontier
to be larger than the slope of the mean-$\mu_4$ frontier?
This is an important question as it conditions the relative costs
in terms of the panel of risks under increasing specified returns.
To address this question, consider figure \ref{fig:GEF}. Changing the value of
the risk-free
interest rate amounts to move the intercept of the straight lines along the
ordinate axis so as to keep them tangent to the
efficient frontiers without risk-free asset. Therefore, it is easy to see that,
in the situation depicted in figure \ref{fig:GEF}, the slope of the four
straight lines will always decay with the order of the centered moment.

\begin{figure}[!h]
\begin{center}
\includegraphics[width=15cm]{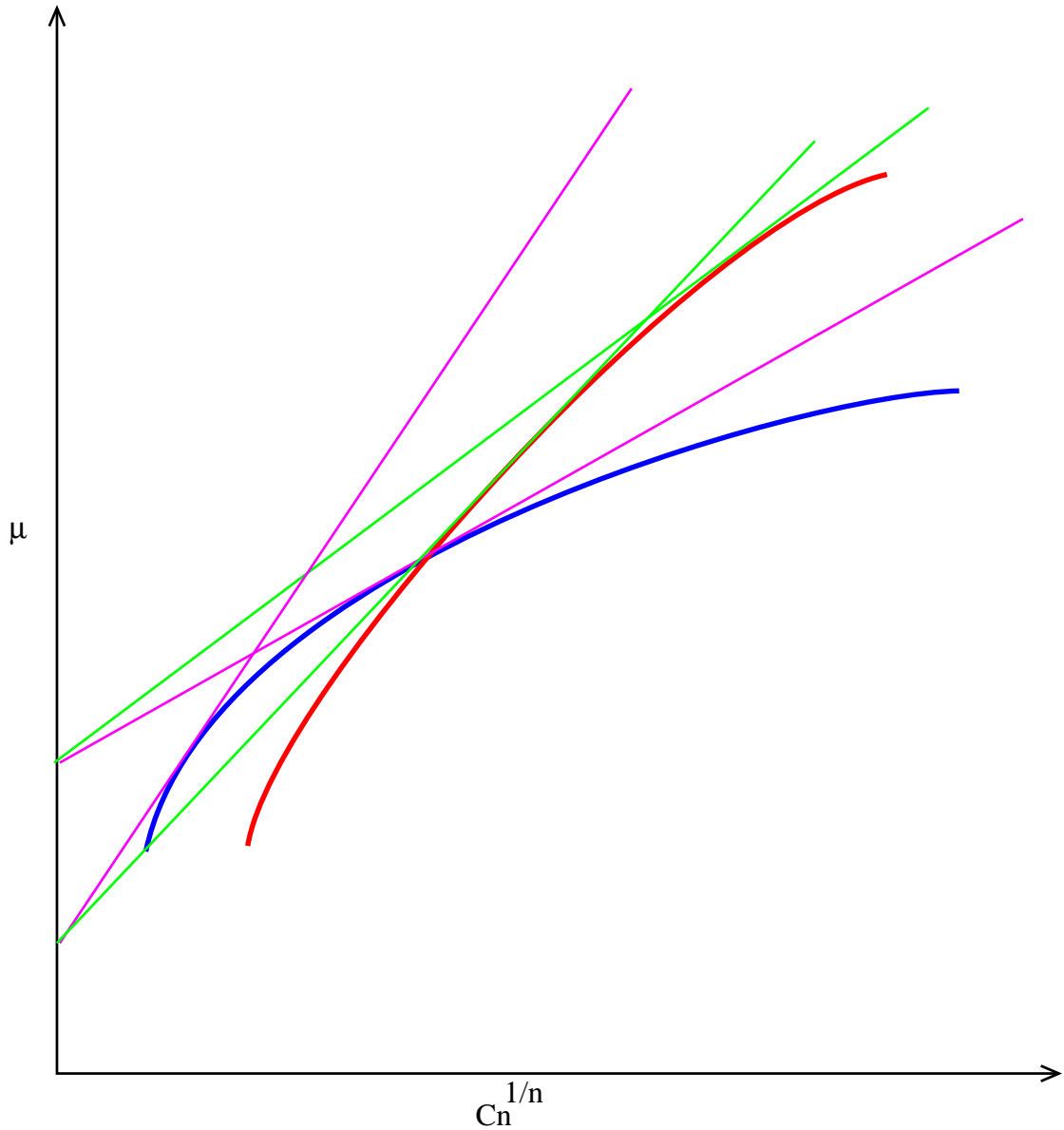}
\end{center}
\caption{\label{fig:intrate} The dark and grey thick curves represent two
efficient frontiers for a portfolio without risk-free interest rate
obtained with two measures of risks.
The dark and grey thin straight lines
represent the efficient frontiers in the presence of a
risk-free asset, whose value is given by the intercept of the straight lines
with the ordinate axis. This illustrates the existence of an
inversion of the dependence of the slope of the efficient frontier with
risk-free asset
as a function of the order $n$ of the measures of risks, which can occur only
when the
efficient frontiers without risk-free asset cross each other.}
   \end{figure}

In order to observe an inversion in the order of the slopes, it is necessary
and sufficient that the efficient frontiers without risk-free asset cross each
other. This assertion is proved by visual inspection of
figure \ref{fig:intrate}. Can we observe such crossing of efficient frontiers?
In the most general case of risk measure, nothing forbids this occurence.
Nonetheless, we think that this kind of behavior is not realistic
in a financial context since, as said above, it would mean that the large risks
could increase at a slower rate than the small risks, implying an irrational
behavior of the economic agents.

\section{Classification of the assets and of portfolios}

Let us consider two assets or portfolios $X_1$ and $X_2$ with different
expected returns $\mu(1)$, $\mu(2)$ and different levels of risk
measured by $\rho_\alpha(X_1)$ and $\rho_\alpha(X_2)$. An important question is
then to be able to compare these two assets or portfolios. The most general way
to perform such a comparison is to refer to decision theory and to
calculate the utility of each of them. But, as already said, the utility
function of an agent is generally not known, so that other approaches have to
be developed. The simplest solution is to consider that the couple (expected
return, risk measure) fully characterizes the behavior of the economic agent
and thus provides a sufficiently good approximation for her utility function.

In the \cite{Markovitz}'s world for instance, the preferences of the agents are
summarized by the two first moments of the distribution of assets returns.
Thus, as shown by \cite{S66,S94} a simple way to synthetize these two
parameters, in order to get a measure of the performance of the assets or
portfolios, is to build the ratio of the expected return $\mu$ (minus the risk
free interest rate) over the standard deviation $\sigma$:
\be
\label{eq:Sharpe}
S=\frac{\mu - \mu_0}{\sigma},
\ee
which is the so-called Sharpe ratio and simply represents the
amount of expected return per unit of risk, measured by the standard deviation.
It is an increasing function of the expected return and a decreasing function
of the level of risk, which is natural for risk-averse or prudential agent.

\subsection{The risk-adjustment approach}

This approach can be generalized to any type of risk measures (see
\cite{Dowd00}, for instance) and thus allows for the comparison of assets whose
risks are not well accounted for by the variance (or the standard deviation).
Indeed, instead of considering the variance, which only accounts for the small
risks, one can build the ratio of the expected return over any risk measure.
In fact, looking at the equation (\ref{eq:eff}) in appendix \ref{app:GEF},
the expression
\be
\label{eq:rar}
\frac{\mu-\mu_0}{\rho_\alpha(X)^{1/\alpha}},
\ee
naturally arises and is constant for every efficient portfolios. In this
expression, $\alpha$ denotes the coefficient of homogeneity of the risk
measure. It is nothing but a simple generalisation of the usual Sharpe ratio.
Indeed, when $\rho_\alpha$ is given by the variance $\sigma^2$, the expression
above recovers the Sharpe ratio. Thus, once the portfolio manager has chosen
his measure of fluctuations $\rho_\alpha$, he can build a consistent
risk-adjusted performance measure, as shown by (\ref{eq:rar}).

As just said, these generalized Sharpe ratios are constant for every efficient
portfolios. In fact, they are not only constant but also maximum for every
efficient portfolios, so that looking for the portfolio with maximum
generalized Sharpe ratio yields the same optimal portfolios as those found with
the whole optimization program solved in the previous section.

As an illutration, table \ref{table:rar} gives the
risk-adjusted performance of the set of seventeen assets already
studied, for several risk measures. We
have considered the three first even order centered moments (columns 2 to 4)
and the three first even order cumulants (columns 2, 5 and 6) as fluctuation
measures. Obviously the second order centered moment and the second order
cumulant are the same, and give again the usual Sharpe ratio
(\ref{eq:Sharpe}). The assets have been sorted with respect to their
Sharpe Ratio.

The first point to note is that the rank of an asset in terms of
risk-adjusted perfomance strongly depends on the risk measure under
consideration. The case of MCI Worldcom is very striking in this respect.
Indeed, according to the usual Sharpe ratio, it appears in the 12$^{th}$
position with a value larger than $0.04$ while according to the other 
measures it
is the last asset of our selection with a value lower than $0.02$.

The second interesting point is that, for a given asset, the
generalize Sharpe ratio is always a decreasing function of the order of the
considered centered moment. This is not particular to our set of assets since
we can prove that
\be
\left( \E \left[ |X|^p \right] \right)^{1/p} \ge
\left( \E \left[ |X|^q \right] \right)^{1/q},
\ee
so that
\be
\forall p>q,~~~ \frac{\mu-\mu_0}{ \left( \E \left[ |X|^p \right]
\right)^{1/p}} \le \frac{\mu-\mu_0}{\left( \E \left[ |X|^q \right]
\right)^{1/q}}. \ee
On the contrary, when the cumulants are used as risk
measures, the generalized Sharpe ratios are not monotonically decreasing, as
exhibited by Procter \& Gamble for instance. This can be surprising
in view of our previous remark that
the larger is the order of the moments involved in a risk
measure, the larger are the fluctuations it is accounting for. Extrapolating
this property to cumulants, it would
mean that Procter \& Gamble presents less large risks according to $C_6$
than according to $C_4$, while according to the centered moments, the reverse
evolution is observed.

Thus, the question of the coherence of the cumulants as measures of 
fluctuations
may arise. And if we accept that such measures are coherent, what are the
implications on the preferences of the agents employing such measures ? To
answer this question, it is informative to express
the cumulants as a function of the moments.
For instance, let us consider the fourth order cumulant
\bea
C_4 &=& \mu_4 - 3 \cdot {\mu_2}^2, \\
&=& \mu_4 - 3 \cdot {C_2}^2~.
\eea
An agent assessing the fluctuations of an asset with respect
to $C_4$ presents aversion for the fluctuations quantified by the fourth
central moment $\mu_4$ -- since $C_4$ increases with $\mu_4$ -- but 
is attracted
by the fluctuations measured by the variance - since $C_4$ decreases with
$\mu_2$. This behavior is not irrational since it remains globally
risk-averse. Indeed, it depicts an agent which tries to avoid the larger risks
but is ready to accept the smallest ones.

This kind of behavior is characteristic of any agent using the cumulants as
risk measures. It thus allows us to understand why Procter \& Gamble is more
attractive for an agent sentitive to $C_6$ than for an agent sentitive to
$C_4$. From the expression of $C_6$, we remark that the agent sensitive to
this cumulant is risk-averse with respect to the fluctuations mesured by
$\mu_6$  and $\mu_2$ but is risk-seeker with respect to the fluctuations
mesured by $\mu_4$ and $\mu_3$. Then, is this particular case, the later ones
compensate the former ones.

It also allows us to understand from a behavioral stand-point why it is
possible to ``have your cake and eat it too'' in the sense of
\cite{AndersenRisk}, that is, why, when the cumulants are choosen as 
risk measures,
it may be possible to increase the expected return of a portfolio while
lowering its large risks, or in other words, why its generalized Sharpe ratio
may increase when one consider larger cumulants to measure its risks. We will
discuus this point again in section \ref{sec:9}.

\subsection{Marginal risk of an asset within a portofolio}

Another important question that arises is the contribution of a given asset to
the risk of the whole portfolio. Indeed, it is crucial to know whether the risk
is homogeneously shared by all the assets of the portfolio or if it is only
held by a few of them. The quality of the diversification is then at stake.
Moreover, this also allows for the sensitivity analysis of the risk of
the portfolio with respect to small changes in its composition\footnote{see
\cite{GLS00, Scaillet00} for a sensitivity analysis of the Value-at-Risk and
the expected shortfall.}, which is of practical interest since it can 
prevent us
from recalculating the whole risk of the portfolio after a small 
re-adjustment of its
composition.

Due to the homogeneity property of the fluctuation measures and to 
Euler's theorem
for homogeneous functions, we can write that
\be
\rho(\{w_1, \cdots, w_N\}) = \frac{1}{\alpha} \sum_{i_1}^N w_i \cdot
\frac{\partial \rho}{\partial w_i},
\ee
provided the risk measure $\rho$ is differentiable which will be assumed in
all the sequel. In this expression, the coefficient $\alpha$ again 
denotes the degree
of homogeneity of the risk measure $\rho$

This relation simply shows that the amount of risk brought by one unit of
the asset $i$ in the portfolio is given by the first derivative of the risk of
the portfolio with respect to the weight $w_i$ ot this asset. Thus,
$\alpha^{-1} \cdot \frac{\partial \rho}{\partial w_i}$ represents the marginal
amount of risk of asset $i$ in the portfolio. It is then easy to check that, in
a portfolio with minimum risk, irrespective of the expected return, the weight
of each asset is such that the marginal risks of the assets in the
portfolio are equal.

\section{A new equilibrum model for asset prices}

Using the portfolio selection method explained in the two
previous sections, we now present an equilibrium model generalizing the
original Capital Asset Pricing Model developed by \cite{S64,L65,M66}. Many
generalizations have already been proposed to account for the fat-tailness of
the assets return distributions, which
led to the multi-moments CAPM. For instance \cite{R73} and \cite{KL76}
or \cite{Lim89} and \cite{HS00} have
underlined and tested the role of the asymmetry in the risk premium by
accounting for the skewness
of the distribution of returns.  More recently, \cite{FL97} and
\cite{HS99} have
introduced a four-moments CAPM to take into account the letpokurtic behavior
of the assets return distributions. Many other extentions have been presented
such as the VaR-CAPM (see \cite{AB02}) or the Distributional-CAPM by
\cite{P02}. All these generalization become more and more complicated and not
do not provide necessarily more accurate prediction of the expected returns.

Here, we will assume that the relevant risk measure is given by any measure of
fluctuations previously presented that obey the axioms I-IV
of section 2. We will also relax the usual assumption of
an homogeneous market to give to the economic agents the choice of 
their own risk
measure: some of them may choose a risk measure which put the emphasis on
the small fluctuations while others may prefer those which account for the
large ones. We will show that, in such an heterogeneous market, an equilibrium
can still be reached and that the excess returns of individual stocks remain
proportional to the market excess return.

   For this, we need the
following assumptions about the market:
\begin{itemize}
\item{H1:} We consider a one-period market, such that all the
    positions held at the begining of a period are cleared at the end
    of the same period.
\item{H2:} The market is perfect, {\it i.e.}, there are no transaction
    cost or taxes, the market is efficient and the investors can lend and borrow
    at the same risk-free rate $\mu_0$.
\end{itemize}

We will now add another assumption that specifies the behavior of the
agents acting on the market, which will lead us to make the
distinction between homogeneous and heterogeneous markets.

\subsection{Equilibrium in a homogeneous market}

The market is said to be homogeneous if all the agents acting on this
market aim at fulfilling the same objective. This means that:
\begin{itemize}
\item{H3-1:} all the agents want to maximize the expected return of
    their portfolio at the end of the period under a given constraint of
    measured risk, using the same measure of risks $\rho_\alpha$ for all of
them.
  \end{itemize}
In the special case where $\rho_\alpha$ denotes the variance, all the agents
follow a Markovitz's optimization procedure, which leads to the CAPM
equilibrium, as proved by \cite{S64}. When $\rho_\alpha$ represents the
centered moments, we will be led to the market equilibrium described by
\cite{R73}. Thus, this approach allows for a generalization of the most popular
asset pricing in equilibirum market models.

When all the agents have the same risk function $\rho_\alpha$, whatever
$\alpha$ may be, we can assert that they have all a fraction of their capital
invested in the same portfolio $\Pi$, whose composition is given in
appendix \ref{app:GEF}, and the remaining in the risk-free asset. The amount of
capital invested in the risky fund only depends on their risk 
aversion or on the
legal margin requirement they have to fulfil.

Let us now assume that the market is at equilibrium, i.e.,
supply equals demand. In such a
case, since the optimal portfolios can be any linear combinations of the
risk-free asset and of the risky portfolio $\Pi$, it is straightforward to show
(see appendix \ref{app:MP}) that the market
portfolio, made of all traded assets in proportion of their market
capitalization, is nothing but the risky portfolio $\Pi$. Thus, as
shown in appendix \ref{app:GCAPM}, we can state that, whatever the risk measure
$\rho_\alpha$ chosen by the agents to perform their optimization, the excess
return of any asset over the risk-free interest rate is proportional to the
excess return of the market portfolio $\Pi$ over the risk-free interest rate:
\be
\label{eq:GCAPM}
\mu(i) - \mu_0 = \beta_\alpha^i \cdot (\mu_\Pi - \mu_0),
\ee
where
\be
\label{eq:Gbeta}
\beta_\alpha^i =\cdot \left. \frac{\partial \ln \left( {\rho_\alpha}^
\frac{1}{\alpha}\right)}{\partial w_i} \right|_{w_1^*, \cdots, w_N^*}~,
\ee
where $w_1^*, \cdots, w_N^*$ are defined in appendix D.
When $\rho_\alpha$ denotes the variance, we recover the usual $\beta^i$ given
by the mean-variance approach:
\be
\beta^i = \frac{\Cov(X_i, \Pi)}{\Var(\Pi)}.
\ee
Thus, the relations (\ref{eq:GCAPM}) and (\ref{eq:Gbeta}) generalize
the usual CAPM formula, showing that the specific choice of the risk measure is
not very important, as long as it follows the axioms I-IV characterizing the
fluctuations of the distribution of asset returns.

\subsection{Equilibrium in a heterogeneous market}

Does this result hold in the more realistic situation of
an heterogeneous market?
A market will be said to be heterogeneous if
the agents seek to fulfill different objectives.  We thus consider the
following
assumption:
\begin{itemize}
\item{H3-2:} There exists N agents. Each agent $n$ is characterized
by her choice of a risk measure $\rho_\alpha(n)$ so that she invests only in
the mean-$\rho_\alpha(n)$ efficient portfolios.
\end{itemize}

According to this hypothesis, an agent $n$  invests a fraction of her wealth
in the risk-free asset and the remaining in $\Pi_n$, the mean-$\rho_\alpha(n)$
efficient portfolio, only made of risky assets. The fraction of wealth invested
in the risky fund depends on the risk aversion of each agents, which may vary
from an agent to another one.

The composition of the market portfolio for such a heterogenous market is
derived in appendix \ref{app:MP}. We find that the market portfolio $\Pi$
is nothing but the weighted sum of the mean-$\rho_\alpha(n)$ optimal portfolio
$\Pi_n$: \be
\Pi = \sum_{n=1}^N \gamma_n \Pi_n,
\ee
where $\gamma_n$ is the fraction of the total wealth invested in the fund
$\Pi_n$ by the n$^{th}$ agent.

Appendix \ref{app:GCAPM} demonstrates that, for every asset $i$ and for any
mean-$\rho_\alpha(n)$ efficient portfolio $\Pi_n$, for all $n$, the
following equation holds
\be
\mu(i) - \mu_0 = \beta_n^i \cdot (\mu_{\Pi_n} - \mu_0)~.
\ee
Multiplying these equations by $\gamma_n/ \beta_n^i$, we get
\be
\frac{\gamma_n}{\beta_n^i} \cdot (\mu(i) - \mu_0) = \gamma_n \cdot (\mu_{\Pi_n}
- \mu_0) ,
\ee
for all $n$, and summing over the different agents, we obtain
\be
\left( \sum_n  \frac{\gamma_n}{\beta_n^i} \right)  \cdot (\mu(i) - \mu_0) =
\left( \sum_n \gamma_n \cdot \mu_{\Pi_n} \right)  - \mu_0,
\ee
so that
\be
\mu(i) -\mu_0 = \beta^i \cdot ( \mu_\Pi - \mu_0),
\ee
with
\be
\beta^i = \left( \sum_n  \frac{\gamma_n}{\beta_n^i} \right)^{-1}~.
\ee
This allows us to conclude that, even in a heterogeneous market, the
expected excess return of each individual stock is directly proportionnal to
the expected excess return of the market portfolio, showing that the
homogeneity of the market is not a key property necessary for 
observing a linear
relationship between individual excess asset returns and the market excess
return.

\section{Estimation of the joint probability distribution of returns
of several assets}

{\it A priori}, one of the main practical advantage of
\cite{Markovitz}'s method and
its generalization presented above is that one does not need
the multivariate probability distribution function of the assets
returns, as the analysis solely relies on the coherent measures $\rho(X)$
defined in section 2, such as the centered moments or the
cumulants of all orders that can in principle be estimated empirically.
Unfortunately, this apparent advantage maybe an
illusion. Indeed, as underlined by
\cite{SO} for instance, the error of the empirically estimated moment of order
$n$ is proportional to the moment of order $2n$, so that the error
becomes quickly of the same order as the estimated moment itself. 
Thus, above $n=6$ (or
may be $n=8$) it is not reasonable to estimate the moments and/or
cumulants directly. Thus, the knowledge of the multivariate
distribution of assets returns
remains necessary. In addition, there is a current of thoughts that provides
evidence that marginal distributions of returns may be regularly varying
with index $\mu$ in the range 3-4 \cite{Lux, P96, Gopikrishnan}, 
suggesting the non-existence of
asymptotically defined moments and cumulants of order equal to or 
larger than $\mu$.

In the standard Gaussian framework, the multivariate distribution
takes the form
of an exponential of minus a quadratic form $X' \Omega^{-1} X$, where
$X$ is the unicolumn
of asset returns and $\Omega$ is their covariance matrix. The beauty
and simplicity
of the Gaussian case is that the essentially impossible task
of determining a large multidimensional function is reduced into
the very much simpler
one of calculating the $N(N+1)/2$ elements of the symmetric covariance matrix.
Risk is then uniquely and completely embodied by the variance of the
portfolio return,
which is easily determined from the covariance matrix. This is the basis of
Markovitz's portfolio theory \cite{Markovitz} and of the CAPM (see
for instance \cite{Merton}).

However, as is well-known, the variance (volatility) of portfolio returns
provides at
best a limited
quantification of incurred risks, as the empirical distributions of
returns have ``fat  tails''
\cite{Lux,Gopikrishnan} and
the dependences between assets are only imperfectly accounted for by
the covariance
matrix \\ \cite{Litterman}.

In this section, we present a novel approach based on 
\cite{Sornette1} to attack
this problem
in terms of the parameterization of the multivariate
distribution of
returns involving two steps: (i) the projection of the empirical
marginal distributions
onto Gaussian laws via nonlinear mappings; (ii) the use of an entropy
maximization
to construct the corresponding most parsimonious
representation of the multivariate distribution.

\subsection{A brief exposition and justification of the method}

We will use the method of determination of multivariate distributions
introduced by \cite{Karlen} and \cite{Sornette1}. This
method consists in two steps: (i) transform each return $x$ into a
Gaussian variable $y$ by a nonlinear monotonous increasing
mapping; (ii) use the principle of entropy
maximization to construct the corresponding multivariate distribution
of the transformed variables $y$.

The first concern to address before going any further
is whether the nonlinear transformation, which is
in principle different for each asset return, conserves the structure of the
dependence. In what sense is the dependence between the transformed
variables $y$
the same as the dependence between the asset returns $x$? It turns out
that the notion of ``copulas'' provides a general and rigorous answer which
justifies the procedure of \cite{Sornette1}.

For completeness and use later on, we briefly recall the definition 
of a copula (for further
details about the concept of copula see \cite{Nelsen}).  A
function  $C$ : $[0,1]^n \longrightarrow [0,1]$ is a $n$-copula if it
enjoys the
following properties~:
\begin{itemize}
\item $\forall u \in [0,1] $, $C(1,\cdots, 1, u,
1 \cdots, 1)=u$~,
\item $\forall u_i \in [0,1] $, $C(u_1, \cdots, u_n)=0$ if
at least one of the $u_i$ equals zero~,
\item $C$ is grounded and $n$-increasing, {\it i.e.}, the $C$-volume of every
boxes whose
vertices lie in $[0,1]^n$ is positive.
\end{itemize}

Skar's Theorem then states that, given an $n$-dimensional
distribution function  $F$ with continuous marginal distributions
$F_1, \cdots, F_n$, there
exists a unique $n$-copula $C$ : $[0,1]^n \longrightarrow [0,1]$ such
that~: \be
F(x_1, \cdots, x_n) = C(F_1(x_1), \cdots, F_n(x_n))~.
\ee
This elegant result shows that the study of the dependence of random variables
can be performed independently of the behavior of the marginal distributions.
Moreover, the following result shows that copulas are intrinsic measures of
dependence. Consider $n$ continuous random variables  $X_1, \cdots ,
X_n$ with copula $C$.
Then, if $g_1(X_1), \cdots, g_n(X_n)$ are
strictly increasing on the ranges of $X_1, \cdots , X_n$, the random variables
$Y_1=g_1(X_1), \cdots, Y_n=g_n(X_n)$ have exactly the same copula $C$
\cite{Lindskog}.
The copula is
thus invariant under strictly increasing tranformation of the variables.
This provides a powerful way
of studying scale-invariant measures of associations.
It is also a natural starting point for construction of
multivariate distributions and provides the theoretical
justification of the method of determination of mutivariate distributions
that we will use in the sequel.

\subsection{Transformation of an arbitrary random variable into a Gaussian
variable}
\begin{figure}
\begin{center}
\includegraphics[width=15cm]{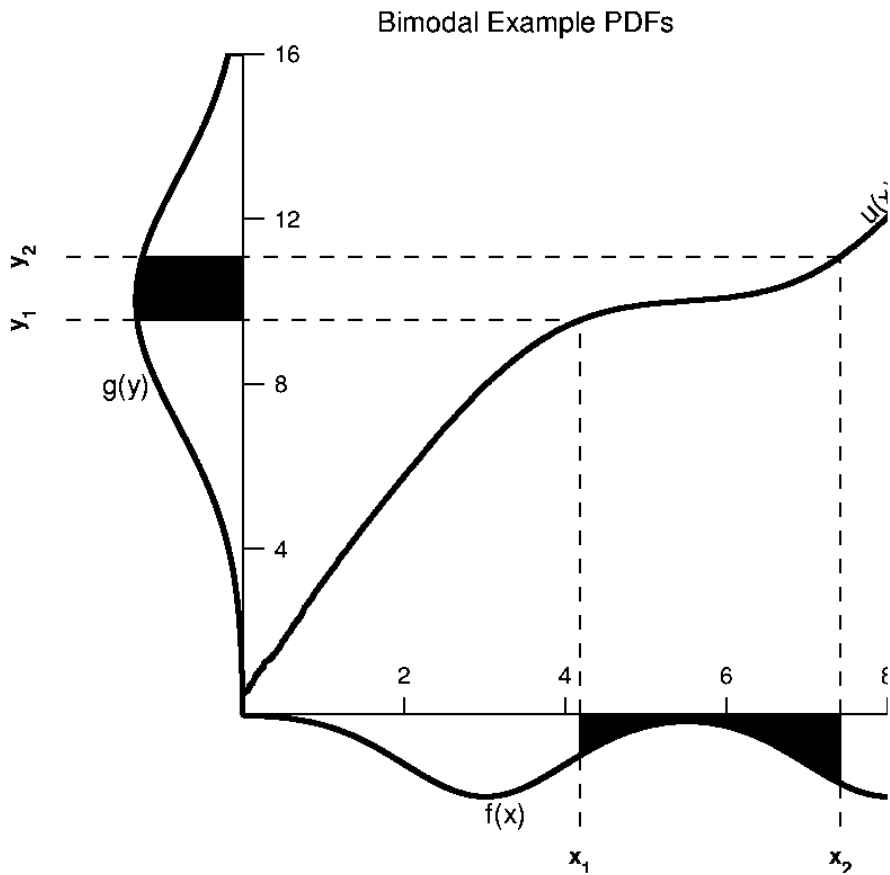}
\end{center}
\caption{\label{fig1} Schematic representation of the nonlinear
mapping $Y = u(X)$
that allows one to transform a variable $X$ with an arbitrary distribution
into a variable $Y$ with a Gaussian distribution.
The probability densities for $X$ and $Y$ are plotted
outside their respective axes. Consistent with the conservation of
probability, the shaded regions have equal area. This conservation of
probability
determines the nonlinear mapping.}
\end{figure}

Let us consider the return $X$, taken as a random variable
characterized by the probability density $p(x)$.
The transformation $y(x)$ which obtains a standard normal variable $y$ from
$x$ is
determined by the conservation of probability:
\be
p(x) dx =\frac{1}{\sqrt{2 \pi}} e^{-\frac{y^2}{2}} dy\; .
\ee
Integrating this equation from $- \infty$ and $x$, we obtain:
\be
F(x)=\frac{1}{2} \left[ 1+  \mbox{erf} \left( \frac{y}{\sqrt{2}} \right)
\right] \; ,
\ee
where $F(x)$ is the cumulative distribution of $X$:
\be
F(x) = \int_{-\infty}^x dx' p(x') \; .
\ee
This leads to the following transformation $y(x)$:
\be
\label{eq7}
y=\sqrt{2}\; \mbox{erf}^{-1}(2F(x)-1)~,
\ee
which is obvously an increasing function of $X$ as required for the
application of the
invariance property of the copula stated in the previous section.
An illustration of the nonlinear transformation (\ref{eq7}) is shown in figure
\ref{fig1}.  Note that it does not require any special hypothesis on the
probability density $X$, apart from being non-degenerate.

In the case where the pdf of $X$ has only one maximum, we may use a
simpler expression
equivalent to (\ref{eq7}). Such a pdf can be written under the
so-called Von Mises parametrization \cite{Embrecht97}~:
\be
\label{eq1}
p(x) = C \frac{f'(x)}{\sqrt{|f(x)|}} e^{-\frac{1}{2}f(x)} \; ,
\ee
where $C$ is a constant of normalization. For $f(x)/x^2 \rightarrow  0$
when $|x| \rightarrow +\infty$, the pdf has a ``fat tail,'' i.e., it
decays slower than a Gaussian at large $|x|$.

Let us now define the change of variable
\be
\label{eq2}
y=sgn(x) \sqrt{|f(x)|} \;.
\ee
Using the relationship $p(y)=p(x) \frac{dx}{dy}$, we get:
\be
\label{eq4}
p(y)=\frac{1}{\sqrt{2 \pi}} e^{-\frac{y^2}{2}} \; .
\ee
It is important to stress the presence of the sign function $sgn(x)$
in equation (\ref{eq2}), which is essential in order to
correctly quantify dependences between random variables.
This transformation (\ref{eq2}) is equivalent to (\ref{eq7})
but of a simpler implementation and will be used in the sequel.

\subsection{Determination of the joint distribution~: maximum entropy
and Gaussian copula}

Let us now consider $N$ random variables $X_i$ with marginal distributions
$p_i(x_i)$. Using the transformation
(\ref{eq7}), we define $N$ standard normal variables $Y_i$. If these variables
were independent, their joint distribution would simply be the product of
the marginal distributions. In many situations, the variables are not
independent and it is necessary to study their dependence.

The simplest approach is to construct their covariance matrix. Applied to
the variables $Y_i$, we are certain that the covariance matrix exists and
is well-defined since their marginal distributions are Gaussian.
In contrast, this is not ensured for the variables $X_i$.
Indeed, in many situations in nature, in economy, finance and in social
sciences,
pdf's are found to have power law tails  $\sim \frac{A}{x^{1+\mu}}$
for large $|x|$. If  $\mu  \le 2$, the variance and the covariances can not be
defined. If  $2 < \mu  \le 4$, the variance and the covariances exit
in principle
but their sample estimators converge poorly.

We thus define the covariance matrix:
\be
V=E[{\bf y}{\bf  y^t}] \; ,
\ee
where ${\bf y}$ is the vector of variables $Y_i$ and the operator $E[\cdot]$
represents the mathematical expectation.
A classical result of information theory \cite{Rao} tells us that,
given the covariance matrix $V$, the best joint
distribution (in the sense of entropy maximization) of the
$N$ variables $Y_i$ is the multivariate Gaussian:
\be
P({\bf y}) = \frac{1}{(2 \pi)^{N/2} \sqrt{\det(V)}} \exp \left(-\frac{1}{2}
{\bf y^t}V^{-1}{\bf y}
\right) \; .
\ee
Indeed, this distribution implies the minimum additional information or
assumption, given the covariance matrix.

Using the joint distribution of the variables $Y_i$, we obtain the joint
distribution of the variables $X_i$:
\be
P({\bf x}) = P({\bf y}) \left| \frac{ \partial y_i}{\partial x_j} \right| \; ,
\ee
where $\left| \frac{ \partial y_i}{\partial x_j} \right|$ is the Jacobian
of the transformation.
Since
\be
\frac{ \partial y_i}{\partial x_j} =\sqrt{2\pi} p_j(x_j) e^{\frac{1}{2} y_i^2}
\delta_{ij} \; , \ee
we get
\be
\left| \frac{ \partial y_i}{\partial x_j} \right| = (2\pi)^{N/2} \prod_{i=1}^N
p_i(x_i) e^{\frac{1}{2}  y_i^2} \; .
\ee
This finally yields
\be
P({\bf x}) = \frac{1}{\sqrt{\det(V)}} \exp \left(-\frac{1}{2} {\bf
y_{(x)}^t}(V^{-1}-I){\bf y_{(x)}}\ \right)\prod_{i=1}^N p_i(x_i) \; .
\label{jfjma}
\ee
As expected, if the variables are independent, $V=I$, and $P({\bf x})$
becomes the product of the marginal distributions of the variables $X_i$.

Let $F({\bf x})$ denote the cumulative distribution function of the 
vector {\bf x} and
$F_i(x_i), i=1, ..., N$ the $N$ corresponding marginal distributions.
The copula $C$ is then such that
\be
F(x_1, \cdots, x_n) = C(F_1(x_1), \cdots, F_n(x_n))~.
\ee

Differentiating with respect to $x_1, \cdots, x_N$ leads to
\be
P(x_1, \cdots, x_n) = \frac{\partial F(x_1, \cdots, x_n)}{\partial x_1 \cdots
\partial x_n} = c(F_1(x_1), \cdots, F_n(x_n))\prod_{i=1}^N p_i(x_i)~,
\ee
where
\be
c(u_1,\cdots, u_N) = \frac{\partial C(u_1,\cdots, u_N)}{\partial u_1 \cdots
\partial u_N}  \label{vfjnqloaq}
\ee
is the density of the copula $C$.

Comparing (\ref{vfjnqloaq}) with (\ref{jfjma}), the density of the
copula is given
in the present case by
\be
c(u_1,\cdots, u_N)= \frac{1}{\sqrt{\det(V)}} \exp \left(-\frac{1}{2} {\bf
y_{(u)}^t}(V^{-1}-I){\bf y_{(u)}}\ \right)~,
\ee
which is the ``Gaussian copula'' with covariance matrix ${\bf V}$. This result
clarifies and justifies the method of \cite{Sornette1} by showing that it
essentially amounts to assume arbitrary marginal distributions with Gaussian
copulas. Note that the Gaussian copula results directly from the
transformation to Gaussian marginals together with the
choice of maximizing the Shannon entropy under the constraint of a 
fixed covariance
matrix. Under differents
    constraint, we would have found another maximum entropy
    copula. This is not unexpected in analogy with
the standard result that the Gaussian law is maximizing the Shannon
entropy at fixed
given variance. If we were to extend this formulation by considering
more general
expressions of the entropy, such that Tsallis entropy \cite{Tsallisfund},
we would have found other copulas.

\subsection{Empirical test of the Gaussian copula assumption}

We now present some tests of the hypothesis of Gaussian copulas
between returns of financial assets. This presentation is only for
illustration purposes, since testing the gaussian copula hypothesis is a
delicate task which has been addressed elsewhere (see \cite{MS01}). Here, as an
example, we propose two simple standard methods.

\begin{figure}
\begin{center}
\includegraphics[width=15cm]{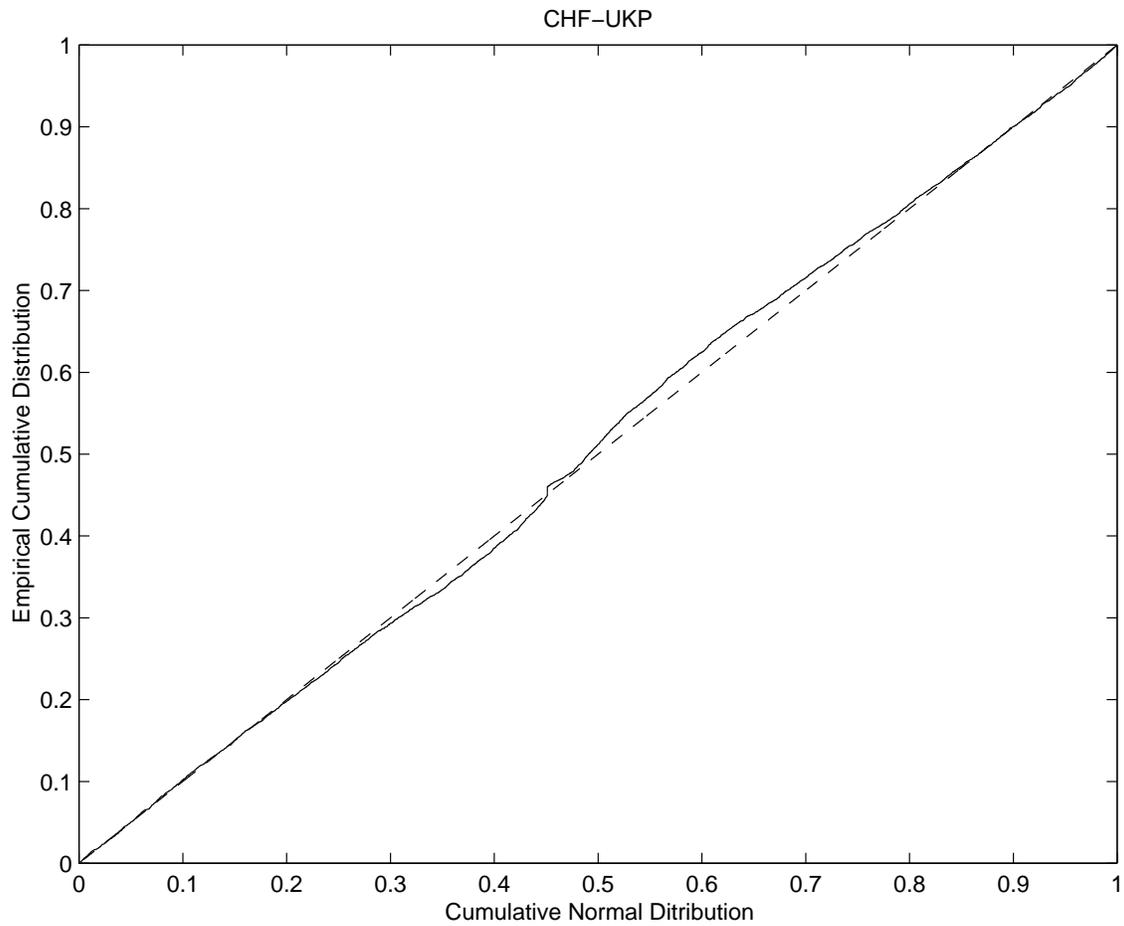}
\end{center}
\caption{\label{fig:sum_chf_ukp} Quantile of the normalized sum of the
Gaussianized returns of the Swiss Franc and The British Pound versus
the quantile of the Normal distribution, for the time interval from Jan. 1971
to Oct. 1998. Different weights in the sum give similar results.} \end{figure}

\begin{figure}
\begin{center}
\includegraphics[width=15cm]{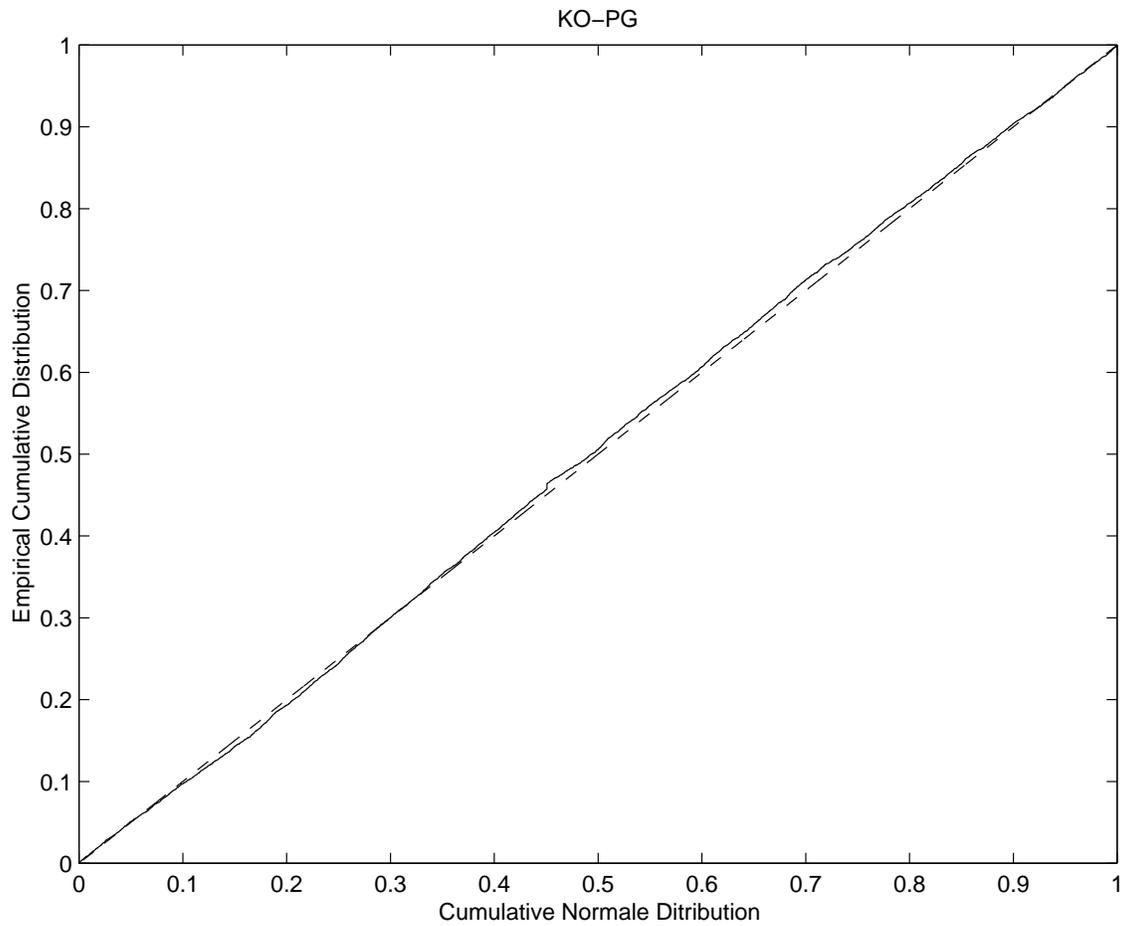}
\end{center}
\caption{\label{fig:sum_ko_pg} Quantile of the normalized sum of the
Gaussianized returns of Coca-Cola and Procter\&Gamble  versus
the quantile of the Normal distribution, for the time interval from Jan. 1970
to Dec. 2000. Different weights in the sum give similar results.}
\end{figure}

\begin{figure}
\begin{center}
\includegraphics[width=15cm]{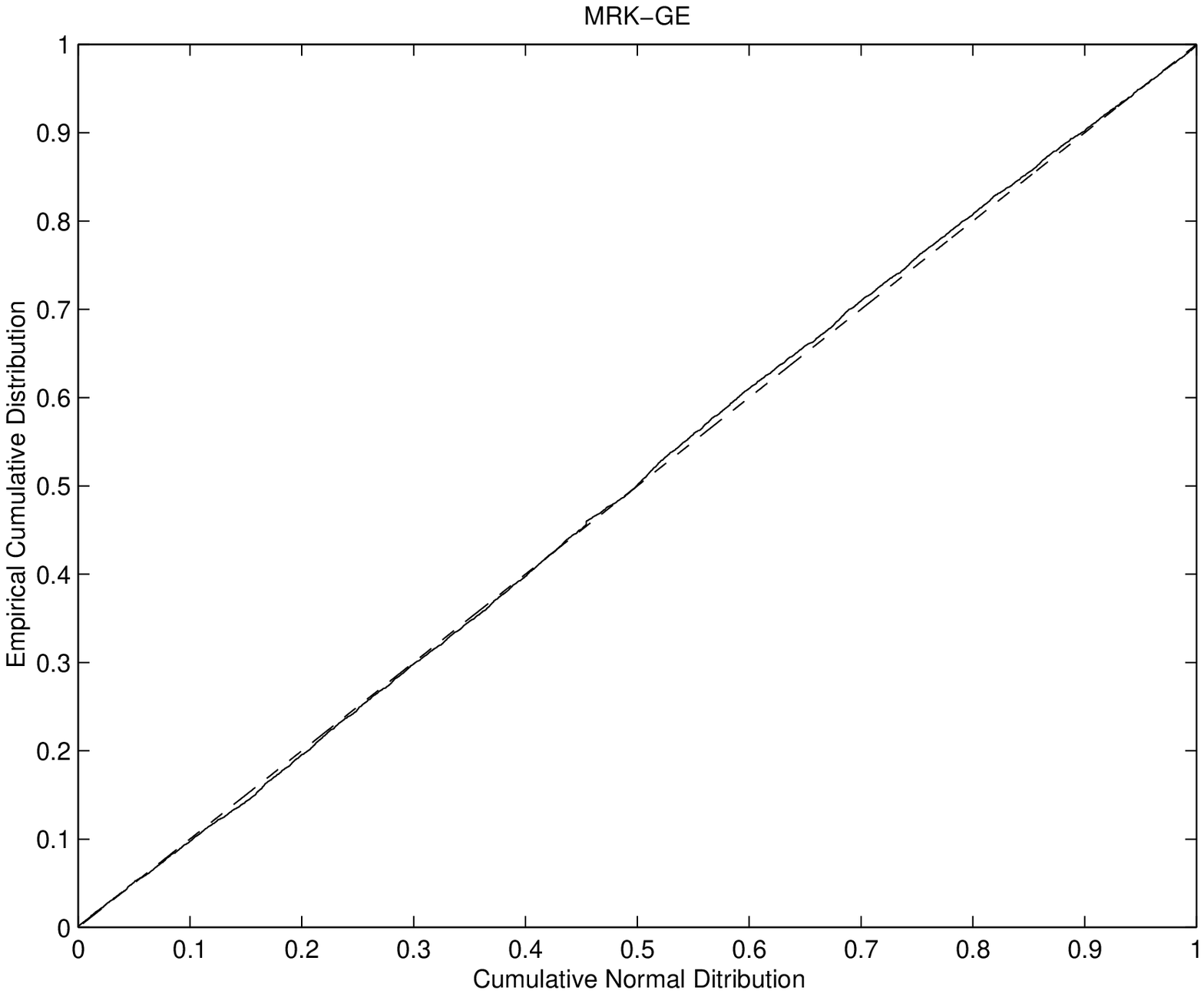}
\end{center}
\caption{\label{fig:sum_mrk_ge} Quantile of the normalized sum of the
Gaussianized returns of Merk and General Electric  versus
the quantile of the Normal distribution, for the time interval from Jan. 1970
to Dec. 2000. Different weights in the sum give similar results.}
\end{figure}

The first one consists in using the property that Gaussian variables 
are stable in
distribution under
addition. Thus, a (quantile-quantile or $Q-Q$) plot of the cumulative
distribution of the sum
$y_1+\cdots+y_p$  versus the cumulative Normal distribution with the same
estimated variance should give a straight line in order to qualify a
multivariate Gaussian
distribution (for the transformed $y$ variables). Such tests on empirical data
are presented in figures \ref{fig:sum_chf_ukp}-\ref{fig:sum_mrk_ge}.

\begin{figure}
\begin{center}
\includegraphics[width=15cm]{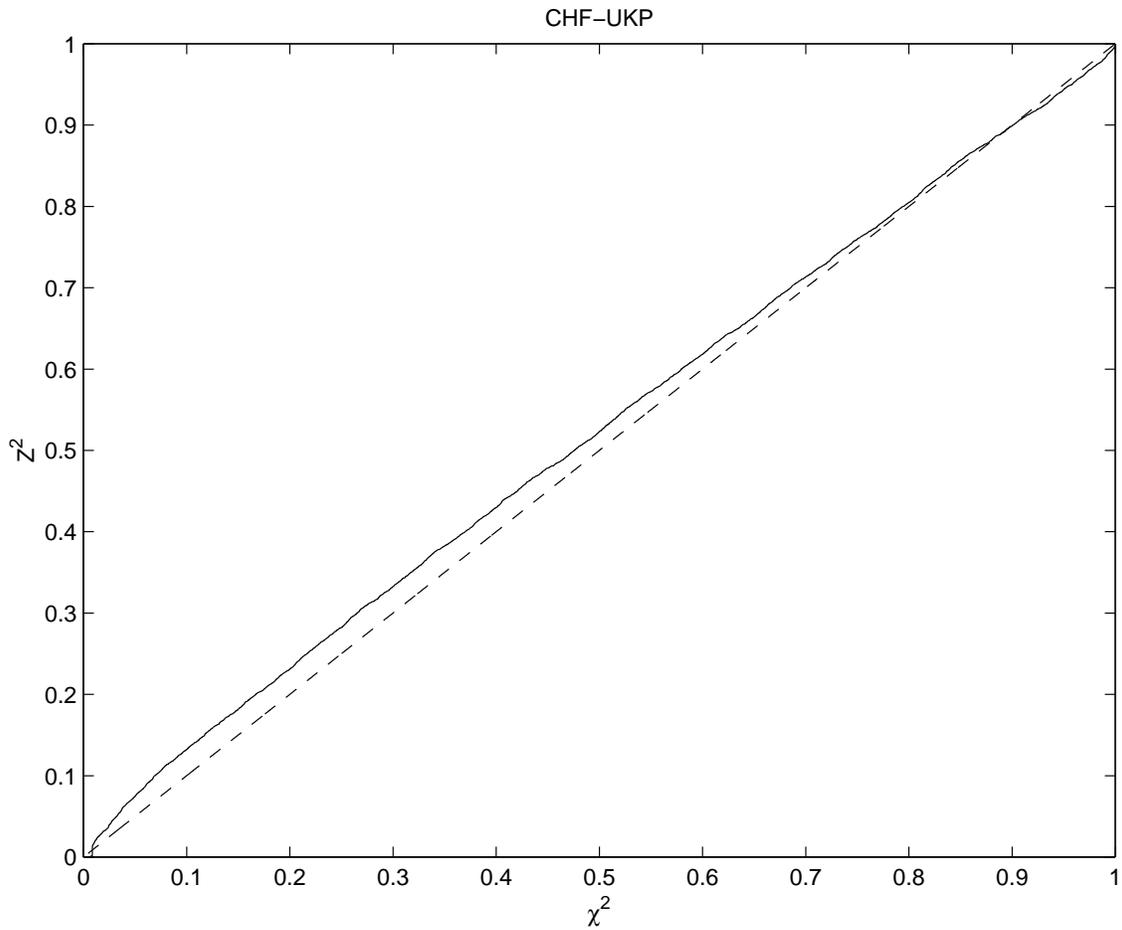}
\end{center}
\caption{\label{fig:chi2_chf_ukp} Cumulative distribution of
$ z^2={\bf y^t V^{-1} y }$ versus the cumulative distribution of
chi-square (denoted
$\chi^2$) with two degrees of freedom for the couple Swiss Franc / British
Pound, for the time interval from Jan. 1971 to Oct. 1998. This
$\chi^2$ should not be confused with
the characteristic scale used in the definition of the modified
Weibull distributions.}
\end{figure}

\begin{figure}
\begin{center}
\includegraphics[width=15cm]{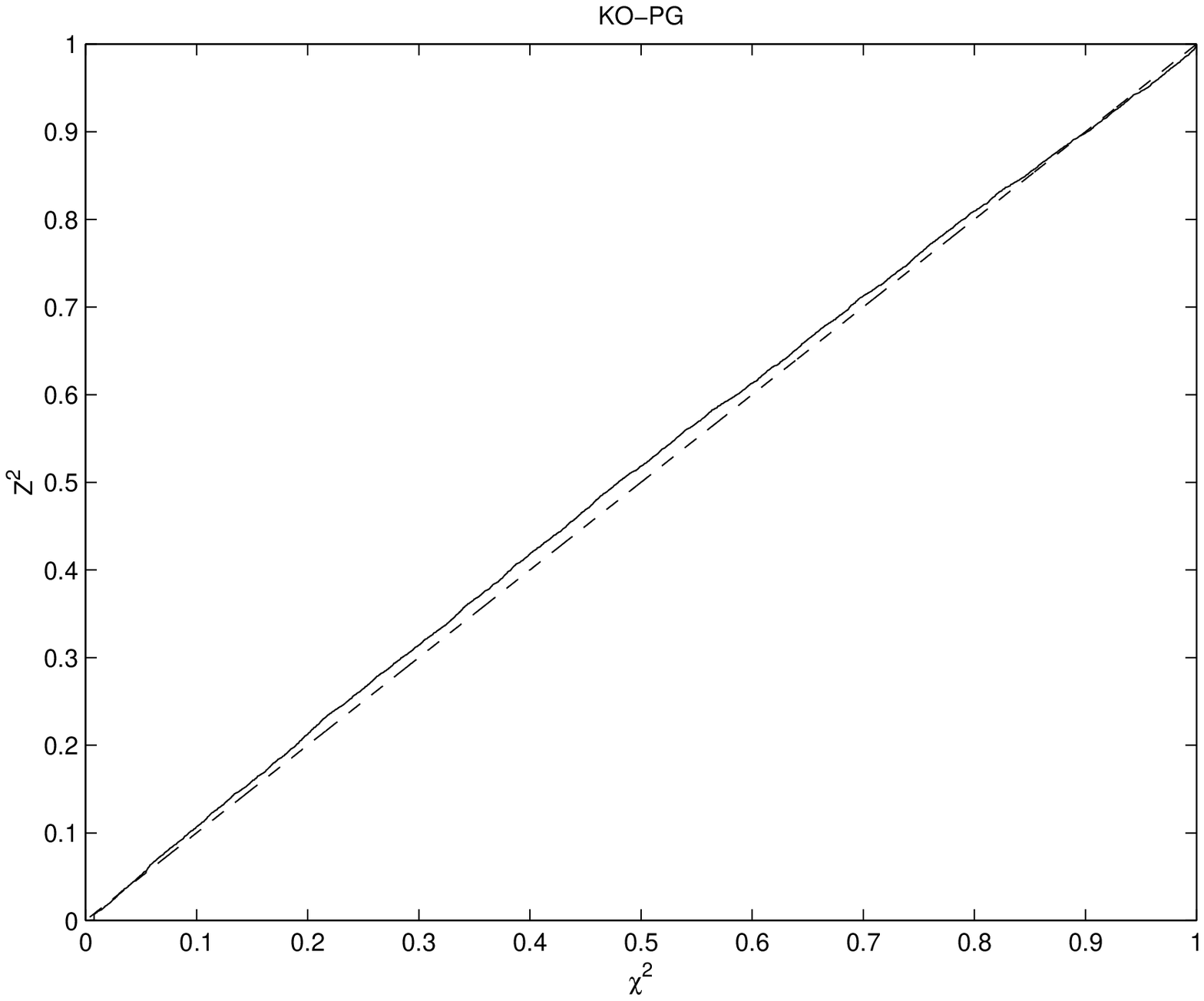}
\end{center}
\caption{\label{fig:chi2_ko_pg} Cumulative distribution of
$z^2={\bf y^t V^{-1} y}$ versus the cumulative distribution
of the chi-square $\chi^2$
with two degrees of freedom for the couple Coca-Cola /
Procter\&Gamble, for the time interval from Jan. 1970
to Dec. 2000. This $\chi^2$ should not be confused with
the characteristic scale used in the definition of the modified
Weibull distributions.}
\end{figure}

\begin{figure}
\begin{center}
\includegraphics[width=15cm]{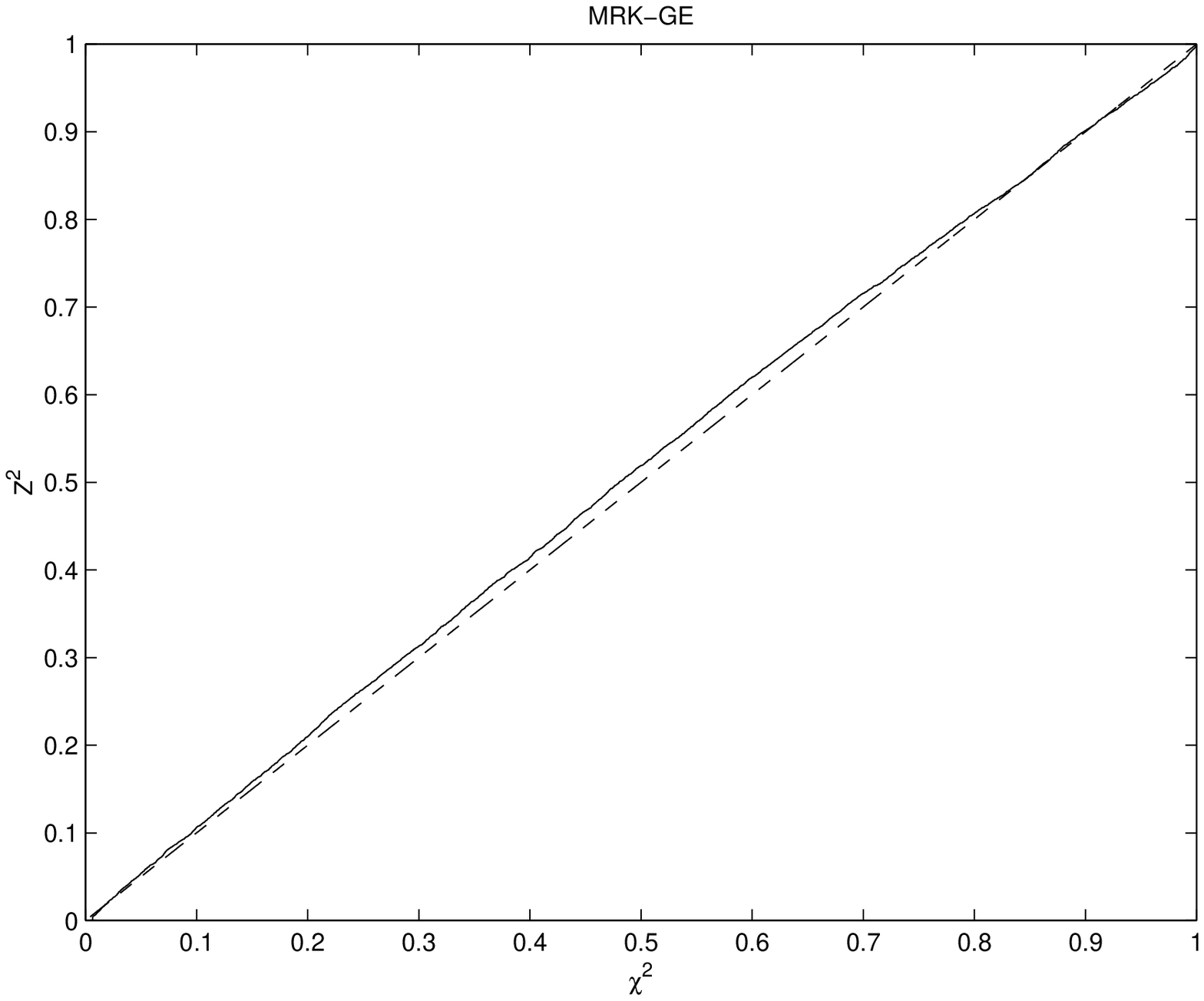}
\end{center}
\caption{\label{fig:chi2_mrk_ge} Cumulative distribution of $z^2={\bf
y^t V^{-1} y}$ versus the cumulative distribution of the
chi-square $\chi^2$
with two degrees of freedom for the couple Merk / General Electric,  for the
time interval from Jan. 1970 to Dec. 2000. This $\chi^2$ should not
be confused with
the characteristic scale used in the definition of the modified
Weibull distributions.}
\end{figure}

The second test amounts to estimating the covariance matrix ${\bf V}$  of the
sample we consider. This step is simple since, for fast decaying pdf's, robust
estimators of the covariance matrix are available. We can then estimate the
distribution of the variable $z^2={\bf y^t V^{-1} y}$. It is well
known that $z^2$ follows a $\chi^2$ distribution if ${\bf y}$
is a Gaussian random vector. Again, the empirical cumulative
distribution of $z^2$ versus the $\chi^2$ cumulative distribution
should give a straight line in order to qualify a multivariate Gaussian
distribution (for the transformed $y$ variables). Such tests on empirical data
are presented in figures \ref{fig:chi2_chf_ukp}-\ref{fig:chi2_mrk_ge}.

First, one can observe that the Gaussian copula
hypothesis appears better for stocks than for currencies. As
    discussed in \cite{MS01}, this result is quite general. A plausible
    explanation lies in the stronger dependence between
    the currencies compared with that between stocks,
    which is due to the
    monetary policies limiting the fluctuations between the
    currencies of a group of countries, such as was the
    case in the European Monetary System before the unique Euro currency.
   Note also that
the test of aggregation seems systematically more in favor of the Gaussian
copula hypothesis than is the $\chi^2$
test, maybe due to its smaller sensitivity. Nonetheless, the very good
performance of the Gaussian hypothesis under
the aggregation test bears good news for a porfolio theory based on it,
since by definition a portfolio corresponds to asset aggregation.
Even if sums of the transformed returns are not equivalent to sums
of returns (as we shall see in the sequel), such sums qualify the collective
behavior whose properties are controlled by the copula.

Notwithstanding some deviations from linearity in
figures \ref{fig:sum_chf_ukp}-\ref{fig:chi2_mrk_ge}, it appears
that, for our purpose of developing a generalized portfolio theory,
the Gaussian copula hypothesis is a good approximation.
A more systematic test of this goodness of fit requires the
quantification of a confidence level, for instance using the Kolmogorov test,
that would allow us to accept or reject the Gaussian copula
hypothesis. Such a test has been performed in \cite{MS01},
where it is shown that this test is sensitive enough only in the
bulk of the distribution, and that
an Anderson-Darling test is preferable for the tails of the distributions.
Nonetheless, the quantitative
conclusions of these tests are identical to the qualitative results
presented here. Some other tests would be useful, such as
the multivariate Gaussianity test presented by \cite{RS93}.

\section{Choice of an exponential family to parameterize the marginal
distributions}
\subsection{The modified Weibull distributions}

We now apply these constructions to a class of distributions with fat tails,
that have been found to provide a convenient and flexible parameterization of
many phenomena found in nature and in the social sciences
\cite{Laherrere}. These so-called stretched exponential distributions
can be seen to be general forms of the extreme tails of product of
random variables \cite{FS97}.

Following \cite{Sornette1}, we postulate the following marginal
probability distributions
of returns:
\be
\label{eq:weibull}
p(x)=\frac{1}{2\sqrt{\pi}} \frac{c}{\chi^{\frac{c}{2}}}
|x|^{\frac{c}{2}-1} e^{-\left( \frac{|x|}{\chi}\right)^c} \; ,
\ee
where $c$ and $\chi$ are the two key parameters. A more general
parameterization taking
into account a possible asymmetry between negative and positive
returns (thus leading to possible non-zero average return) is
\bea
\label{eq:pasym1}
p(x)&=&\frac{Q}{\sqrt{\pi}} \frac{c_+}{\chi_+^{\frac{c_+}{2}}}
|x|^{\frac{c_+}{2}-1} e^{-\left(
\frac{|x|}{\chi_+}\right)^{c_+}}~~~~\mbox{if}~x\ge 0\\
\label{eq:pasym2}
p(x)&=&\frac{1-Q}{\sqrt{\pi}} \frac{c_-}{\chi_-^{\frac{c_-}{2}}}
|x|^{\frac{c_-}{2}-1} e^{-\left(
\frac{|x|}{\chi_-}\right)^{c_-}}~~~~\mbox{if}~x<0~,
\eea
where $Q$ (respectively $1-Q$) is the fraction of positive
(respectively negative) returns. In the sequel, we will only
consider the case $Q=\frac{1}{2}$, which is the only analytically
tractable case. Thus the pdf's asymmetry will be only accounted for by the
exponents $c_+$,  $c_-$ and the scale factors $\chi_+$, $\chi_-$.

We can
note that these expressions are close to the Weibull distribution, with the
addition of a power law prefactor to the exponential such that the Gaussian law
is retrieved for $c=2$. Following \cite{Sornette1,SorAnderSim2,AndersenRisk},
we call (\ref{eq:weibull}) the modified Weibull distribution. For $c<1$, the
pdf is a stretched exponential, also called sub-exponential. The exponent $c$
determines the shape of the distribution, which is fatter than an exponential
if $c<1$. The parameter $\chi$
controls the scale or characteristic width of the distribution. It plays
a role analogous to the standard deviation of the Gaussian law.
See chapter 6 of \cite{S2000} for a recent review on maximum likelihood
and other estimators of such generalized Weibull distributions.

\subsection{Transformation of the modified Weibull pdf into a Gaussian Law}

One advantage of the class of distributions (\ref{eq:weibull}) is that the
transformation into a Gaussian is particularly simple. Indeed, the expression
(\ref{eq:weibull}) is of the form (\ref{eq1}) with
\be
f(x)=2 \left( \frac{|x|}{\chi} \right) ^c~.
\ee
Applying the change of variable (\ref{eq2}) which reads
\be
   \label{eq:yx}
y_i=sgn(x_i)~\sqrt{2}~\left(\frac{ |x_i|}{\chi_i}\right)^{\frac{c_i}{2}} \;,
\ee
leads automatically to a Gaussian distribution.

These variables $Y_i$ then allow us to obtain the covariance matrix $V$ :
\be
V_{ij}=\frac{2}{T} \sum_{n=1}^T sgn(x_ix_j) \left( \frac{|x_i|}{\chi_i}
\right)^{\frac{c_i}{2}} \left( \frac{|x_j|}{\chi_j} \right)^{\frac{c_j}{2}}
\; ,
\ee
and thus the multivariate distributions $P({\bf y})$ and $P({\bf x})$ :
\be
\label{eq:px}
P(x_1, \cdots , x_N) = \frac{1}{2^N \pi^{N/2}\sqrt{V}}
~\exp\left[-\sum_{i,j}
V_{ij}^{-1}\left(\frac{|x_i|}{\chi_i}\right)^{c/2}
\left(\frac{|x_j|}{\chi_j}\right)^{c/2} \right]~
\left( \prod_{i=1}^N
\frac{c_i |x_i|^{c/2-1}}{\chi_i^{c/2}} ~e^{-\left( 
\frac{|x_i|}{\chi}\right)^c} \right).
\ee
Similar transforms hold, {\it mutatis mutandis}, for the asymmetric
case. Indeed, for asymmetric assets of interest for financial risk managers,
the equations (\ref{eq:pasym1}) and (\ref{eq:pasym2}) yields the following
    change of variable:
\bea
y_i&=&\sqrt{2}~\left(\frac{x_i}{\chi_i^+}\right)^{\frac{c_i^+}{2}}~~
{\rm and} ~~ x_i \ge 0	, \\
y_i&=&-\sqrt{2}~\left(\frac{|x_i|}{\chi_i^-}\right)^{\frac{c_i^-}{2}}
~~ {\rm and} ~~ x_i <0~.
\eea
This allows us to define the correlation matrix $V$ and to
obtain the multivariate distribution $P({\bf x})$, generalizing
equation (\ref{eq:px}) for asymmetric assets. Since this expression is
rather cumbersome and nothing but a straightforward generalization of
(\ref{eq:px}), we do not write it here.

\subsection{Empirical tests and estimated parameters}

\begin{figure}
\begin{center}
\includegraphics[width=15cm]{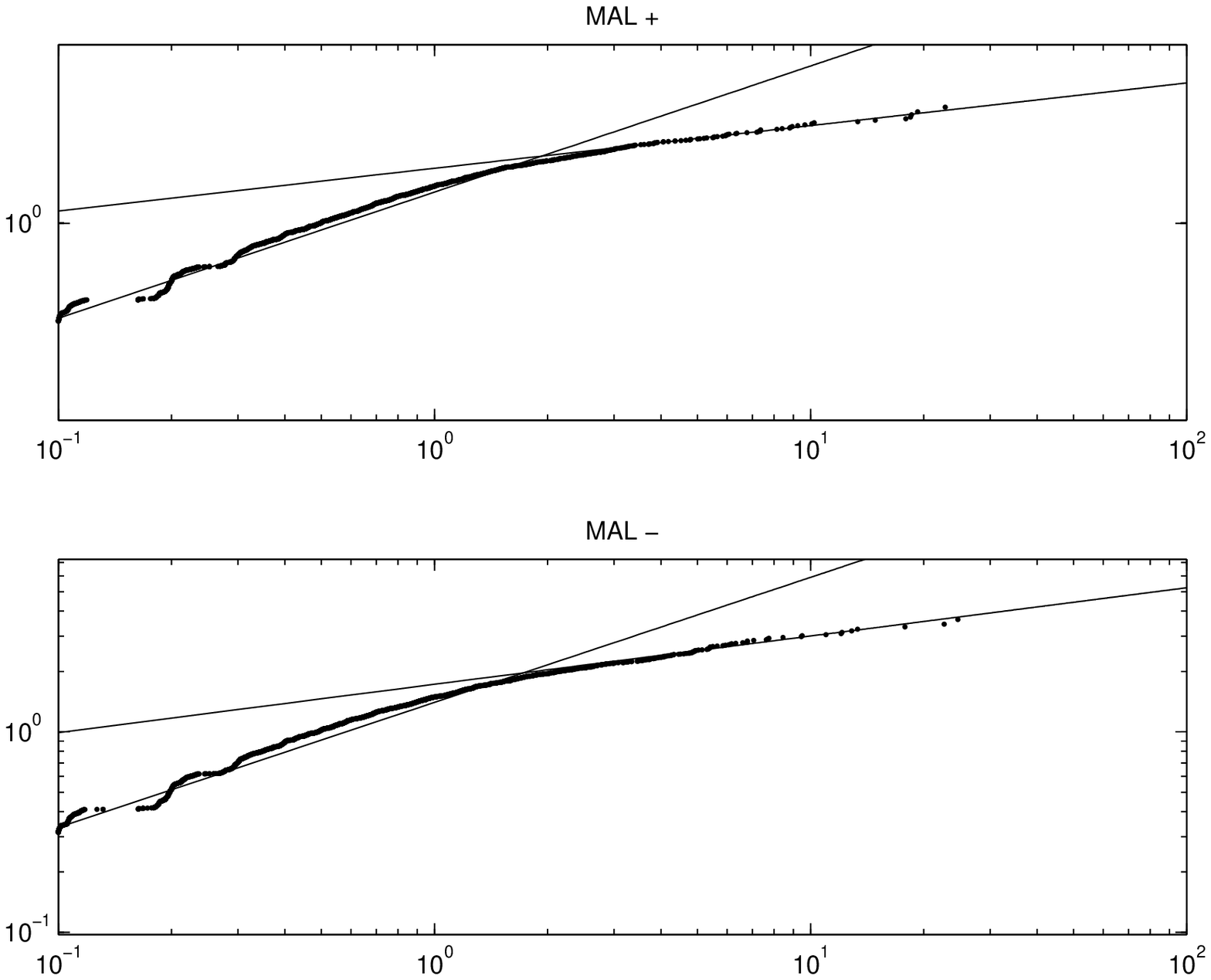}
\end{center}
\caption{\label{figYMAL} Graph of Gaussianized Malaysian Ringgit returns versus
Malaysian Ringgit returns, for the time interval from Jan. 1971
to Oct. 1998. The upper graph gives the positive tail and the
lower one the negative tail. The two straight lines represent the curves
$y=\sqrt{2} \left( \frac{x}{\langle \chi_\pm \rangle} \right)^{\langle c_\pm
\rangle}$ and $y=\sqrt{2} \left( \frac{x}{\chi_\pm} \right)^{c_\pm}$}
\end{figure}

\begin{figure}
\begin{center}
\includegraphics[width=15cm]{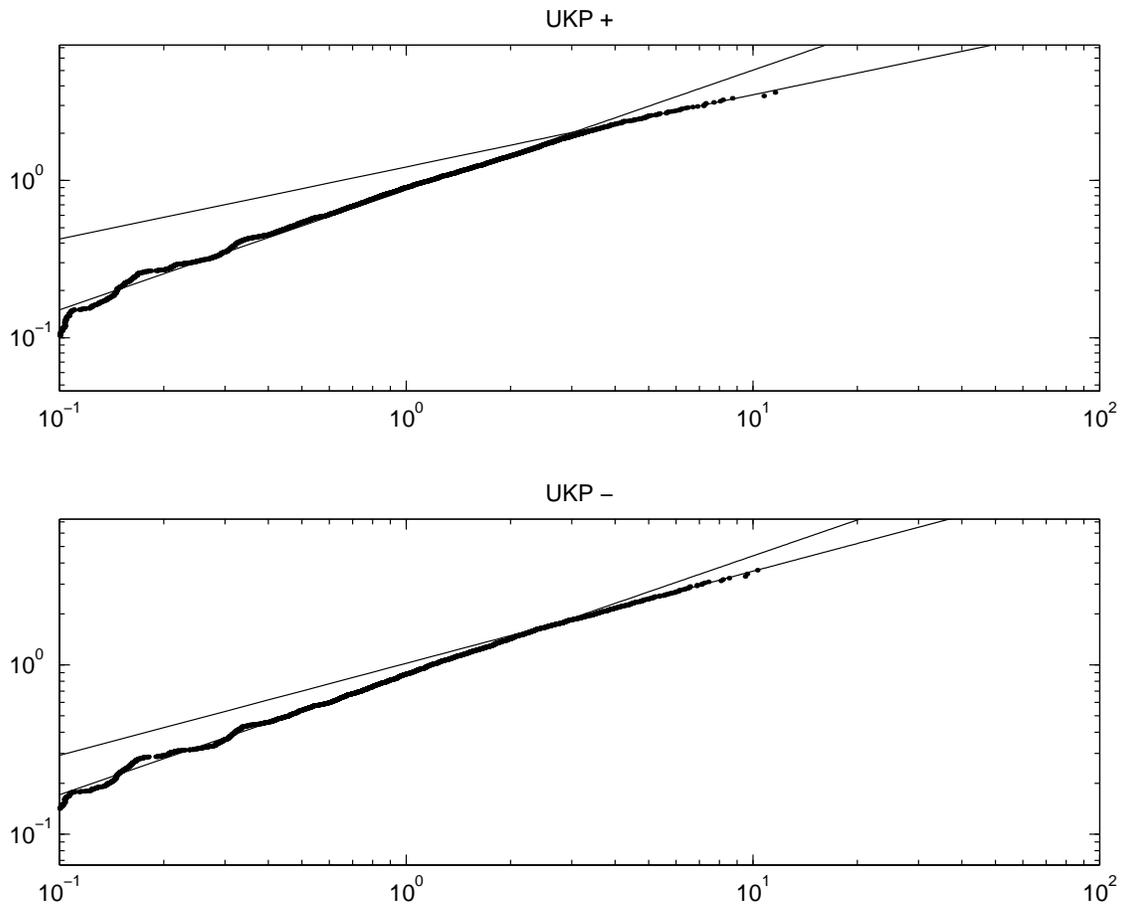}
\end{center}
\caption{\label{figYUKP} Graph of Gaussianized British Pound returns versus
British Pound returns, for the time interval from Jan. 1971
to Oct. 1998. The upper graph gives the positive tail and the lower
one the negative tail. The two straight lines represent the curves $y=\sqrt{2}
\left( \frac{x}{\langle \chi_\pm \rangle} \right)^{\langle c_\pm \rangle}$ and
$y=\sqrt{2} \left( \frac{x}{\chi_\pm} \right)^{c_\pm}$}
\end{figure}

\begin{figure}
\begin{center}
\includegraphics[width=15cm]{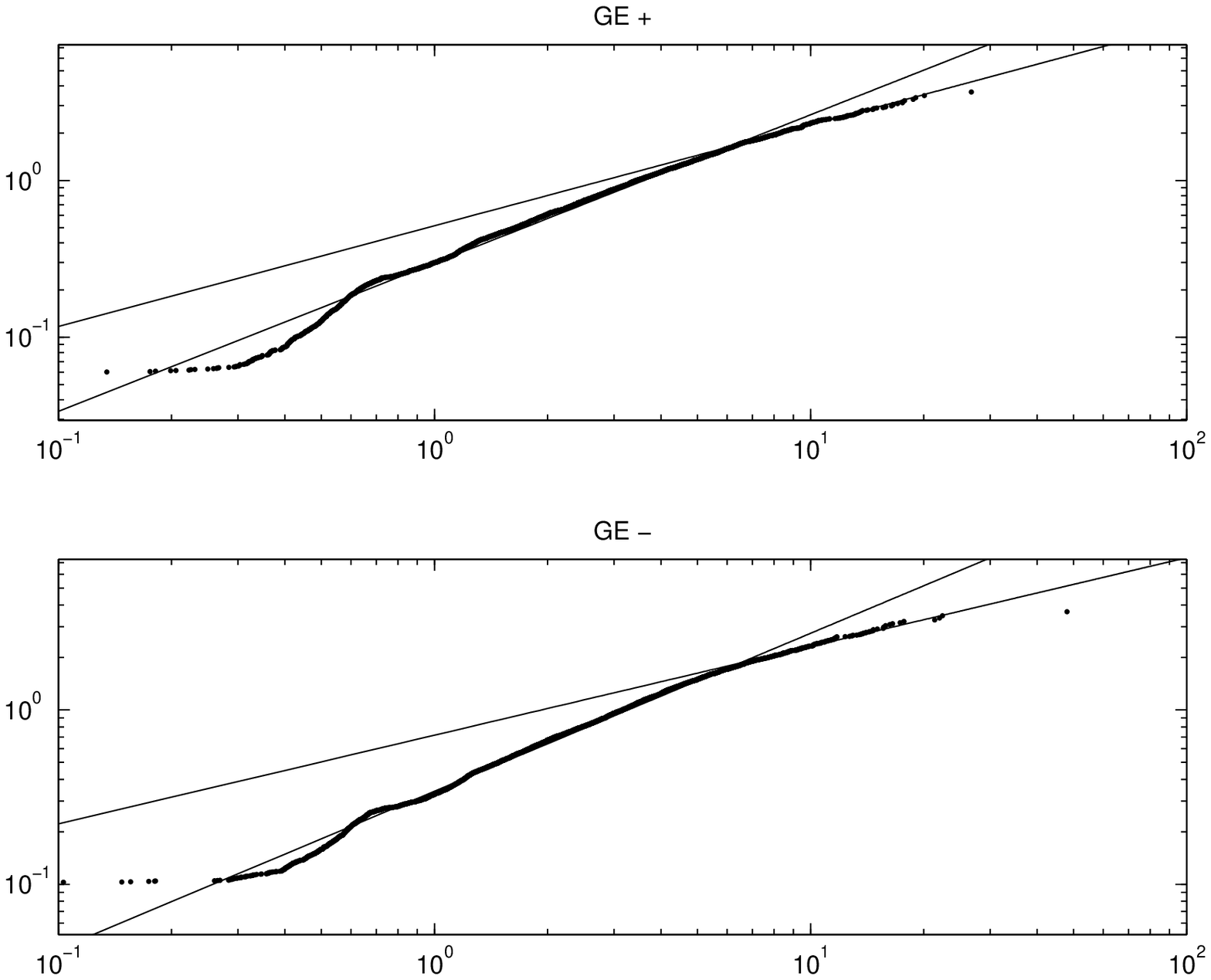}
\end{center}
\caption{\label{figYGE} Graph of Gaussianized General Electric returns versus
General Electric returns, for the time interval from Jan. 1970
to Dec. 2000. The upper graph gives the positive tail and the lower
one the negative tail. The two straight lines represent the curves $y=\sqrt{2}
\left( \frac{x}{\langle \chi_\pm \rangle} \right)^{\langle c_\pm \rangle}$ and
$y=\sqrt{2} \left( \frac{x}{\chi_\pm} \right)^{c_\pm}$}
\end{figure}

\begin{figure}
\begin{center}
\includegraphics[width=15cm]{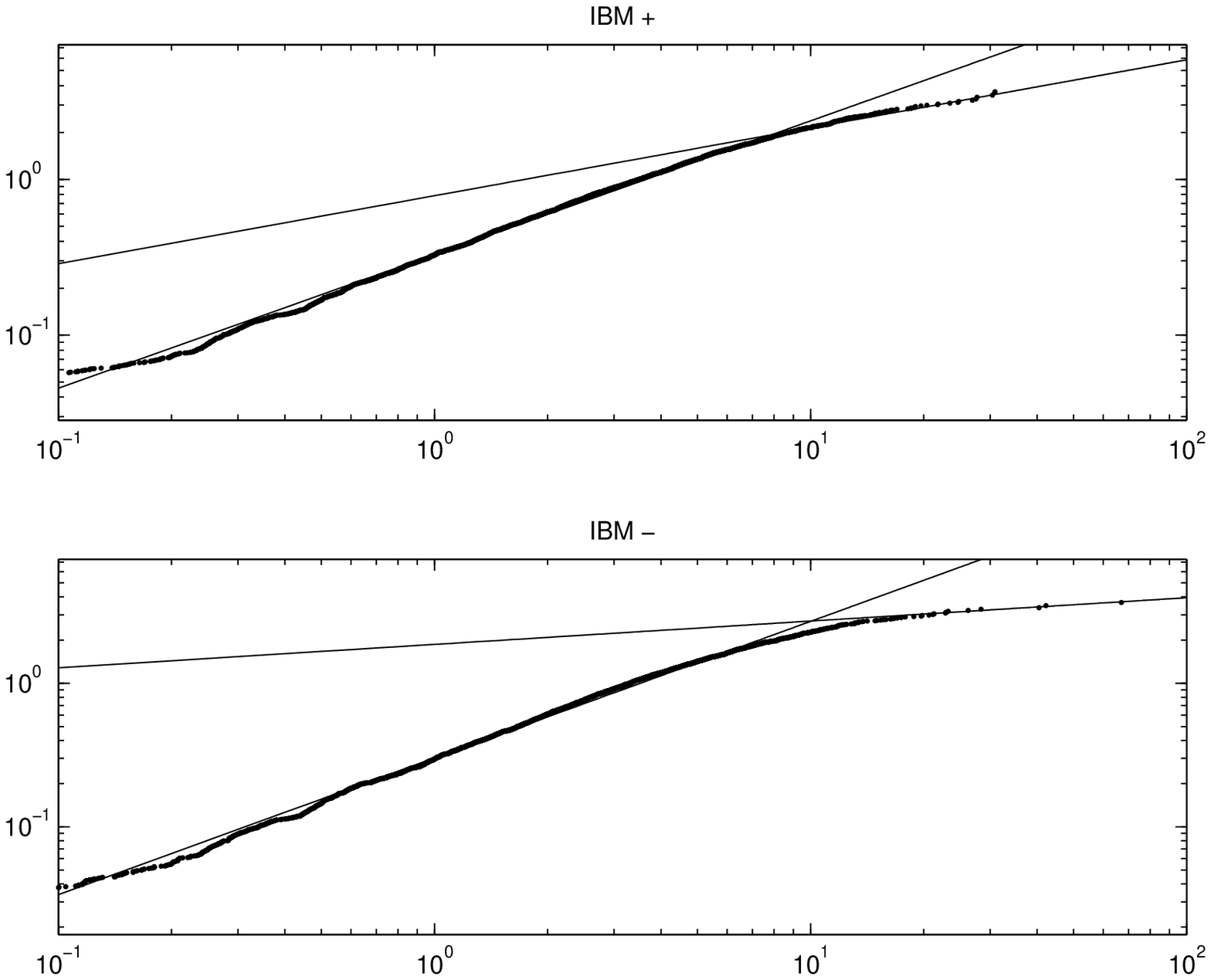}
\end{center}
\caption{\label{figYIBM} Graph of Gaussianized IBM returns versus
IBM returns, for the time interval from Jan. 1970
to Dec. 2000. The upper graph gives the positive tail and the lower
one the negative tail. The two straight lines represent the curves $y=\sqrt{2}
\left( \frac{x}{\langle \chi_\pm \rangle} \right)^{\langle c_\pm \rangle}$ and
$y=\sqrt{2} \left( \frac{x}{\chi_\pm} \right)^{c_\pm}$}
\end{figure}

\begin{figure}
\begin{center}
\includegraphics[width=15cm]{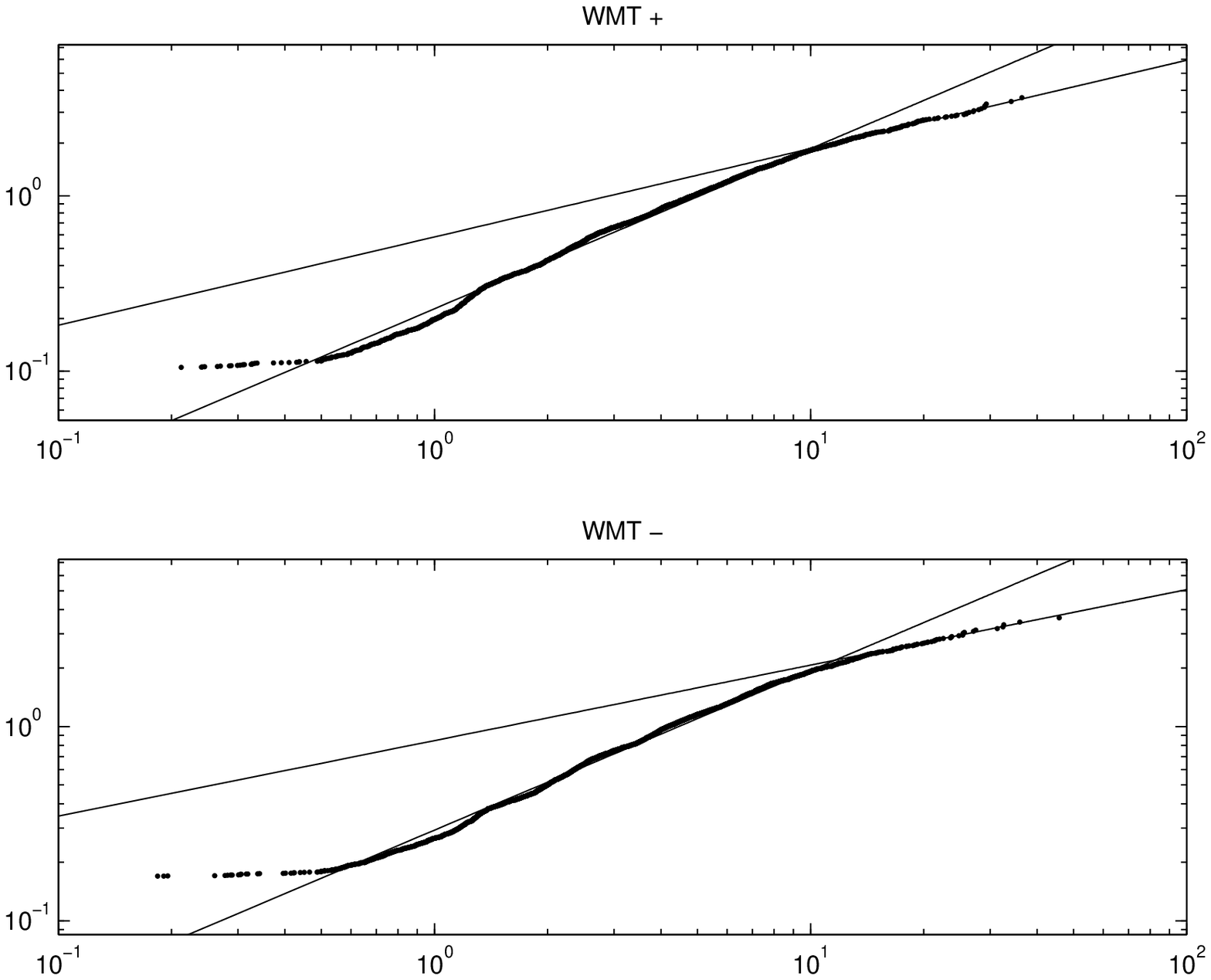}
\end{center}
\caption{\label{figYWMT} Graph of Gaussianized Wall Mart returns versus
Wall Mart returns, for the time interval from Sep. 1972
to Dec. 2000. The upper graph gives the positive tail and the lower
one the negative tail. The two straight lines represent the curves $y=\sqrt{2}
\left( \frac{x}{\langle \chi_\pm \rangle} \right)^{\langle c_\pm \rangle}$ and
$y=\sqrt{2} \left( \frac{x}{\chi_\pm} \right)^{c_\pm}$}
\end{figure}

In order to test the validity of our assumption, we have studied a
large basket of
financial assets including currencies and stocks. As an example, we present
in figures \ref{figYMAL} to \ref{figYWMT}  typical
log-log  plot of the transformed return variable $Y$ versus the
return variable $X$
for a certain number of assets. If our
assumption was right, we should observe a single straight line whose
slope is given by
$c/2$.  In contrast, we observe in general two approximately linear
regimes separated by
a cross-over. This means that the marginal distribution of returns
can be approximated
by two modified Weibull distributions, one for small returns which is
close to a Gaussian
law and one for large returns with a fat tail. Each regime is
depicted by its corresponding
straight line in the graphs. The exponents $c$ and the scale factors
$\chi$ for the different assets we have studied are given in tables
\ref{tab:1} for currencies and \ref{tab:2} for stocks. The
coefficients within brackets are the
coefficients estimated for small returns while the non-bracketed coefficients
correspond to the second fat tail regime.

The first point to note is the difference between currencies and
stocks. For small as
well as for large returns,
the exponents $c_-$ and $c_+$ for
currencies (excepted Poland and Thailand) are all close to each other.
Additional tests are required to establish whether their relatively
small differences
are statistically significant. Similarly, the scale factors are also
comparable.
In contrast, many stocks exhibit a large asymmetric behavior for large returns
with $c_+ - c_- \gtrsim 0.5$ in about one-half of the investigated stocks.
This means that the tails of the large negative returns (``crashes'')~
are often much fatter
than those of the large positive returns (``rallies'').

The second important point is that, for small returns, many stocks
have an exponent
$\langle c_+ \rangle \approx \langle c_- \rangle \simeq 2$ and thus
have a  behavior not far from a
pure Gaussian in the bulk of the distribution, while the average 
exponent for currencies is about
$1.5$ in the same ``small
return'' regime.
Therefore, even for small returns, currencies exhibit
a strong departure from Gaussian behavior.

In conclusion, this empirical study
shows that the modified Weibull parameterization, although not exact on
the entire range of variation of the returns $X$, remains consistent within
each of the two regimes of small versus large returns, with a sharp
transition between
them. It seems especially relevant in the tails of the return
distributions, on which we
shall focus our attention next.

\section{Cumulant expansion of the portfolio return distribution}
\label{sec:cumsep}
\subsection{link between moments and cumulants}

Before deriving the main result of this section, we recall a standard
relation between moments and cumulants that we need below.

The moments $M_n$ of the distribution $P$ are defined by
\be
\hat P(k) = \sum_{n=0}^{+ \infty} \frac{(ik)^n}{n!}M_n~,
\ee
where $\hat P$ is the characteristic function, i.e., the Fourier transform of
$P$~:
\be
\hat P(k) = \int_{-\infty}^{+ \infty} dS~P(S) e^{ikS}~.
\ee
Similarly, the cumulants $C_n$ are given by
\be
\hat P(k) = \exp \left(\sum_{n=1}^{+ \infty} \frac{(ik)^n}{n!}C_n \right)~.
\label{ghlqjfjgf}
\ee

Differentiating $n$ times the equation
\be
\ln \left(\sum_{n=0}^{+ \infty} \frac{(ik)^n}{n!}M_n \right) = \sum_{n=1}^{+
\infty} \frac{(ik)^n}{n!}C_n \; ,
   \ee
we obtain the following recurrence relations between the moments and the
cumulants~:
\bea
M_n & = & \sum_{p=0}^{n-1}  {{n-1} \choose{p}} M_p C_{n-p} \; ,\\
\label{eq:cum}
C_n & = &  M_n - \sum_{p=1}^{n-1} {{n-1} \choose {n-p}} C_p M_{n-p} \;.
\eea

In the sequel, we will first evaluate the moments, which turns out to
be easier, and
then using eq (\ref{eq:cum}) we will be able to calculate the
cumulants.

\subsection{Symmetric assets}

We start with the expression of the distribution of
the weighted
sum of $N$ assets~:
\be
P_S(s)=\int_{R^N} d{\bf x} \; P({\bf x}) \delta(\sum_{i=1}^N w_i x_i-s) \; ,
\label{eq:defmom}
\ee
where $\delta(\cdot)$ is the Dirac distribution.
Using the change of variable (\ref{eq2}), allowing us to go from the
asset returns
$X_i$'s to the transformed returns $Y_i$'s, we get
\be
P_S(s) = \frac{1}{(2 \pi)^{N/2}
\sqrt{\det(V)}} \int_{R^N} d{\bf y} \;   e^{-\frac{1}{2} {\bf
y^t}V^{-1}{\bf y} }
\; \delta(\sum_{i=1}^N w_i sgn(y_i)f^{-1}(y_i^2)-s) \; .
\ee
Taking its Fourier transform $\hat P_S(k) = \int ds P_S(s) e^{iks}$, we obtain
\be
\label{eqq}
\hat P_S(k) = \frac{1}{(2 \pi)^{N/2} \sqrt{\det(V)}}  \int_{R^N} d{\bf y}
\;   e^{-\frac{1}{2}
{\bf y^t}V^{-1}{\bf y} + ik \sum_{i=1}^N w_i sgn(y_i)f^{-1}(y_i^2)} \; ,
\ee
where $\hat P_S$ is the characteristic function of $P_S$.

In the particular case of interest here where the marginal distributions
of the variables $X_i$'s are the modified Weibull pdf,
\be
f^{-1}(y_i)=\chi_i |\frac{y_i}{\sqrt{2}}|^{q_i}
\ee
with
\be
q_i=2/c_i~,
\ee
the equation
(\ref{eqq}) becomes
\be
\label{eq2.5}
\hat P_S(k) = \frac{1}{(2 \pi)^{N/2} \sqrt{\det(V)}}  \int_{R^N} d{\bf y}
\;   e^{-\frac{1}{2}
{\bf y^t}V^{-1}{\bf y} + ik \sum_{i=1}^N w_i sgn(y_i) \chi_i
|\frac{y_i}{\sqrt{2}}|^{q_i}} \; .
\ee
The task in front of us is to evaluate this expression through the
determination
of the moments and/or cumulants.

\subsubsection{Case of independent assets}
In this case, the cumulants can be obtained explicitely \cite{Sornette1}.
Indeed, the expression (\ref{eq2.5}) can be expressed as a product of integrals
of the form
\be
\int_0^{+\infty} du \;e^{-\frac{u^2}{2}+ik w_i \chi_i \left(
\frac{u}{\sqrt{2}}\right)^{q_i}} \; .
\ee
We obtain
\be
\label{eq:sumcum}
C_{2n}=\sum_{i=1}^N c(n,q_i) (\chi_i w_i)^{2n} \; ,
\ee
and
\be
c(n,q_i) = (2n)! \left\{ \sum_{p=0}^{n-2} (-1)^n \frac{\Gamma
\left(
q_i(n-p)+\frac{1}{2} \right)}{(2n-2p)! \pi^{1/2}} \left[ \frac{\Gamma \left(
q_i+\frac{1}{2} \right)}{2! \pi^{1/2}} \right]^p - \frac{(-1)^n}{n}\left[
\frac{\Gamma \left(
q_i+\frac{1}{2} \right)}{2! \pi^{1/2}} \right] ^n \right\} \; .
\ee

Note that the coefficient $c(n,q_i)$ is the cumulant of order $n$ of the
marginal distribution (\ref{eq:weibull}) with $c=2/q_i$ and $\chi=1$. The
equation (\ref{eq:sumcum}) expresses simply the fact that the cumulants
of the sum of independent variables is the sum of the cumulants of each
variable.
The odd-order cumulants are zero due to the symmetry of the distributions.

\subsubsection{Case of dependent assets}

Here, we restrict our exposition to the case of two random variables. The
case with $N$ arbitrary can be treated in a similar way but involves
rather complex formulas. The equation (\ref{eq2.5}) reads
\bea
\hat P_S(k) = \frac{1}{2 \pi \sqrt{1-\rho^2}} \int dy_1 dy_2~\exp \left[
-\frac{1}{2} y^t V^{-1} y + ik \left( \chi_1 w_1 sgn(y_1)
\left| \frac{y_1}{\sqrt{2}}\right|^{q_1} + \right. \right. {\nonumber} \\
\left. \left. + \chi_2 w_2 sgn(y_2) \left| \frac{y_2}{\sqrt{2}}
\right|^{q_2} \right)
\right]~,
\eea
and we can show (see appendix \ref{app:B}) that the moments read
\be
\label{eq:amom}
M_n = \sum_{p=0}^n {n \choose p}  w_1^p w_2^{n-p}
\gamma_{q_1 q_2}(n,p)~,
\ee
with
\bea
\label{eq:gammasym}
\gamma_{q_1 q_2}(2n,2p) &=& \chi_1^{2p} \chi_2^{2(n-p)}\frac{\Gamma \left( q_1p
+\frac{1}{2} \right) \Gamma \left(q_2(n-p)+\frac{1}{2}
\right)}{\pi}~
{_2F_1} \left(-q_1p, -q_2(n-p) ; \frac{1}{2} ; \rho^2 \right) \; , \\
\gamma_{q_1 q_2}(2n,2p+1) &=&2\chi_1^{2p+1} \chi_2^{2(n-p)-1} \frac{\Gamma
\left( q_1p+1+\frac{q_1}{2} \right) \Gamma \left(q_2(n-p)+1-\frac{q_2}{2}
\right)}{\pi} \rho~{_2F_1}\left( -q_1p - \frac{q_1-1}{2},\right.\nonumber \\
&& \left. ,  -q_2(n-p)+ \frac{q_2+1}{2} ;
\frac{3}{2} ;\rho^2 \right) ~,
\eea
where ${_2F_1}$ is an hypergeometric function.

These two relations allow us to calculate the moments and cumulants for
any possible values of $q_1=2/c_1$ and $q_2=2/c_2$. If one of the
$q_i$'s is an integer, a
simplification occurs and the coefficients $\gamma(n,p)$ reduce to
polynomials. In the simpler case where all the $q_i$'s are odd integer the
expression of moments becomes~:
\be
\label{eq:mom2}
M_n=\sum_{p=0}^n {n \choose p} \left( w_1 {\chi_1} \right)^p \left( w_2
{\chi_2} \right)^{n-p}  \sum_{s=0}^{min\{q_1p,q_2(n-p)\}} {\rho}^s~
s!~ a_{s}^{(q_1p)} a_s^{(q_2(n-p))} ~,
\ee
with
\bea
a_{2p}^{(2n)} & = & (2p)! {2(n-p) \choose 2n}~
(2(n-p)-1)!!=\frac{(2n)!}{2^{n-p}~  (n-p)!}  \; ,\\
a_{2p+1}^{(2n)} & = & 0  \; ,\\
a_{2p}^{(2n+1)} & = & 0  \; ,\\
a_{2p+1}^{(2n+1)} & = & (2p+1)! {2(n-p) \choose 2n+1}~
(2(n-p)-1)!!= \frac{(2n+1)!}{2^{n-p} ~ (n-p)!} \; .
\eea

\subsection{Non-symmetric assets}

In the case of asymmetric assets, we have to consider the formula
(\ref{eq:pasym1}-\ref{eq:pasym2}), and using the same notation as in the
previous section, the moments are again given by (\ref{eq:amom}) with
the coefficient $\gamma(n,p)$ now equal to~:
\bea
\gamma(n,p)=\frac{(-1)^n (\chi_1^{-})^p (\chi_2^{-})^{n-p}}{4 \pi} \left[
\Gamma\left( \frac{q_1^- p +1}{2} \right) \Gamma\left(\frac{q_2^-(n-p)+1}{2}
\right)~_2F_1 \left( - \frac{q_1^- p}{2} , - \frac{q_2^- (n-p)}{2} ;
\frac{1}{2} ; \rho^2 \right) +  \right. \nonumber \\
\left. + 2 \Gamma\left(
\frac{q_1^- p}{2} + 1 \right) \Gamma\left(\frac{q_2^-(n-p)}{2} + 1
\right)\rho ~_2F_1 \left( - \frac{q_1^- p - 1}{2} , - \frac{q_2^- (n-p) - 1}{2}
; \frac{3}{2} ; \rho^2 \right) \right] +  \nonumber \\
+ \frac{ (-1)^p(\chi_1^{-})^p (\chi_2^{+})^{n-p}}{4 \pi} \left[ \Gamma\left(
\frac{q_1^- p +1}{2} \right) \Gamma\left(\frac{q_2^+(n-p)+1}{2}
\right)~_2F_1 \left( - \frac{q_1^- p}{2} , - \frac{q_2^+ (n-p)}{2} ;
\frac{1}{2} ; \rho^2 \right) +  \right. \nonumber \\
\left . - 2 \Gamma\left(
\frac{q_1^- p}{2} + 1 \right) \Gamma\left(\frac{q_2^+(n-p)}{2} + 1
\right)\rho ~_2F_1 \left( - \frac{q_1^- p - 1}{2} , - \frac{q_2^+ (n-p) - 1}{2}
; \frac{3}{2} ; \rho^2 \right) \right] +  \nonumber \\
+ \frac{(-1)^{n-p}(\chi_1^{+})^p (\chi_2^{-})^{n-p}}{4 \pi} \left[ \Gamma\left(
\frac{q_1^+ p +1}{2} \right) \Gamma\left(\frac{q_2^-(n-p)+1}{2}
\right)~_2F_1 \left( - \frac{q_1^+ p}{2} , - \frac{q_2^- (n-p)}{2} ;
    \frac{1}{2} ; \rho^2 \right) +  \right. \nonumber \\
\left . - 2 \Gamma\left(
\frac{q_1^+ p}{2} + 1 \right) \Gamma\left(\frac{q_2^-(n-p)}{2} + 1
\right)\rho ~_2F_1 \left( - \frac{q_1^+ p - 1}{2} , - \frac{q_2^- (n-p) -
1}{2} ;   \frac{3}{2} ; \rho^2 \right) \right]+  \nonumber \\
+\frac{(\chi_1^{+})^p (\chi_2^{+})^{n-p}}{4 \pi} \left[ \Gamma\left(
\frac{q_1^+ p +1}{2} \right) \Gamma\left(\frac{q_2^+(n-p)+1}{2}
\right)~_2F_1 \left( - \frac{q_1^+ p}{2} , - \frac{q_2^+ (n-p)}{2} ;
    \frac{1}{2} ; \rho^2 \right) +  \right. \nonumber \\
\left . + 2 \Gamma\left(
\frac{q_1^+ p}{2} + 1 \right) \Gamma\left(\frac{q_2^+(n-p)}{2} + 1
\right)\rho ~_2F_1 \left( - \frac{q_1^+ p - 1}{2} , - \frac{q_2^+ (n-p) - 1}{2}
;   \frac{3}{2} ; \rho^2 \right) \right] \nonumber~.\\
\eea
This formula is obtained in the same way as for the formulas given in the
symmetric case. We retrieve the formula (\ref{eq:gammasym}) as it should if the
coefficients with index '+' are equal to the coefficients with index '-'.

\subsection{Empirical tests}

Extensive tests have been performed for currencies under the assumption that
the distributions of asset returns are symmetric \cite{Sornette1}.

As an exemple, let us consider
the Swiss franc and the Japanese Yen against the US dollar. The calibration of
the modified Weibull distribution to the tail of the empirical histogram of
daily returns give $(q_{CHF}=1.75 , c_{CHF}=1.14, \chi_{CHF}=2.13)$ and
$(q_{JPY}=2.50 , c_{JPY}=0.8, \chi_{JPY}=1.25)$ and their correlation
coefficient is
$\rho=0.43$.

\begin{figure}
\begin{center}
\includegraphics[width=15cm]{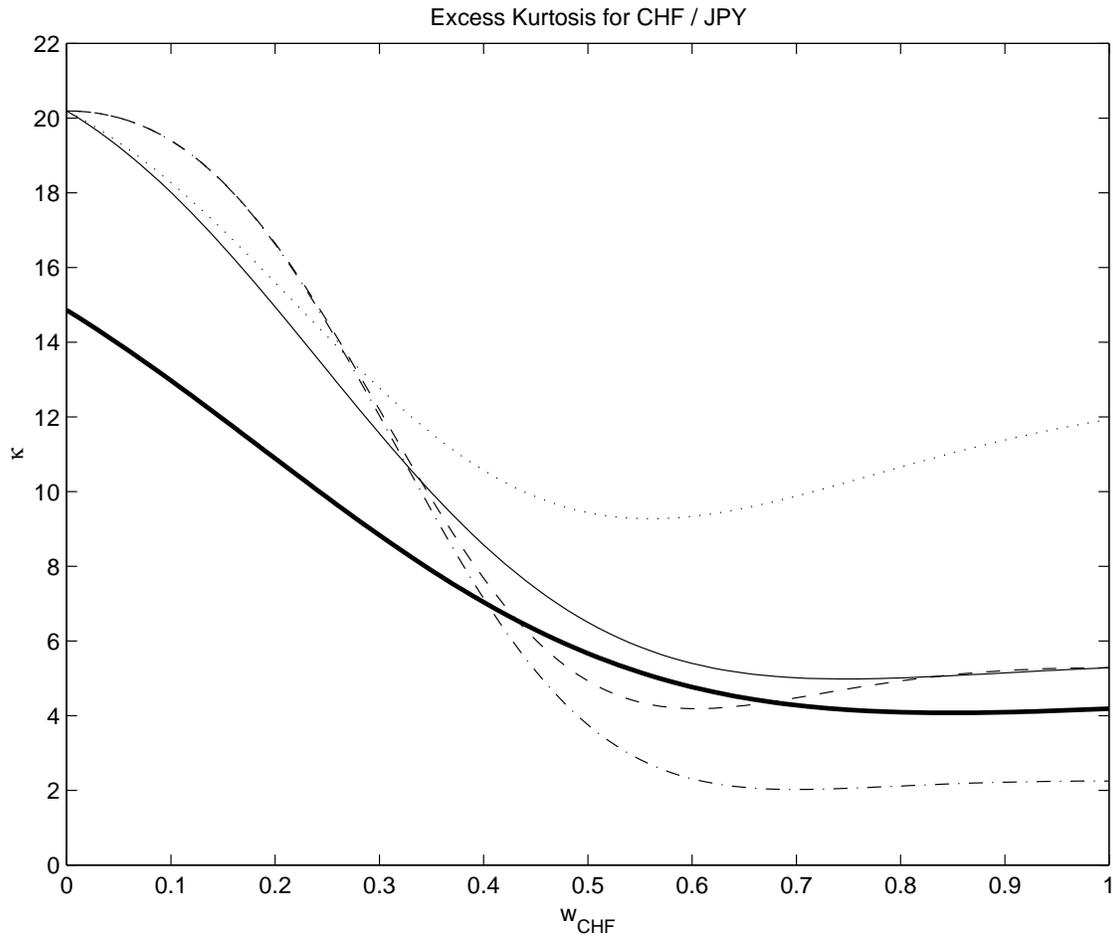}
\end{center}
\caption{\label{fig2.2} Excess kurtosis of the distribution of the price
variation $w_{CHF} x_{CHF} + w_{JPY}
x_{JPY}$ of the portfolio made of
a fraction $w_{CHF}$ of Swiss franc and a fraction
$w_{JPY}=1-w_{CHF}$ of the Japanese Yen against the US dollar, as a function of
$w_{CHF}$.
Thick solid line : empirical curve, thin solid line : theoretical curve, dashed
line : theoretical curve with $\rho=0$ (instead of $\rho=0.43$), dotted line:
theoretical curve with $q_{CHF}=2$ rather than $1.75$ and dashed-dotted line:
theoretical curve with $q_{CHF}=1.5$. The excess kurtosis has been evaluated
for the time interval from Jan. 1971 to Oct. 1998.}
\end{figure}

Figure \ref{fig2.2} plots the excess kurtosis of the sum
$w_{CHF} x_{CHF} + w_{JPY} x_{JPY}$ as a function of $w_{CHF}$,
with the constraint $w_{CHF} + w_{JPY} =1$.
The thick solid line is determined empirically, by direct calculation
of the kurtosis from the data.
The thin solid line
is the theoretical prediction using our theoretical formulas with the
empirically determined exponents $c$ and characteristic scales $\chi$
given above.
While there is a non-negligible difference,
the empirical and theoretical excess kurtosis have essentially the same
behavior with their minimum reached almost at the same value
of $w_{CHF}$.

Three origins of the discrepancy between theory and empirical data
can be invoked. First, as already pointed out in the preceding section,
the modified Weibull distribution with constant exponent and scale parameters
describes accurately only the tail of the empirical distributions while, for
small returns, the empirical distributions are close to a Gaussian law.
While putting a strong emphasis on large fluctuations,
cumulants of order $4$ are still significantly sensitive to the bulk
of the distributions. Moreover, the excess kurtosis is normalized
by the square second-order cumulant, which is almost exclusively
sensitive to the bulk
of the distribution.
Cumulants of higher order should thus be better
described by the modified Weibull distribution. However, a careful
comparison between theory and data would then be hindered by the
difficulty in estimating reliable empirical cumulants of high
order. This estimation problem is often invoked as a criticism
against using high-order moments or cumulants. Our approach
suggests that this problem can be in large part circumvented
by focusing on the estimation of a reasonable parametric expression for the
   probability density or distribution function of the assets returns.
The second possible origin of the discrepancy between theory and data
is the existence of a weak asymmetry of the empirical distributions,
particularly of the Swiss franc, which has not been taken into account.
The figure also suggests that an error in the determination of the
exponents $c$
can also contribute to the discrepancy.

In order to investigate the sensitivity with respect to the choice of
the parameters $q$ and $\rho$, we have also constructed the
dashed line corresponding to the theoretical curve with $\rho=0$
(instead of $\rho=0.43$)
and the dotted line corresponding to the
theoretical curve with $q_{CHF}=2$ rather than $1.75$. Finally,
the dashed-dotted line corresponds to the
theoretical curve with $q_{CHF}=1.5$.
We observe that the dashed line remains rather close to the
thin solid line while the dotted line departs significantly when 
$w_{CHF}$ increases.
Therefore, the most sensitive parameter is $q$, which is natural
because it controls directly the extend of the fat tail of the
distributions.

\begin{figure}
\begin{center}
\includegraphics[width=15cm]{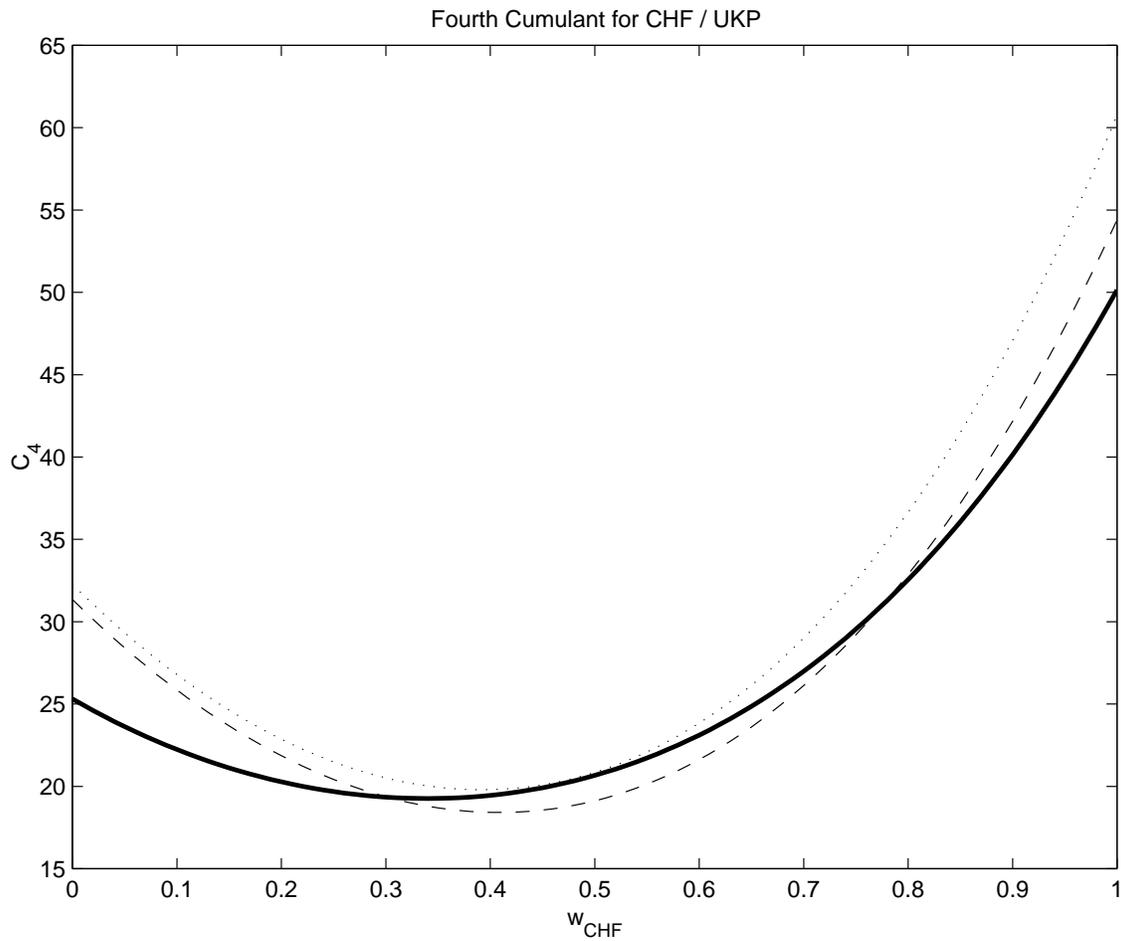}
\end{center}
\caption{\label{fig:c4} Fourth cumulant for a portfolio made of a fraction
$w_{CHF}$ of Swiss Franc and $1-w_{CHF}$ of British Pound. The thick solid line
represents the empirical cumulant while the dotted line represents the
theoretical cumulant under the symmetric assumption. The dashed line shows the
theoretical cumulant when the
slight asymmetry of the assets has been taken into account. This
cumulant has been evaluated
for the time interval from Jan. 1971 to Oct. 1998.}
\end{figure}

In order to account for the effect of asymmetry, we have plotted the fourth
cumulant of a portfolio composed of Swiss Francs and British Pounds. On figure
\ref{fig:c4}, the solid line represents the empirical cumulant while the dashed
line shows the theoretical cumulant. The agreement between the two curves is
better than under the symmetric asumption. Note once again that an accurate
determination of the parameters is the key point to obtain a good agreement
between empirical data and theoretical prediction. As we can see in figure
\ref{fig:c4}, the paramaters of the Swiss Franc seem well adjusted since the
theoretical and empirical cumulants are both very close when $w_{CHF}
\simeq 1$, i.e., when
the Swiss Franc is almost the sole asset in the portfolio, while when $w_{CHF}
\simeq 0$, the theoretical cumulant is far from the empirical one, i.e., the
parameters of the Bristish Pound are not sufficiently well-adjusted.

\section{Can you have your cake and eat it too ?}
\label{sec:9}

Now that we have shown how to accurately estimate the multivariate
distribution fonction of the assets return, let us come back to the
portfolio selection problem. In figure \ref{fig:GEF}, we can see that the
expected return of the portfolios with minimum risk according to $C_n$
decreases when $n$ increases. But, this is not the general situation.

\begin{figure}
\begin{center}
\includegraphics[width=15cm]{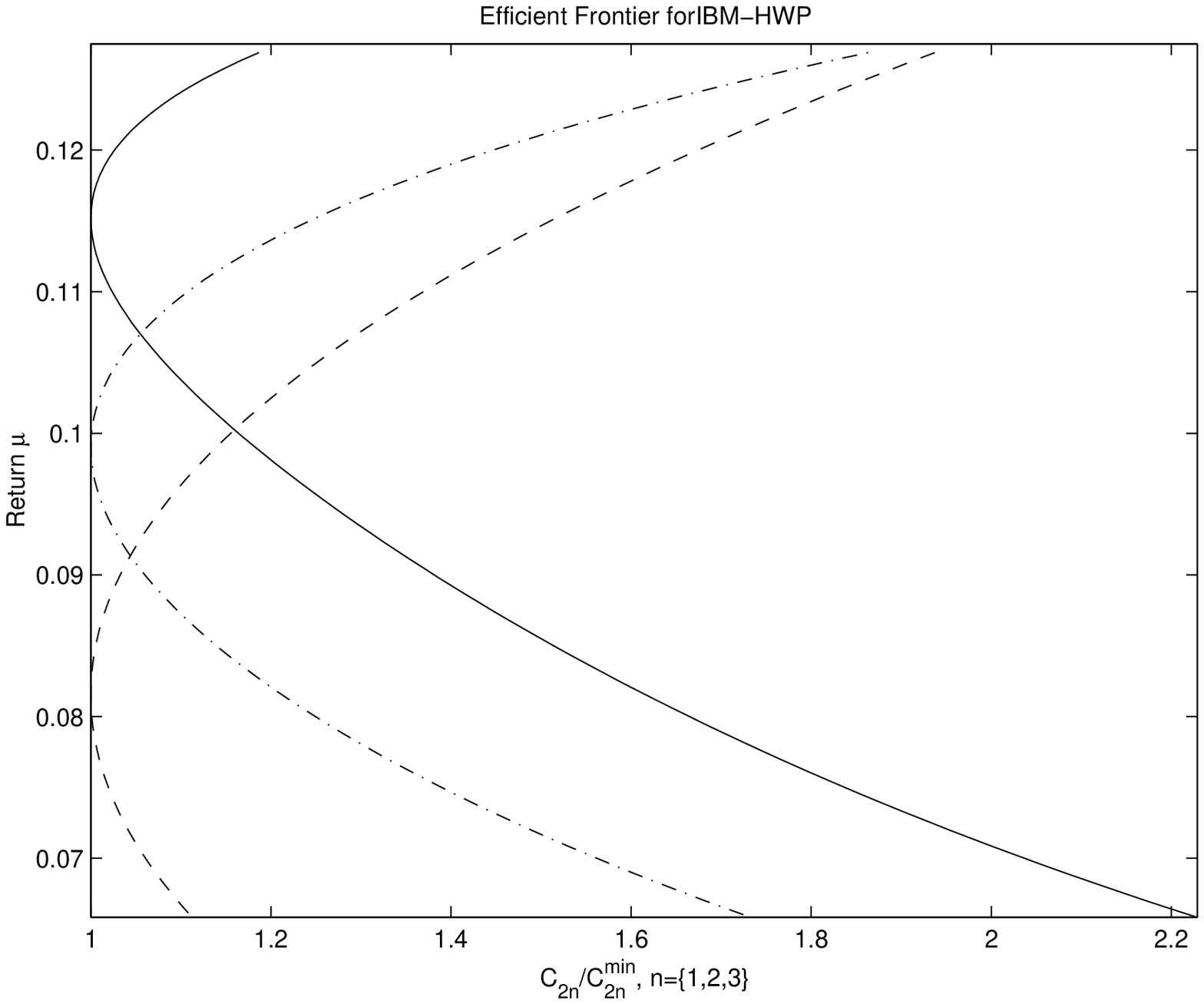}
\end{center}
\caption{\label{fig:ef1} Efficient frontier for a portfolio composed of two
stocks: IBM and Hewlett-Packard. The dashed line represents the efficient
frontier with respect to the second cumulant, i.e., the standard Markovitz
efficient frontier, the dash-dotted line represents the efficient frontier
with respect to the fourth cumulant and the solid line is the 
efficient frontier
with respect to the sixth cumulant. The data set used covers the time interval
from Jan. 1977 to Dec 2000.}
\end{figure}

\begin{figure}
\begin{center}
\includegraphics[width=15cm]{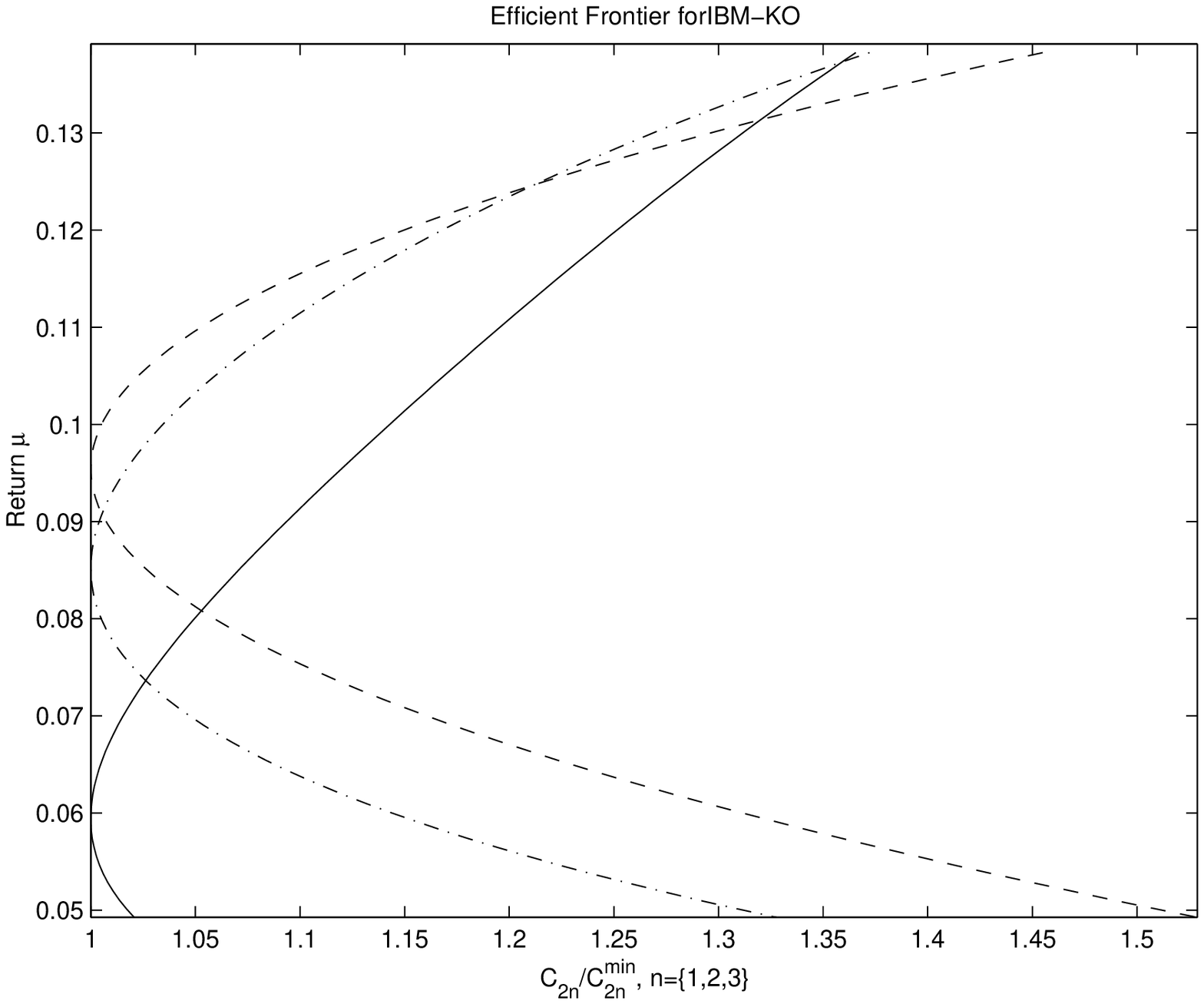}
\end{center}
\caption{\label{fig:ef2} Efficient frontier for a portfolio composed of two
stocks: IBM and Coca-Cola. The dashed line represents the efficient
frontier with respect to the second cumulant, i.e., the standard Markovitz
efficient frontier, the dash-dotted line represents the efficient frontier
with respect to the fourth cumulant and the solid line the efficient frontier
with repect to the sixth cumulant. The data set used covers the time interval
from Jan. 1970 to Dec 2000.}
\end{figure}

Figure \ref{fig:ef1} and \ref{fig:ef2} show the generalized efficient
frontiers using $C_2$ (Markovitz case), $C_4$ or $C_6$ as relevant measures of
risks, for two portfolios composed of two stocks : IBM and Hewlett-Packard in
the first case and IBM and Coca-Cola in the second case.

Obviously, given a certain amount of risk, the mean return of the portfolio
changes when the cumulant considered changes. It is interesting to note that,
in figure \ref{fig:ef1}, the minimisation of large risks, i.e., with
respect to $C_6$, increases the average return  while, in figure
\ref{fig:ef2}, the minimisation of large risks lead to decrease the average
return.

This allows us to make precise and quantitative the previously reported
empirical observation that it is possible to ``have your cake and eat it too''
\cite{AndersenRisk}.
We can indeed give a general criterion to determine under which values of the
parameters (exponents $c$ and characteristic scales $\chi$ of the distributions
of the asset returns) the average return of the
portfolio may increase while the large risks decrease {\it at the
same time}, thus
allowing one to gain on both account (of course, the small risks
quantified by the variance
will then increase).
For two independent
assets, assuming that the cumulants of order $n$ and $n+k$ of the portfolio
admit a minimum in the interval $]0,1[$, we can show that
\be
\mu_n^* < \mu_{n+k}^*
\ee
if and only if
\be
\left(\mu(1) - \mu(2) \right) ~\cdot ~
\left[ \left( \frac{C_n(1)}{C_n(2)} \right)^{\frac{1}{n-1}} -
\left( \frac{C_{n+k}(1)}{C_{n+k}(2)} \right)^{\frac{1}{n+k-1}} \right] > 0~,
\ee
where $\mu_n^*$ denotes the return of the portfolio evaluated with respect to
the minimum of the cumulant of order $n$ and $C_n(i)$ is the cumulant of order
$n$ for the asset $i$.

The proof of this result and its generalisation to $N>2$ are given in appendix
\ref{app:C}.  In fact, we have observed that when the exponent $c$ of the
assets remains sufficiently different, this result still holds in presence of
dependence between assets. This last empirical observation in the presence
of dependence between assets has not been proved mathematically.
It seems reasonable for assets with moderate
dependence while it may fail when the dependence becomes too
strong as occurs for comonotonic assets.

For the assets considered above, we have found $\mu_{IBM} = 0.13$,
$\mu_{HWP}=0.07$, $\mu_{KO}=0.05$ and
\bea
\frac{C_2(IBM)}{C_2(HWP)} &= 1.76 >
\left(\frac{C_4(IBM)}{C_4(HWP)}\right)^\frac{1}{3} &= 1.03
>   \left( \frac{C_6(IBM)}{C_6(HWP)}\right)^\frac{1}{5} = 0.89 \\
\frac{C_2(IBM)}{C_2(KO)} &= 0.96 <
\left(\frac{C_4(IBM)}{C_4(KO)}\right)^\frac{1}{3} &= 1.01
< \left( \frac{C_6(IBM)}{C_6(KO)}\right)^\frac{1}{5} = 1.06~,
\eea
which shows that, for the portfolio IBM / Hewlett-Packard, the
efficient return is
an increasing function of the order of the cumulants while, for the portfolio
IBM / Coca-Cola, the inverse phenomenon occurs. This is exactly what 
is shown on
figures \ref{fig:ef1} and \ref{fig:ef2}.

The underlying intuitive mechanism is the following: if a portfolio
contains an asset with a rather fat tail (many ``large''
risks) but narrow waist (few ``small'' risks) with very little return to
gain from it, minimizing the variance $C_2$ of the return portfolio
will overweight
this asset which is wrongly perceived as having little risk due to its small
variance (small waist). In contrast, controlling for the larger risks
quantified by $C_4$ or $C_6$ leads to decrease the
weight of this asset in the portfolio, and correspondingly to
increase the weight of the more
profitable assets. We thus see that the effect of ``both decreasing
large risks and increasing profit'' appears when the asset(s) with
the fatter tails,
and therefore the narrower central part, has(ve) the smaller overall
return(s). A mean-variance
approach will weight them more than deemed appropriate from a prudential
consideration of large risks and consideration of profits.

  From a behavioral point of view, this phenomenon is very
interesting and can probably be linked with the fact that the main
risk measure considered by the agents is the volatility (or the
variance), so that the other dimensions of the risk, measured by
higher moments, are often neglected. This may sometimes offer the
opportunity  of increasing the expected return while lowering large
risks.

\section{Conclusion}

We have introduced three axioms that define a consistent set of
risk measures, in the spirit of \cite{A_etal98,A_etal99}.
Contrarily to the risk measures of \cite{A_etal98,A_etal99}, our consistent risk measures may
account for both-side risks and not only for down-side risks. Thus, they
supplement the notion of coherent measures of risk and are well adapted to the
problem of portfolio risk assessment and optimization. We have shown that
these risk measures, which contain centered moments (and cumulants with some
restriction) as particular examples, generalize them significantly. We have
presented a generalization of previous generalizations of the
efficient frontiers and of the CAPM based on these
risk measures in the cases of homogeneous and heterogeneous agents.
We have then proposed a simple but powerful specific von Mises representation
of multivariate distribution of returns that allowed us to obtain
new analytical results on and empirical
tests of a general framework for a portfolio
theory of non-Gaussian risks with non-linear correlations.
Quantitative tests have been presented on a basket of seventeen stocks
among the largest capitalization on the NYSE.

This work opens several novel interesting avenues for research. One consists
in extending the Gaussian copula assumption, for instance by using
the maximum-entropy
principle with non-extensive Tsallis entropies, known to be the
correct mathematical
information-theoretical representation of power laws. A second line of research
would be to extend the present framework to encompass simultaneously different
time scales $\tau$ in the spirit of \cite{MuzyQF} in the case of a
cascade model of volatilities.

\newpage

\appendix
\section{Description of the data set}
\label{app:DS}

We have considered a set of seventeen assets traded on the New York Stock
Exchange: Applied Material, Coca-Cola, EMC, Exxon-Mobil, General
Electric, General
Motors, Hewlett Packard, IBM, Intel, MCI WorldCom, Medtronic, Merck, Pfizer,
Procter \& Gambel, SBC Communication, Texas Instrument, Wall Mart.
These assets have been choosen since they are among the largest capitalizations
of the NYSE at the time of writing.

The dataset comes from the Center for Research in Security Prices (CRSP)
database and covers the time interval from the end of January 1995 to the end
of December 2000, which represents exactly 1500 trading days. The main
statistical features of the compagnies composing the dataset are presented in
the table \ref{table:stat}. Note the high kurtosis of each
distribution of returns as well as the large values of the observed minimum and
maximum returns compared with the standard deviations, that clearly
underlines the
non-Gaussian behavior of these assets.

\clearpage

\section{Generalized efficient frontier and two funds separation theorem}
\label{app:GEF}
Let us consider a set of $N$ risky assets $X_1, \cdots, X_N$ and a risk-free
asset $X_0$. The problem is to
find the optimal allocation of these assets in the following sense:
\be
\left\{
\begin{array}{l}
\inf_{w_i\in[0,1]} \rho_\alpha(\{w_i\})\\
\sum_{i \ge 0} w_i = 1 \\
\sum_{i \ge 0} w_i \mu(i) = \mu~,
\end{array}
\right.
\ee
In other words, we search for the portfolio ${\cal P}$ with minimum risk as
measured by any risk measure $\rho_\alpha$ obeying axioms I-IV of 
section 2 for a given
amount of expected return $\mu$ and normalized weights $w_i$. Short-sells are
forbidden except for the risk-free asset which can be lent and borrowed at the
same interest rate $\mu_0$. Thus, the weights $w_i$'s are assumed positive for
all $i\ge1$.

\subsection{Case of independent assets when the risk is measured by the
cumulants}

To start with a simple example, let us assume that the risky assets are
independent and that we choose to measure the risk with the cumulants
of their distributions of returns. The case
when the assets are dependent and/or when the risk is measured by any
$\rho_\alpha$ will be considered later. Since the assets are assumed
independent, the cumulant of order $n$ of the pdf of returns of the portfolio
is simply given by \be C_n = \sum_{i=1}^N {w_i}^n ~ C_n(i),
\ee
where $C_n(i)$ denotes the marginal n$^{th}$ order cumulant of the pdf of
returns of the asset $i$.
In order to solve this problem, let us introduce the Lagrangian
\be
\label{eq:L}
{\cal L} = C_n - {\lambda_1} \left( \sum_{i=0}^N w_i~ \mu(i) - \mu \right) -
\lambda_2 \left( \sum_{i=0}^N w_i  - 1\right), \ee
where ${\lambda_1}$ and $\lambda_2$ are two Lagrange multipliers.
Differentiating with respect to $w_0$ yields
\be
\lambda_2 = \mu_0 ~ {\lambda_1},
\ee
which by substitution in equation (\ref{eq:L}) gives
\be
\label{eq:Ls}
{\cal L} = C_n - \lambda_1 \left( \sum_{i=1}^N w_i~ (\mu(i) - \mu_0) - (\mu -
\mu_0) \right).
\ee
Let us now differentiate ${\cal L}$ with respect to $w_i$, $i\ge1$, we obtain
\be
n~ {w_i^*}^{n-1}~ C_n(i) - \lambda_1 (\mu(i) - \mu_0) =0,
\ee
so that
\be
{w_i^*} = {\lambda_1}^\frac{1}{n-1} ~ \left( \frac{\mu(i) - \mu_0}{n~ C_n(i)}
\right)^\frac{1}{n-1}.
\ee
Applying the normalization constraint yields
\be
w_0 +  {\lambda_1}^\frac{1}{n-1} ~\sum_{i=1}^N \left( \frac{\mu(i) -
\mu_0}{n~ C_n(i)} \right)^\frac{1}{n-1} = 1,
\ee
thus
\be
{\lambda_1}^\frac{1}{n-1} = \frac{1-w_0}{\sum_{i=1}^N \left( \frac{\mu(i) -
\mu_0}{n~ C_n(i)} \right)^\frac{1}{n-1}},
\ee
and finally
\be
w_i^* = (1-w_0)~\frac{\left( \frac{\mu(i) - \mu_0}{C_n(i)}
\right)^\frac{1}{n-1}}{\sum_{i=1}^N \left( \frac{\mu(i) -
\mu_0}{C_n(i)} \right)^\frac{1}{n-1}}.
\ee
Let us now define the portfolio $\Pi$ exclusively made of risky assets with
weights
\be
\tilde w_i = \frac{\left( \frac{\mu(i) - \mu_0}{C_n(i)}
\right)^\frac{1}{n-1}}{\sum_{i=1}^N \left( \frac{\mu(i) -
\mu_0}{C_n(i)} \right)^\frac{1}{n-1}}, ~~~~~~~ i \ge 1.
\ee
The optimal portfolio ${\cal P}$ can be split in two funds : the risk-free
asset whose weight is $w_0$ and a risky fund $\Pi$ with weight $(1-w_0)$. The
expected return of the portfolio ${\cal P}$ is thus
\be
\mu = w_0~ \mu_0 + (1-w_0) \mu_\Pi,
\ee
where $\mu_\Pi$ denotes the expected return of portofolio $\Pi$:
\be
\mu_\Pi =  \frac{\sum_{i=1}^N \mu(i)~ \left( \frac{\mu(i) - \mu_0}{C_n(i)}
\right)^\frac{1}{n-1}}{\sum_{i=1}^N \left( \frac{\mu(i) -
\mu_0}{C_n(i)} \right)^\frac{1}{n-1}}.
\ee
The risk associated with ${\cal P}$ and measured by the cumulant $C_n$
of order $n$ is
\be
C_n = (1-w_0)^n ~  \frac{\sum_{i=1}^N C_n(i)~ \left( \frac{\mu(i) -
\mu_0}{C_n(i)} \right)^\frac{n}{n-1}}{\left[\sum_{i=1}^N \left( \frac{\mu(i) -
\mu_0}{C_n(i)} \right)^\frac{1}{n-1} \right]^n}~.
\ee
Putting together the three last equations allows us to obtain the equation
of the efficient frontier:
\be
\mu = \mu_0 + \left[ \sum \frac{(\mu(i) - \mu_0)^\frac{n}{n-1}}{C_n(i)
^\frac{1}{n-1}}  \right]^\frac{n-1}{n} \cdot {C_n}^\frac{1}{n},
\ee
which is a straight line in the plane ${(C_n}^{1/n},~ \mu)$.

\subsection{General case}

Let us now consider the more realistic case when the risky assets are
dependent and/or when the risk is measured by any risk measure 
$\rho_\alpha$ obeying the
axioms I-IV presented in section \ref{sec:cum}, where $\alpha$ denotes the
degres of homogeneity of $\rho_\alpha$. Equation (\ref{eq:Ls}) always holds
(with $C_n$ replaced by $\rho_\alpha$), and the differentiation with respect to
$w_i$, $i \ge 1$ yields the set of equations:
\be \label{eq:EP}
\frac{\partial \rho_\alpha}{\partial w_i}(w_1^*, \cdots, w_N^*) = \lambda_1 ~
(\mu(i) - \mu_0),~~~~~~~i \in\{ 1, \cdots, N\}.
\ee
Since $\rho_\alpha(w_1, \cdots, w_N)$ is a homogeneous function of order
$\alpha$, its first-order derivative with respect to $w_i$ is also a
homogeneous function of order $\alpha-1$. Using this homogeneity
property allows us to write
\bea
{\lambda_1}^{-1} ~ \frac{\partial \rho_\alpha}{\partial w_i}(w_1^*, \cdots,
w_N^*) &=& (\mu(i) - \mu_0),~~~~~~~i \in\{ 1, \cdots, N\}, \\
\frac{\partial \rho_\alpha}{\partial w_i} \left(
{\lambda_1}^{-\frac{1}{\alpha-1}} w_1^*, \cdots,
{\lambda_1}^{-\frac{1}{\alpha-1}} w_N^* \right) &=& (\mu(i) - \mu_0),~~~~~~~i
\in\{ 1, \cdots, N\}~. \eea
Denoting by $\{\hat w_1, \cdots, \hat w_N\}$ the solution of
\be
\frac{\partial \rho_\alpha}{\partial w_i}(\hat w_1, \cdots, \hat w_N) =
(\mu(i) - \mu_0),~~~~~~~i \in\{ 1, \cdots, N\},
\ee
this shows that the optimal weights are
\be
w_i^* = {\lambda_1}^\frac{1}{\alpha-1} \hat w_i.
\ee
Now, performing the same calculation as in the case of independent risky
assets, the efficient portfolio ${\cal P}$ can be realized
by investing a weight $w_0$ of the initial wealth in the risk-free asset and
the weight $(1-w_0)$ in the risky fund $\Pi$, whose weights are given by
\be
\tilde w_i = \frac{\hat w_i}{\sum_{i=1}^N \hat w_i}~.
\label{mgjllss}
\ee

Therefore, the expected return of every efficient portfolio is
\be
\mu = w_0 \cdot \mu_0 + (1-w_0) \cdot \mu_\Pi,
\ee
where $\mu_\Pi$ denotes the expected return of the market portfolio $\Pi$,
while the risk, measured by $\rho_\alpha$
is
\be
\rho_\alpha = (1- w_0)^\alpha \rho_\alpha(\Pi)~,
\ee
so that
\be
\label{eq:eff}
\mu = \mu_0 + \frac{\mu_\Pi - \mu_0}{{\rho_\alpha(\Pi)}^{1/\alpha}} ~
{\rho_\alpha}^{1/\alpha}~. \ee
This expression is the natural generalization of the relation obtained by
\cite{Markovitz} for mean-variance efficient portfolios.

\newpage

\section{Composition of the market portfolio}
\label{app:MP}
In this appendix, we derive the relationship between the composition of the
market portfolio and
the composition of the optimal portfolio $\Pi$ obtained by the
minimization of the risks measured by $\rho_\alpha(n)$.

\subsection{Homogeneous case}
We first consider a homogeneous market, peopled with agents
choosing their optimal portfolio with respect to the same risk measure
$\rho_\alpha$. A given agent $p$ invests a fraction $w_0(p)$ of his wealth
$W(p)$ in the risk-free asset and a fraction $1-w_0(p)$ in the optimal
portfolio $\Pi$. Therefore, the total demand $D_i$ of asset $i$ is the
sum of the demand $D_i(p)$ over all agents $p$ in asset $i$:
\bea
D_i &=& \sum_p D_i(p)~,\\
&=& \sum_p W(p) \cdot (1- w_0(p)) \cdot \tilde w_i~,\\
&=& \tilde w_i \cdot \sum_p W(p) \cdot (1- w_0(p))~,
\eea
where the $\tilde w_i$'s are given by (\ref{mgjllss}).
The aggregated demand $D$ over all assets is
\bea
D &=& \sum_i D_i,\\
&=& \sum_i \tilde w_i \cdot \sum_p W(p) \cdot (1- w_0(p)),\\
&=& \sum_p W(p) \cdot (1- w_0(p)).\\
\eea

By definition, the weight of asset $i$, denoted by $w_i^m$, in the
market portfolio equals the ratio of its capitalization (the supply $S_i$ of
asset $i$) over the total capitalization of the market $S=\sum
S_i$. At the equilibrium, demand equals supply, so that
\be
w_i^m = \frac{S_i}{S} = \frac{D_i}{D} = \tilde w_i.
\ee
Thus, at the equilibrium, the optimal portfolio $\Pi$ is the market
portfolio.

\subsection{Heterogeneous case}
We now consider a heterogenous market, defined such that the agents
choose their optimal portfolio with respect to
different risk measures. Some of them choose the usual mean-variance
optimal portfolios, others prefer any mean-$\rho_\alpha$ efficient portfolio,
and so on. Let us denote by $\Pi_n$ the mean-$\rho_\alpha(n)$ optimal
portfolio made only of risky assets. Let $\phi_n$ be the fraction of
agents who choose the mean-$\rho_\alpha(n)$ efficient portfolios. By
normalization, $\sum_n \phi_n =1$. The demand $D_i(n)$ of asset $i$ from the
agents optimizing with respect to $\rho_\alpha(n)$ is
\bea
D_i(n) &=& \sum_{p \in {\cal S}_n} W(p) \cdot (1- w_0(p)) \cdot
\tilde w_i(n),\\
&=& \tilde w_i(n) \sum_{p \in {\cal S}_n} W(p) \cdot (1-
w_0(p)),
\eea
where ${\cal S}_n$ denotes the set of agents, among all the agents,
who follow the optimization stragtegy with respect to $\rho_\alpha(n)$. Thus,
the total demand of asset $i$ is
\bea
D_i &=& \sum_n {\cal N} \phi_n \cdot D_i(n),\\
&=& {\cal N} \sum_n \phi_n  \cdot \tilde w_i(n)  \sum_{p \in {\cal S}_n} W(p)
\cdot (1-w_0(p)), \eea
where ${\cal N}$ is the total number of agents. This finally yields the total
demand $D$ for all assets and for all agents
\bea
D &=& \sum_i  D_i,\\
&=& {\cal N} \sum_i    \sum_n \phi_n  \cdot \tilde w_i(n)  \sum_{p \in {\cal
S}_n} W(p) \cdot (1-w_0(p)),\\
&=& {\cal N} \sum_n \phi_n  \sum_{p \in {\cal S}_n} W(p)
\cdot (1-w_0(p)),
\eea
since $\sum_i \tilde w_i(n) = 1$, for every $n$.  Thus, setting
\be
\gamma_n = \frac{ \phi_n  \sum_{p \in {\cal S}_n} W(p)
\cdot (1-w_0(p))}{\sum_n \phi_n  \sum_{p \in {\cal S}_n} W(p)
\cdot (1-w_0(p))}~,
\ee
the market portfolio is the weighted sum of the mean-$\rho_\alpha(n)$ optimal
portfolios $\Pi_n$:
\be
w_i^m = \frac{S_i}{S} = \frac{D_i}{D} = \sum_n \gamma_n \cdot \tilde w_i(n)~.
\ee

\newpage

\section{Generalized capital asset princing model}
\label{app:GCAPM}

Our proof of the generalized capital asset princing model is similar to the
usual
demontration of the CAPM.

Let us consider an efficient portfolio ${\cal P}$. It necessarily satisfies
equation (\ref{eq:EP}) in appendix \ref{app:GEF} :
\be
\label{eq:49}
\frac{\partial \rho_\alpha}{\partial w_i}(w_1^*, \cdots, w_N^*) = \lambda_1 ~
(\mu(i) - \mu_0),~~~~~~~i \in\{ 1, \cdots, N\}.
\ee
Let us now choose any portfolio ${\cal R}$ made only of risky assets and let us
denote by $w_i({\cal R})$ its weights. We can thus write
\bea
\sum_{i=1}^N    w_i({\cal R}) \cdot \frac{\partial \rho_\alpha}{\partial
w_i}(w_1^*, \cdots, w_N^*) &=& \lambda_1 ~ \sum_{i=1}^N w_i({\cal R}) \cdot
(\mu(i) - \mu_0),\\
&=& \lambda_1 ~ (\mu_{\cal R} - \mu_0).
\eea
We can apply this last relation to the market portfolio $\Pi$, because
it is only composed of risky assets (as proved in appendix
\ref{app:GEF}). This leads to $w_i({\cal R}) = w_i^*$ and $\mu_{\cal R} =
\mu_\Pi$,
so that
\be
\sum_{i=1}^N    w_i^* \cdot \frac{\partial \rho_\alpha}{\partial w_i}(w_1^*,
\cdots, w_N^*) = \lambda_1 ~ (\mu_\Pi - \mu_0),
\ee
which, by the homogeneity of the risk measures $\rho_\alpha$, yields
\be
\label{eq:51}
\alpha \cdot \rho_\alpha (w_1^*,\cdots, w_N^*) = \lambda_1 ~ (\mu_\Pi -
\mu_0)~. \label{mgmwl}
\ee
Substituting equation (\ref{eq:49}) into (\ref{eq:51}) allows us to obtain
\be
\mu_j-\mu_0 = \beta_\alpha^j \cdot (\mu_\Pi-\mu_0),
\ee
where
\be
\beta_\alpha^j = \frac{ \partial \left( \ln
{\rho_\alpha}^\frac{1}{\alpha} \right)}{ \partial w_j}~, \label{nhnlw}
\ee
calculated at the point $\{w_1^*, \cdots, w_N^* \}$.
Expression (\ref{mgmwl}) with (\ref{nhnlw}) provides our CAPM,
generalized  with respect to the risk measures $\rho_\alpha$.

In the case where $\rho_\alpha$ denotes the variance, the second-order centered
moment is equal to the second-order cumulant and reads
\bea
C_2 &=& w_1^* \cdot \Var[X_1] + 2 w_1^* w_2^*\cdot \Cov(X_1, X_2) +
w_2^* \cdot \Var[X_2],\\
&=& \Var[\Pi]~.
\eea
Since
\bea
\frac{1}{2} \cdot \frac{\partial C_2}{\partial w_1} &=& w_1^* \cdot
\Var[X_1] + w_2^* \cdot \Cov(X_1, X_2)~,\\
&=& \Cov(X_1, \Pi),
\eea
we find
\be
\beta = \frac{\Cov(X_1, X_\Pi)}{\Var[X_\Pi]}~,
\ee
which is the standard result of the CAPM derived from
the mean-variance theory.

\newpage

\section{Calculation of the moments of the distribution of portfolio returns}
\label{app:B}

Let us start with equation (\ref{eq2.5}) in the $2$-asset case~:
\bea
\hat P_S(k) = \frac{1}{2 \pi \sqrt{1-\rho^2}} \int dy_1 dy_2~ exp \left[
-\frac{1}{2} y^t V^{-1} y + ik \left( \chi_1 w_1 sgn(y_1)
\left| \frac{y_1}{\sqrt{2}} \right|^{q_1} +  \right. \right. {\nonumber} \\
\left. \left.  + \chi_2 w_2 sgn(y_2) \left| \frac{y_2}{\sqrt{2}}
\right|^{q_2} \right) \right]
\; .
\eea
Expanding the exponential and using the definition
(\ref{eq:defmom}) of moments, we get
\bea
M_n = \frac{1}{2 \pi \sqrt{1-\rho^2}} \int dy_1 \int  dy_2 
\sum_{p=0}^n {n \choose
p} \chi_1^p \chi_2^{n-p} w_1^p w_2^{n-p} sgn(y_1)^p \left|
\frac{y_1}{\sqrt{2}} \right|^{q_1p} \times
\nonumber \\
\times sgn(y_2)^{n-p} \left| \frac{y_2}{\sqrt{2}}
\right|^{q_2(n-p)}\, e^{-\frac{1}{2} y^t
V^{-1} y } \; .
\eea
Posing
\be
\label{eq:a1}
\gamma_{q_1 q_2}(n,p) = \frac{ {\chi_1}^p {\chi_2}^{n-p}}{2 \pi
\sqrt{1-\rho^2}} \int dy_1 dy_2 \; sgn(y_1)^p \left| \frac{y_1}{\sqrt{2}}
\right|^{q_1p} sgn(y_2)^{n-p} \left| \frac{y_2}{\sqrt{2}} \right|^{q_2(n-p)}\,
e^{-\frac{1}{2} y^t V^{-1} y } \; ,
\ee
this leads to
\be
M_n = \sum_{p=0}^n {n \choose p} w_1^p w_2^{n-p}
\gamma_{q_1 q_2}(n,p) \; .
\ee

Let us defined the auxiliary variables $\alpha$ and $\beta$ such that
\be
\left\{ \begin{array}{rlc}
\alpha &=& (V^{-1})_{11} = (V^{-1})_{22}  =  \frac{1}{1 - \rho^2} \; , \\
\beta   &=& -(V^{-1})_{12} =-(V^{-1})_{21}  =\frac{\rho}{1 - \rho^2 } \; .
\end{array} \right.
\ee
Performing a simple change of variable in (\ref{eq:a1}), we can transform
the integration such that it is defined solely within the first quadrant
($y_1 \ge 0$, $y_2 \ge 0$), namely
\bea
\label{eq:a2}
\gamma_{q_1 q_2}(n,p) ={\chi_1}^p {\chi_2}^{n-p} \frac{1+(-1)^n}{2 \pi
\sqrt{1-\rho^2}} \int_0^{+\infty} dy_1 \int_0^{+\infty} dy_2~
\left(\frac{y_1}{\sqrt{2}}\right)^{q_1p} \left(
\frac{y_2}{\sqrt{2}}\right)^{q_2(n-p)}\, e^{-\frac{\alpha}{2}   (y_1^2+y_2^2)}
\times \nonumber \\
\times  \left( e^{\beta y_1 y_2} + (-1)^p e^{-\beta y_1 y_2}
\right) \; . \eea

This equation imposes that the coefficients $\gamma$ vanish if $n$ is odd.
This leads to the vanishing of the moments of odd orders, as expected for
a symmetric distribution. Then, we expand
$e^{\beta y_1 y_2} + (-1)^p e^{-\beta y_1 y_2}$ in series. Permuting
the sum sign and the integral allows us to decouple the integrations
over the two variables $y_1$ and $y_2$:
\bea
\gamma_{q_1 q_2}(n,p) ={\chi_1}^p {\chi_2}^{n-p} \frac{1+(-1)^n}{2 \pi
\sqrt{1-\rho^2}} \sum_{s=0}^{+\infty}[1+(-1)^{p+s}] \frac{\beta^s}{s!} \left(
\int_0^{+\infty} dy_1~ \frac{y_1^{q_1p+s}}{2^\frac{q_1p}{2}} ~
e^{-\frac{\alpha}{2} y_1^2}  \right) \times \nonumber\\
\times \left(
\int_0^{+\infty}
dy_2~\frac{y_2^{q_2(n-p)+s}}{2^\frac{q_2(n-p)}{2}}~e^{-\frac{\alpha}{2}
y_1^2} \right) \; . \eea
This brings us back to the problem of calculating the same type of integrals
as in the uncorrelated case. Using the expressions of $\alpha$ and
$\beta$, and taking into account the parity of $n$ and $p$, we obtain:
\bea
\label{eq:a3}
\gamma_{q_1 q_2}(2n,2p)&=& {\chi_1}^{2p} {\chi_2}^{2n-2p}\frac{(1-\rho^2)^{q_1p
+ q_2(n - p) + \frac{1}{2}}}{\pi}\sum_{s=0}^{+\infty}
\frac{(2\rho)^{2s}}{(2s)!} \Gamma \left( q_1p+s+\frac{1}{2} \right)\times
\nonumber \\
&&\times  \Gamma \left(  q_2(n-p)+s+\frac{1}{2} \right)  \; ,\\
\label{eq:a4}
\gamma_{q_1 q_2}(2n,2p+1)&=&{\chi_1}^{2p+1} {\chi_2}^{2n-2p-1}
\frac{(1-\rho^2)^{q_1p + q_2(n - p) +
\frac{q_1-q_2+1}{2}}}{\pi}\sum_{s=0}^{+\infty}
\frac{(2\rho)^{2s+1}}{(2s+1)!}
\times \nonumber \\
   && \times \Gamma \left( q_1p+s+1+\frac{q_1}{2} \right)  \Gamma
\left(q_2(n-p)+s+1-\frac{q_2}{2} \right) \; . \eea

Using the definition of the hypergeometric functions $_2F_1$
\cite{AB}, and the relation (9.131) of \cite{Gradshteyn}, we finally obtain
\bea
\gamma_{q_1 q_2}(2n,2p) &=&  {\chi_1}^{2p} {\chi_2}^{2n-2p} \frac{\Gamma
\left( q_1p +\frac{1}{2} \right) \Gamma \left(q_2(n-p)+\frac{1}{2}
\right)}{\pi}~
{_2F_1} \left(-q_1p, -q_2(n-p) ; \frac{1}{2} ; \rho^2 \right) \; , \\
\gamma_{q_1 q_2}(2n,2p+1) &=&{\chi_1}^{2p+1} {\chi_2}^{2n-2p-1} \frac{2\Gamma
\left( q_1p+1+\frac{q_1}{2}   \right) \Gamma \left(q_2(n-p)+1-\frac{q_2}{2}
\right)}{\pi} \rho~ \times \nonumber \\
&&\times~ {_2F_1}\left( -q_1p - \frac{q_1-1}{2},  -q_2(n-p)+ \frac{q_2+1}{2} ;
\frac{3}{2} ;\rho^2 \right) \; .
\eea

In the asymmetric case, a similar calculation follows, with the sole
difference that the results
involves four terms in the integral (\ref{eq:a2}) instead of two.

\newpage

\section{Conditions under which it is possible to increase the return
and decrease large risks simultaneously}
\label{app:C}

We consider $N$ independent assets $\{1 \cdots N\}$, whose returns are denoted
by $\mu(1) \cdots \mu(N)$. We aggregate these assets in a portfolio. Let $w_1
\cdots w_N$ be their weights. We consider that short positions
are forbidden and that $\sum_i w_i=1$. The return $\mu$ of the portfolio is
\be
\mu = \sum_{i=1}^N w_i \mu(i).
\ee
The risk of the portfolio is quantified by the cumulants of the distribution of
$\mu$.

Let us denote $\mu_n^*$ the return of the portfolio evaluated for the asset 
weights
which minimize the cumulant of order $n$.

\subsection{Case of two assets}

Let $C_n$ be the cumulant of order n for the portfolio. The assets
being independent,
we have
\bea
C_n &=& C_n(1){w_1}^n + C_n(2) {w_2}^n ,\\
&=& C_n(1){w_1}^n + C_n(2) (1-w_1)^n .
\eea

In the following, we will drop the subscript $1$ in $w_1$, and only write $w$.
Let us evaluate the value $w=w^*$ at the minimum of $C_n$, $n>2$ :
\bea
\frac{d C_n}{d w} = 0 &\Longleftrightarrow& C_n(1) w^{n-1} - C_n(2)
(1-w)^{n-1}=0 ,\\
&\Longleftrightarrow& \frac{C_n(1)}{C_n(2)} = \left(
\frac{1-w^*}{w^*} \right)^{n-1}, \eea
and assuming that $C_n(1)/C_n(2)>0$, which is satisfied according to our
positivity axiom 1, we obtain
\be
w^* =
\frac{C_n(2)^{\frac{1}{n-1}}}{C_n(1)^{\frac{1}{n-1}}+C_n(2)^{\frac{1}{
n-1}}}.
\ee

This leads to the following expression for $\mu_n^*$ :
\be
\mu_n^* = \frac{\mu(1) \cdot C_n(2)^\frac{1}{n-1} + \mu(2) \cdot
C_n(1)^\frac{1}{n-1}}
{C_n(1)^\frac{1}{n-1} + C_n(2)^\frac{1}{n-1}}.
\ee

Thus, after simple algebraic manipulations, we find
\be
\mu_n^* < \mu_{n+k}^* \Longleftrightarrow (\mu(1)-\mu(2))
\left( C_n(1)^\frac{1}{n-1} C_{n+k}(2)^\frac{1}{n+k-1}
- C_n(2)^\frac{1}{n-1} C_{n+k}(1)^\frac{1}{n+k-1} \right) >0,
\ee
which concludes the proof of the result announced in the main body of the text.

\subsection{General case}
We now consider a portfolio with $N$ independent assets.
Assuming that the cumulants $C_n(i)$ have the same sign for all $i$ (according
to axiom 1), we are going to show that the minimum of $C_n$ is
obtained for a portfolio whose weights are given by
\be
\label{eq:F}
w_i = \frac{\prod_{j \neq i}^N C_n(j)^\frac{1}{n-1}}
{\sum_{j=1}^N C_n(j)^\frac{1}{n-1}} ,
\ee
and we have
\be
\mu_n^* = \frac{\sum_{i=1}^N \left(
\mu(i) \prod_{j \neq i}^N C_n(j)^\frac{1}{n-1} \right)}
{\sum_{j=1}^N C_n(j)^\frac{1}{n-1}}~.  \label{njfkalkaa}
\ee

Indeed, the cumulant of the portfolio is given by
\be
C_n = \sum_{i=1}^N C_n(i)~ w_i^n
\ee
subject to the constraint
\be
\sum_{i=1}^N w_i =1.
\ee
Introducing a Lagrange multiplier $\lambda$, the first order conditions yields
\be
 C_n(i) ~ w_i^{n-1} -\lambda = 0 , ~~~~\forall i \in \{1, \cdots, N\},
\ee
so that
\be
w_i^{n-1} = \frac{\lambda}{C_n(i)}.
\ee
Since all the $C_n(i)$ are positive, we can find a $\lambda$
such that all the $w_i$ are real and positive, which yields the 
announced result (\ref{eq:F}). From here, there is no simple condition
that ensures $\mu_n^* <\mu_{n+k}^*$. The simplest way to compare $\mu_n^*$ and
$\mu_{n+k}^*$ is to calculate diretly these quantities using the formula
(\ref{njfkalkaa}).

\newpage

\begin{table}[h]
\begin{center}

\begin{tabular}{ccccc}
\hline
$\mu$ & ${\mu_2}^{1/2}$ & ${\mu_4}^{1/4}$ & ${\mu_6}^{1/6}$ & ${\mu_8}^{1/8}$
\\ \hline
0.10\%  & 0.92\% & 1.36\%  & 1.79\%  & 2.15\%  \\
0.12\%  & 0.96\% & 1.43\%  & 1.89\%  & 2.28\%  \\
0.14\%  & 1.05\% & 1.56\%  & 2.06\%  & 2.47\%  \\
0.16\%  & 1.22\% & 1.83\%  & 2.42\%  & 2.91\%  \\
0.18\%  & 1.47\% & 2.21\%  & 2.92\%  & 3.55\%  \\
0.20\%  & 1.77\% & 2.65\%  & 3.51\%  & 4.22\%  \\
\hline
\end{tabular}

\caption{\label{table:EF}  This table presents the risk measured by
$\mu_n^{1/n}$ for n=2,4,6,8, for a given value of the expectedd (daily) return
$\mu$.} \end{center}
\end{table}

\begin{table}[h]
\begin{center}

\begin{tabular}{lrrrrr}
\hline
  & $\frac{\mu}{\mu_2^{1/2}}$ & $\frac{\mu}{\mu_4^{1/4}}$ &
$\frac{\mu}{\mu_6^{1/6}}$ & $\frac{\mu}{C_4^{1/4}}$ & $\frac{\mu}{C_6^{1/6}}$
\\ \hline
Wall Mart  & 0.0821 & 0.0555 & 0.0424 & 0.0710 & 0.0557 \\
EMC  & 0.0801 & 0.0552 & 0.0430 & 0.0730 & 0.0612 \\
Intel  & 0.0737 & 0.0512 & 0.0397 & 0.0694 & 0.0532 \\
Hewlett Packard  & 0.0724 & 0.0472 & 0.0354 & 0.0575 & 0.0439 \\
IBM  & 0.0705 & 0.0465 & 0.0346 & 0.0574 & 0.0421 \\
Merck  & 0.0628 & 0.0415 & 0.0292 & 0.0513 & 0.0331 \\
Procter \& Gamble& 0.0590 & 0.0399 & 0.0314 & 0.0510 & 0.0521 \\
General Motors  & 0.0586 & 0.0362 & 0.0247 & 0.0418 & 0.0269 \\
SBC Communication  & 0.0584 & 0.0386 & 0.0270 & 0.0477 & 0.0302 \\
General Electric  & 0.0569 & 0.0334 & 0.0233 & 0.0373 & 0.0258 \\
Applied Material  & 0.0525 & 0.0357 & 0.0269 & 0.0462 & 0.0338 \\
MCI WorldCom  & 0.0441 & 0.0173 & 0.0096 & 0.0176 & 0.0098 \\
Medtronic  & 0.0432 & 0.0278 & 0.0202 & 0.0333 & 0.0237 \\
Coca-Cola  & 0.0430 & 0.0278 & 0.0207 & 0.0335 & 0.0252 \\
Exxon-Mobil  & 0.0410 & 0.0256 & 0.0178 & 0.0299 & 0.0197 \\
Texas Instrument  & 0.0324 & 0.0224 & 0.0171 & 0.0301 & 0.0218 \\
Pfizer  & 0.0298 & 0.0184 & 0.0131 & 0.0213 & 0.0148 \\
\hline
\end{tabular}

\caption{\label{table:rar} This table presents the values of the generalized
Sharpe ratios for the set of seventeen assets listed in the first
column. The assets are ranked with respect to
their Sharpe ratio, given in the second column.
The third and fourth columns give the generalized Sharpe ratio calculated
with respect to the fourth and sixth centered moments $\mu_4$ and $\mu_6$ while
the fifth and sixth columns give the generalized Sharpe ratio calculated
with respect to the fourth and sixth cumulants $C_4$ and $C_6$.  }

\end{center}
\end{table}

\begin{table}[!h]
\begin{center}
\begin{tabular} {cccccccccc}
\hline
& \multicolumn{4}{c}{Positive Tail} & &  \multicolumn{4}{c}{Negative Tail} \\
\cline{2-5}
\cline{7-10}
&       $<\chi_+>$&     $<c_+>$&  $\chi_+$&     $c_+$& & $<\chi_->$&
$<c_->$&    $\chi_-$& $c_-$\\
\hline
CHF&    2.45&   1.61&   2.33&   1.26& &  2.34&   1.53&   1.72&   0.93\\
DEM&    2.09&   1.65&   1.74&   1.03& &  2.01&   1.58&   1.45&   0.91\\
JPY&    2.10&   1.28&   1.30&   0.76& &  1.89&   1.47&   0.99&   0.76\\
MAL&    1.00&   1.22&   1.25&   0.41&&   1.01&   1.25&   0.44&   0.48\\
POL&    1.55&   1.02&   1.30&   0.73& &  1.60&   2.13&   1.25&   0.62\\
THA&    0.78&   0.75&   0.75&  0.54& &   0.82&   0.73&   0.30& 0.38\\
UKP&    1.89&   1.52&   1.38&   0.92&&   2.00&   1.41&   1.82&   1.09\\
\hline
\end{tabular}
\end{center}
\caption{\label{tab:1} Table of the exponents $c$ and the scale
parameters $\chi$ for different currencies. The subscript ''+'' or
''-'' denotes the positive or negative part of the distribution of returns and
the terms between brackets refer to parameters estimated in the bulk
of the distribution while naked parameters refer to the tails of the
distribution.}
\end{table}

\begin{table}[!h]
\begin{center}
\begin{tabular} {lccccccccc}
\hline

& \multicolumn{4}{c}{Positive Tail} & &  \multicolumn{4}{c}{Negative Tail} \\
\cline{2-5}
\cline{7-10}
&       $<\chi_+>$&     $<c_+>$&  $\chi_+$&     $c_+$&&  $<\chi_->$&
$<c_->$&    $\chi_-$& $c_-$\\
\hline
&&&&&&&&&\\
Applied Material&   12.47&   1.82&   8.75&   0.99& &  11.94&   1.66&   8.11
&   0.98 \\
Coca-Cola&     5.38&   1.88&   4.46&   1.04& &  5.06&   1.74&   2.98&   0.78\\
EMC&    13.53&   1.63&   13.18&   1.55&&   11.44&   1.61&   3.05&   0.57  \\
General Electric&     5.21&   1.89&   1.81&   1.28& &  4.80&   1.81&   4.31&
1.16\\
General Motors&     5.78&   1.71&   0.63&   0.48& &  5.32&   1.89&   2.80&
0.79\\
Hewlett Packart&    7.51&   1.93&   4.20&   0.84& &  7.26&   1.76&   1.66&
0.52\\
IBM&    5.46&   1.71&   3.85&   0.87& &  5.07&   1.90&   0.18&   0.33\\
Intel&   8.93&   2.31&   2.79&   0.64& &  9.14&   1.60&   3.56&   0.62\\
MCI WorldCom&9.80&      1.74&   11.01&  1.56&&   9.09&   1.56&   2.86&   0.58\\
Medtronic&    6.82&   1.95&   6.09&   1.11&&   6.49&   1.54&   2.55&   0.67\\
Merck&    5.36&   1.91&   4.56&   1.16&&   5.00&   1.73&   1.32&   0.59\\
Pfizer&    6.41&   2.01&   5.84&   1.27& &  6.04&   1.70&   0.26&   0.35\\
Procter \& Gambel&     4.86&   1.83&   3.53&   0.96& &  4.55&   1.74&   2.96&
0.82\\
SBC Communication&    5.21&   1.97&   1.26&   0.59&&   4.89&   1.59&   1.56&
0.60\\
Texas Instrument&    9.06&   1.78&   4.07&   0.72&&   8.24&   1.84&   2.18&
0.54\\
Wall Mart&    7.41&   1.83&   5.81&   1.01& &  6.80&   1.64&   3.75&   0.78\\
\hline
\end{tabular}
\end{center}
\caption{\label{tab:2} Table of the exponents $c$ and the scale
parameters $\chi$ for different stocks. The subscript ''+'' or
''-'' denotes the positive or negative part of the distribution and
the terms between brackets refer to parameters estimated in the bulk
of the distribution while naked parameters refer to the tails of the
distribution.}
\end{table}

\begin{table}[h]
\begin{center}

\begin{tabular}{lcccccc}
\hline
   & {\textbf Mean ($10^{-3}$)} & {\textbf Variance ($10^{-3}$)} & {\textbf
Skewness} & {\textbf Kurtosis} & {\textbf min} & {\textbf max} \\ \hline
   &  &  &  &  &  &  \\
Applied Material & 2.11 & 1.62 & 0.41 & 4.68 & -14\% & 21\% \\
Coca-Cola & 0.81 & 0.36 & 0.13 & 5.71 & -11\% & 10\% \\
EMC & 2.76 & 1.13 & 0.23 & 4.79 & -18\% & 15\% \\
Exxon-Mobil & 0.92 & 0.25 & 0.30 & 5.26 & -7\% & 11\% \\
General Electric & 1.38 & 0.30 & 0.08 & 4.46 & -7\% & 8\% \\
General Motors & 0.64 & 0.39 & 0.12 & 4.35 & -11\% & 8\% \\
Hewlett Packard & 1.17 & 0.81 & 0.16 & 6.58 & -14\% & 21\% \\
IBM & 1.32 & 0.54 & 0.08 & 8.43 & -16\% & 13\% \\
Intel & 1.71 & 0.85 & -0.31 & 6.88 & -22\% & 14\% \\
MCI WorldCom & 0.87 & 0.85 & -0.18 & 6.88 & -20\% & 13\% \\
Medtronic & 1.70 & 0.55 & 0.23 & 5.52 & -12\% & 12\% \\
Merck & 1.32 & 0.35 & 0.18 & 5.29 & -9\% & 10\% \\
Pfizer & 1.57 & 0.46 & 0.01 & 4.28 & -10\% & 10\% \\
Procter\&Gambel & 0.90 & 0.41 & -2.57 & 42.75 & -31\% & 10\% \\
SBC Communication & 0.86 & 0.39 & 0.06 & 5.86 & -13\% & 9\% \\
Texas Instrument & 2.20 & 1.23 & 0.50 & 5.26 & -12\% & 24\% \\
Wall Mart & 1.35 & 0.52 & 0.16 & 4.79 & -10\% & 9\% \\
\hline
\end{tabular}

\caption{\label{table:stat} This table presents the main statistical features
of the daily returns of the set of seventeen assets studied here
over the time interval from the end
of January 1995 to the end of December 2000.} \end{center}
\end{table}

\newpage

\clearpage

\clearpage

\clearpage

\clearpage

\newpage

\newpage

\newpage

\newpage

\newpage

\newpage

\newpage

\newpage

\newpage

\newpage

\newpage

\newpage

\newpage

\clearpage

\clearpage

\clearpage

\end{document}